\documentclass[12pt]{article}
\PassOptionsToPackage{hyphens}{url}
\usepackage[margin=1in]{geometry}
\usepackage[T1]{fontenc}
\usepackage[utf8]{inputenc}
\usepackage{newtxtext,newtxmath}
\usepackage{amsmath}
\usepackage{setspace}
\usepackage{booktabs,longtable,array}
\usepackage{threeparttable}
\usepackage{tabularx}
\usepackage{adjustbox}
\usepackage{pdflscape}
\usepackage{graphicx}
\usepackage{placeins}
\newcommand{\notesize}{\footnotesize}
\usepackage{caption}
\usepackage{float}

\usepackage{xcolor}
\usepackage{microtype}
\usepackage{fvextra}
\usepackage{tikz}
\usetikzlibrary{arrows.meta,positioning}
\usepackage[authoryear,round]{natbib}
\setcitestyle{aysep={}}

\usepackage[hidelinks]{hyperref}
\graphicspath{{figures/}}
\numberwithin{equation}{section}
\begin{document}

\hypersetup{pageanchor=false}
\begin{titlepage}
\renewcommand{\thefootnote}{\fnsymbol{footnote}}
\begin{center}
{\Large\bfseries Advancing AI Capabilities and Evolving Labor Outcomes\footnote{We thank Bharat Chandar, Jeff Campbell, Anton Korinek, Cassandra Merritt, and participants of the CEPR Paris Symposium, the Wharton AI and the Future of Work Conference, Meta's Generative AI, Labor, and Upskilling Workshop, the Korean Economic Association Annual Conference, and the University of Notre Dame Applied Microeconomics Brownbag for helpful comments. Lee is grateful for financial support from the Notre Dame Ethics Initiative.}\footnote{AI use statement: The occupational AI exposure scores are produced by three large language models, Claude Haiku 4.5 (Anthropic), GPT-5.6-luna (OpenAI), and Gemini 3.1 Flash-Lite (Google) by rating occupational tasks. The prompts are in Appendix A. The authors used Claude Code (Fable models) to write the Python scripts that assemble the tables and create figures, and to produce the replication package and repository on GitHub (\href{https://github.com/folablab/aaicelo-replication}{github.com/folablab/aaicelo-replication}). In the writing process, Claude Code was used for converting and formatting drafts to LaTeX, putting together the Appendix, and editing text. The authors take responsibility for all text, code, and results.}\par} 
\vspace{1.8em}
{\large
Jacob Dominski\footnote{Department of Economics, University of Maryland, College Park; dominski@umd.edu.} \qquad
Tyler Boes\footnote{Department of Economics, University of Notre Dame; tboes2@nd.edu.} \qquad
Yong Suk Lee\footnote{Corresponding author. Keough School of Global Affairs, University of Notre Dame; yong.s.lee@nd.edu.}
\par}
\vspace{1.2em}
{November 2026\par}
\vspace{1.8em}
\begin{minipage}{0.88\textwidth}
\begin{spacing}{1}
\small
\noindent\textbf{Abstract.}
This paper examines how labor market outcomes changed across occupations as AI capabilities advanced from chatbots to reasoning models and AI agents.
We construct an occupational AI exposure score for each capability stage and relate its change to the Current Population Survey through June 2026.
In aggregate, occupations whose exposure rose more did not lose employment or see higher unemployment over three years.
However, the number of workers under 30 in these occupations fell by about 8 percent per standard deviation of the exposure change, and the decline is concentrated among young workers without a college degree and in office and customer service occupations rather than in technical occupations.
Their unemployment did not rise, consistent with less entry rather than more layoffs.
College graduates aged 30 to 49 gained full-time positions in the same occupations, so the workforce in exposed occupations became older and more educated.
The same comparisons for 2017 to 2022 show no such patterns.

\vspace{1em}


\end{spacing}
\end{minipage}
\end{center}
\end{titlepage}
\hypersetup{pageanchor=true}
\setcounter{footnote}{0}

\section{Introduction}
\label{sec:introduction}

Whether AI is replacing jobs, and if so whose jobs it is replacing, are questions that have been hotly debated since the release of ChatGPT.
Research on the topic has expanded rapidly, and public concern and interest remain high.
Nearly four years after the public release of generative AI chatbots, aggregate employment has not fallen in the way that many had feared.
Nonetheless, concerns that AI will displace workers persist.
The most consistent finding so far is that employment of young workers in AI-exposed occupations has declined \citep{brynjolfsson2026canaries, hosseinimaasoum2025seniority, kleinteeselink2025uk, orr2026graduating, atkinson2026young}.
However, whether the decline can be attributed to AI is contested \citep{imas2026has, emanuel2026remote, lambert2026ladder, frank2026deteriorated, fairlie2026graduates, kauhanen2026finland, facius2026consequences, hernaes2026norway}.

\begin{figure}[t]
\centering
\includegraphics[width=\linewidth]{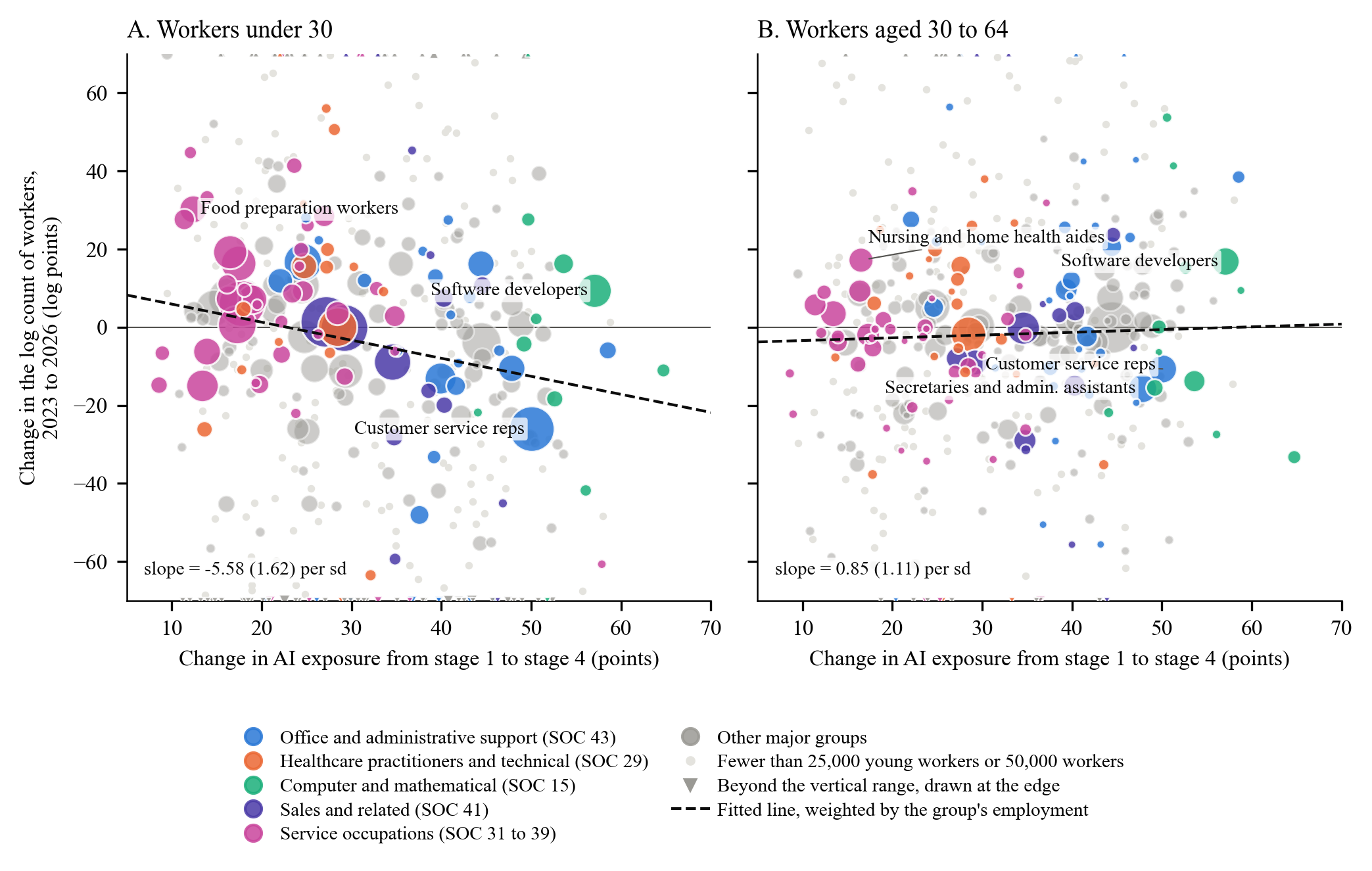}
\caption{Change in employment and change in AI exposure across occupations, 2023 to 2026}
\label{fig:intro-raw}
\begin{minipage}{0.96\textwidth}
\notesize
Notes: Each bubble is an occupation. The horizontal axis is the change in the occupation's AI exposure score from stage 1, before generative AI, to stage 4, AI agents, in points on the 100-point scale (Section~\ref{sec:measurement}); the vertical axis is the change in the log count of the group's workers in the occupation between the first half of 2023 and the first half of 2026, in log points, 2026Q1 vs.\ 2023Q1 and 2026Q2 vs.\ 2023Q2 averaged (Section~\ref{sec:data}). Bubble area is proportional to the group's employment in the occupation in the base period. Colors mark four major groups of the Standard Occupational Classification, office and administrative support (SOC 43), healthcare practitioners and technical (29), computer and mathematical (15), and sales and related (41), and the service occupations (31 to 39); the other major groups are gray. The dashed line is the fitted line weighted by that employment. Labels mark the three occupations that contribute most to the slope in each panel, in absolute value, and software developers and customer service representatives in both panels; occupations with fewer than 25,000 young workers or 50,000 workers are small gray dots, and occupations beyond the vertical range are triangles at the edge.
\end{minipage}
\end{figure}

This paper aims to shed light on this debate by examining how labor market outcomes have evolved across occupations as AI capabilities advanced, from early 2023, soon after the release of ChatGPT, through the first half of 2026.
Figure~\ref{fig:intro-raw} plots, for each occupation, the change in employment in the Current Population Survey over this period against the change in the occupation's AI exposure as capabilities advanced.
Among workers under 30, employment fell more in the occupations whose exposure rose more, whereas among workers aged 30 to 64 there is no such relation.

However, attributing such a pattern to AI is difficult for several reasons.
The first is that the technology is advancing rapidly.
New models and capabilities have been released every few months, so any assessment of what AI can do at one point in time is soon out of date.
The second is measurement.
AI is software, delivered through many models and products, and its use and diffusion are hard to quantify.
Most representative labor statistics do not record whether a worker uses AI, and the surveys that do take time to collect, so the available evidence often describes a technology and a labor market that have already moved on.

The third is that many other forces have shaped the labor market in recent years.
The COVID-19 pandemic was followed in 2021 and 2022 by a hiring boom and a wave of quits \citep{autor2023compression}, a surge in inflation \citep{bernanke2025inflation}, and the sharpest rise in interest rates in decades.
The technology sector, which had expanded rapidly during the pandemic, began to contract in late 2022, just as ChatGPT was released, and the labor market for AI-exposed jobs had started to weaken earlier that year \citep{frank2026deteriorated}.
All of these affected employment, hours, and wages across occupations, which makes isolating the effect of AI challenging.

We address these challenges in three ways.
First, we construct an occupational AI exposure score that evolves with AI capabilities.
Although models are updated continually, the advances since the era of machine learning can be grouped into three breakthroughs: first-generation large language models such as ChatGPT (hereafter, chatbots), reasoning models, and agentic AI \citep{korinek2025agents, bengio2026safety, metz2024openai}.
For every occupation, we measure the share of its work that AI could perform at each stage, using task descriptions in O*NET and language models from three leading AI companies.
We then relate changes in an occupation's labor market outcomes to the increase in its exposure since before the release of chatbots.
This change in exposure strongly correlates with the occupation-level adoption rates that workers report in the Real-Time Population Survey of \citet{bick2026work}.
Unlike adoption, however, exposure can be measured for every occupation as soon as a new model is released, and it reflects what the technology can do, which is decided by AI developers, rather than the choices of the workers and firms whose outcomes we study.

Second, we use the Current Population Survey (CPS), the monthly labor force survey of the U.S. Census Bureau and the Bureau of Labor Statistics, whose microdata are released within weeks of collection.
We focus on the period from the first quarter of 2023, soon after the release of ChatGPT, through the second quarter of 2026, which is long enough to observe how the labor market has changed as AI has spread.

Third, we address the other forces that have shaped the labor market.
Our comparisons are across occupations within the same period and control for the composition of each occupation's workforce, so changes that affected all occupations alike do not enter our estimates, although changes that affected occupations differently could.
Hence, we estimate the same regressions on the years 2017 to 2022 as a placebo test, and we examine the changes since 2022 that could line up with the change in exposure, such as remote work and the contraction of the technology sector.
Our estimates are differences in how outcomes changed between occupations whose exposure rose more and occupations with similar workforces whose exposure rose less, and we do not claim that they identify the causal effect of AI.

Our empirical analysis uses two complementary designs.
The first compares changes over time.
We relate the change in an occupation's outcomes over three years to the increase in its AI exposure, controlling for the age, gender, race, education, and region of its workforce in the base period, so that the comparison is between occupations that employ similar workers but differ in how much their exposure grew.
With more than three years since the release of ChatGPT, the three-year horizon would allow the margins of adjustment within an occupation to play out.
The second design uses the timing of each capability jump.
We examine each outcome against AI exposure in each quarter with occupation and time fixed effects, so that the estimates come from changes within an occupation as AI moves from one stage to the next.
We estimate this model at the occupation level and the individual level. The two lets us examine changes in the composition of an occupation's workforce as well as changes in the outcomes of a given type of worker.
We examine not only the number of workers in an occupation but also the unemployment rate, hours, part-time status, and earnings, since firms may adjust these before headcount.
For young workers in particular, examining both the number of workers in an occupation and the unemployment rate, which conditions on having held a job in the occupation, indicates whether the decline came through unemployment.

What do we find?
Over the three years from the first half of 2023 to the first half of 2026, occupations whose AI exposure increased more did not lose employment and did not see higher overall unemployment relative to occupations whose exposure increased less. 
We also find no change in the years before generative AI.

However, underlying the stable aggregate, there are substantial changes across groups of workers.
The number of workers under 30 in exposed occupations fell by about 8 percent for each standard deviation increase in exposure, where one standard deviation is about 12 points on the 100-point scale, roughly the gap between the average office and administrative occupation and the average computer occupation.
The decline is largest in the most recent periods.
The three-year windows ending in 2025 give a decline of 3 to 5 percent, and the windows ending in 2026 give about 8 percent.
The occupation fixed-effects design, which uses every quarter from 2023 to 2026, gives about 5 percent.
The decline is concentrated among young workers without a college degree.
The decline is not accompanied by higher unemployment among young workers who have held jobs in these occupations, which is consistent with fewer young workers entering these occupations rather than with incumbent workers being laid off.
In contrast, college graduates aged 30 to 49 in the same occupations gained full-time employment.
A decomposition of the under-30 estimate by occupation shows that the decline comes mostly from office and administrative support occupations, and from customer service work in particular, relative to healthcare and service occupations, whose exposure rose little and where the number of young workers held up.
The technical occupations in which AI capabilities advanced the most, and which have drawn the most public attention, contribute little to the estimate.

The change in who works in exposed occupations also shows up in other margins.
Among full-time workers, the share under 30 fell by about 1.6 percentage points and the share with a bachelor's degree rose by about 1.8 percentage points per standard deviation.
Part-time work, already less common in exposed occupations, declined further, by about 0.7 percentage points, among workers aged 30 and over.

Our paper contributes to several strands of the literature.
The first measures occupational exposure to AI, from progress on AI benchmarks \citep{felten2023language}, from assessments of whether language models can perform occupational tasks \citep{eloundou2024gpts, gmyrek2025refined}, and more recently from usage data on AI platforms \citep{handa2025tasks, tomlinson2025working, massenkoff2026labor}.
Several of these measures are built to reflect change in the technology \citep{engberg2024unboxed, oecd2026exposure}.
\citet{yin2026unstable} show that exposure scores differ substantially across the models that produce them and caution against treating them as fixed.
Our measure and analysis differ in a few ways.
First, our measure combines three ratings of each task, how much time AI could save, whether AI could complete the task without a person, and the share of the task that remains once the work on which people retain a comparative advantage is set aside, averaged into one score from language models of three AI companies.
Second, we rate the same tasks against dated stages of AI capability, which gives us variation within occupations over time, so that we can ask whether outcomes change when an occupation's exposure increases, and at different capabilities. 
Relatedly, this allows us to examine labor market changes in long differences which is better suited to capture general equilibrium effects of shocks at more aggregate levels, as in trade shocks \citep{autor2013china} or technology shocks \citep{graetz2018robots, acemoglu2020robots, chung2023robots}.
Because our exposure measure can be extended to each new generation of models, it can be updated as the technology and the labor market continues to evolve and be adopted in different countries and regions.

The second strand examines the early labor market effects of generative AI.
Several studies find little aggregate effect on employment, unemployment, or earnings in exposed occupations \citep{chandar2025tracking, gimbel2025evaluating, humlum2025small, hartley2026effects, massenkoff2026labor}.
\citet{chandar2025tracking} further finds that, within the most exposed quartile, occupations with a high share of college graduates, such as software development, grew while those with a low share, such as customer service, declined, in the CPS through early 2025.
Others find sizable employment declines among young workers in exposed occupations, in payroll data \citep{brynjolfsson2026canaries}, in résumé and job posting data \citep{hosseinimaasoum2025seniority, kleinteeselink2025uk}, and in administrative records on recent graduates \citep{orr2026graduating}.
However, this evidence is contested, as \citet{fairlie2026graduates} find no rise in unemployment among recent college graduates through the summer of 2026, administrative data from the Nordic countries show little effect on young workers \citep{kauhanen2026finland, facius2026consequences, hernaes2026norway}, and others argue that the decline began before ChatGPT or reflects the rise of remote work \citep{frank2026deteriorated, emanuel2026remote, lambert2026ladder}.
Our results speak to this debate in four ways.
First, in nationally representative US data through mid-2026, the number of young workers in exposed occupations has fallen, while total employment shows no significant change.
The total employment estimate is about what the decline among young workers alone implies, since they are about a quarter of employment in these occupations.
College graduates aged 30 to 49 gained full-time positions in the same occupations, but unemployment among graduates also rose by about half a percentage point, which points to more churn among graduates rather than only more hiring.
Second, the decline is concentrated among young workers without a college degree and is not accompanied by higher unemployment, which is consistent with \citet{fairlie2026graduates} and with evidence that the adjustment operates through reduced entry rather than separations \citep{atkinson2026young, brynjolfsson2026canaries}.
In particular, our findings shift attention from recent college graduates, who have received considerable public attention, to young workers without a degree.
The placebo years before generative AI show no such pattern.
Third, the decline comes from clerical and customer-facing occupations rather than from the technical occupations in which AI capabilities advanced the most. 
This is consistent with the divergence that \citet{chandar2025tracking} documents within the most exposed quartile, and we extend it with data through mid-2026, the split by age and education, and a decomposition that attributes the decline in young workers to individual occupations.
Fourth, we test the remote-work explanation directly with the CPS telework questions.
Holding fixed the change in each occupation's telework share over the period leaves the decline among young workers almost unchanged.
Holding fixed its level as well reduces the decline by up to about half and makes it less precise, but at least half of the decline remains.

The third strand concerns the division of labor between less and more experienced workers or different levels of expertise.
The task framework \citep{autor2003skill, acemoglu2018race, acemoglu2022tasks} implies that the effect of a new technology on employment in an occupation is ambiguous, because the displacement of workers from some tasks is offset by higher productivity in others.
Recent papers of expertise and knowledge work predict that AI affects less experienced workers first, since the tasks it can perform are the ones on which they learn the job \citep{autor2025expertise, ide2025knowledge, garicano2025training}.
Our findings of stable aggregate employment, fewer young workers, and gains for experienced and educated workers are consistent with these predictions.
We also add evidence on part-time work, margins that could be predicted to adjust first and on which the existing evidence is limited and mixed \citep{jiang2025workday, humlum2025small, chung2023robots}.

The paper proceeds as follows.
Section~\ref{sec:measurement} describes the exposure score, Section~\ref{sec:data} the CPS data and the empirical strategy, Section~\ref{sec:findings} the results for all workers and by age and education, Section~\ref{sec:robustness} alternative explanations, and Section~\ref{sec:conclusion} concludes.

\section{Measuring AI Capability Change by Occupation}
\label{sec:measurement}

To examine how AI affects the labor market, researchers have constructed AI exposure scores that capture the extent to which AI can perform the tasks that make up an occupation.
An occupation can be seen as a bundle of tasks \citep{autor2003skill}, where each task differs in how important it is to the occupation and in the degree to which AI can perform it, either by completing the task or by reducing the time it takes.
We construct our own AI exposure score, and we use language models as the raters, for the reasons set out in Appendix A.

Exposure measures the potential of AI to perform the tasks of an occupation.
It is a statement about capability, not a measure of adoption, which depends on the beliefs and choices of workers and firms.
Exposure by itself does not say whether AI displaces workers or complements them.
If AI can do all of a worker's tasks it may replace the worker, but if it can do only some, the worker may concentrate on the others.
It could also depend on how those tasks are allocated to workers of different ages, experience, and skill.
Younger and less experienced workers tend to concentrate on a narrower set of tasks, whereas experienced workers cover a wider span of tasks, some of which AI can readily do and some of which still require human judgment or are reserved for people by institutional rules.
Hence, the same exposure score can mean different things for different workers even within the same occupation.
In short, AI exposure is a measure of capability and potentiality. How AI exposure translates into employment, hours, and wages is an empirical question which this paper examines.

\subsection{Constructing the AI Exposure Score}
\label{sec:construction}

We first define four stages of AI capability and map each to a date.
Although models have been updated and released frequently in the nearly four years since ChatGPT, three breakthroughs characterize the technology so far: first-generation large language models such as the earlier versions of ChatGPT and Claude (henceforth, chatbots), reasoning models, and agentic AI \citep{korinek2025agents, bengio2026safety}.
We treat each as a jump in capability, with pre-generative machine learning as the baseline.
We assign two dates to each stage, the release date, when a leading AI company first released a widely available system of that kind, and the inflection date, when those capabilities started to be widely used at work.
The inflection dates are March 2023 for chatbots, January 2025 for reasoning models, and January 2026 for AI agents.
Figure~\ref{fig:ratermeans} lists the four stages with both dates, and Appendix~\ref{app:stages} gives the evidence behind them.

Second, we construct and rate subtasks rather than whole tasks.
The O*NET database \citep{onet2019database} lists 19,281 tasks across nearly 900 occupations, and the literature has taken these tasks as the unit of analysis, but many tasks contain work of different kinds.
For example, one task for tax preparers is ``prepare or assist in preparing simple to complex tax returns for individuals or small businesses.'' This would involve gathering and reviewing the client's documents, computing taxable income, entering the data, reviewing the return, communicating with the client, and authorizing the return for submission. AI may handle some of these better than others.
Hence, we first split each task into three to six subtasks, about five on average, and assign each subtask a share of the task's time, which yields 99,157 subtasks.
The subtasks are generated once, by one language model, and held fixed, so that every model at every stage rates the same subtasks and the variation in the scores comes from the models and the stages rather than from the subtask construction.\footnote{Appendix A describes the prompt, the activity type attached to each subtask, and their distributions, and compares the subtask-based score with scoring each task as a whole; the whole-task scores are lower and coarser, rank tasks similarly, and track adoption slightly less well.}

Third, we score each subtask in three ways.
The first asks how much of the time currently spent on the subtask AI at the given stage could save, and the time-weighted average over a task's subtasks gives \emph{the percent of the task's time that AI could save}, $A_t(s)$.
The second asks whether AI at that stage could complete the subtask to a professional standard without a person, and the time-weighted average of the yes-or-no answers, scaled to 100, gives \emph{the share of the task that AI could complete on its own}, $B_t(s)$.
The third asks whether a person retains the advantage on the subtask because it requires physical presence, interpersonal trust, institutional or legal authority, or tacit expert judgment, and the time-weighted average gives \emph{the share of the task on which people maintain an advantage}, $C_t(s)$.
For the first two questions the model answers twice, once for the capability of the technology alone and once taking into account the economic, regulatory, and organizational constraints on using AI for the subtask, and we use the more conservative answer.
In practice, each task-level rating is the time-weighted average of the subtask ratings:
\begin{equation}
X_t(s) = \sum_{i} \omega_{it}\, x_{it}(s), \qquad \sum_i \omega_{it} = 1,
\label{eq:taskrating}
\end{equation}
where $x_{it}(s)$ is one of the three ratings for subtask $i$ of task $t$ at stage $s$, $\omega_{it}$ is the subtask's share of the task's time, $X$ stands for $A$ (time saved), $B$ (completion), or $C$ (human advantage), and yes-or-no ratings are scaled to 100.
The task's exposure score is the average of the three ratings, with the complement of $C$ so that every component increases with capability:
\begin{equation}
E_t(s) = \tfrac{1}{3}\left[ A_t(s) + B_t(s) + \bigl(100 - C_t(s)\bigr) \right]
\label{eq:composite}
\end{equation}
The models we use are Claude Haiku 4.5 (Anthropic), GPT-5.6-luna (OpenAI), and Gemini 3.1 Flash-Lite (Google), chosen after a pilot of seven models on 1,000 tasks.
Each model rates every subtask against the same prompts and stage descriptions.
Some models are systematically more generous than others about what AI can do. We average the three models' task scores, so that no one model's tendency drives the score.

Finally, an occupation's exposure score at each stage is the average of its tasks' scores, weighted by the O*NET relevance weights, i.e., the share of incumbents who report the task as part of their job:
\begin{equation}
E_o(s) = \frac{\sum_{t \in o} r_t\, \bar{E}_t(s)}{\sum_{t \in o} r_t},
\label{eq:occscore}
\end{equation}
where the bar denotes the three-model mean and $r_t$ is the task's relevance weight.\footnote{Simple, importance-weighted, and frequency-weighted averages give nearly identical scores (Appendix A).}
The O*NET occupations are crosswalked to 513 census occupation codes, and to 473 harmonized codes for the 2017 to 2026 panel, and the crosswalk covers over 99 percent of employment.
Appendix A provides the prompts, the output formats, the checks applied to the ratings, and the cost of the run, and Appendix Figure~\ref{fig:pipeline} summarizes the steps.

\subsection{What the Scores Show}
\label{sec:scores}

The three models agree closely on the ranking of tasks and occupations.
The correlations of the score between pairs of models are 0.87 to 0.91 at the task level and above 0.9 at the occupation level, and rerunning one model on 1,000 tasks gives task-level scores that correlate 0.83 to 0.98 across runs, which implies a reliability above 0.98 for the occupation scores.
However, there is considerable variance in the level of exposure, as \citet{yin2026unstable} also find.
This matters less for our analysis, which uses the change in exposure scores.
A model that is more generous rates nearly every occupation higher, so that the ranking of occupations and the gaps between them change little.
Our regressions examine whether occupations with larger increases in exposure had different outcomes from occupations with smaller increases, and hence depend mainly on how occupations are ordered by their change in exposure rather than on the level of exposure.

\begin{figure}[htbp]
\centering
\includegraphics[width=\linewidth]{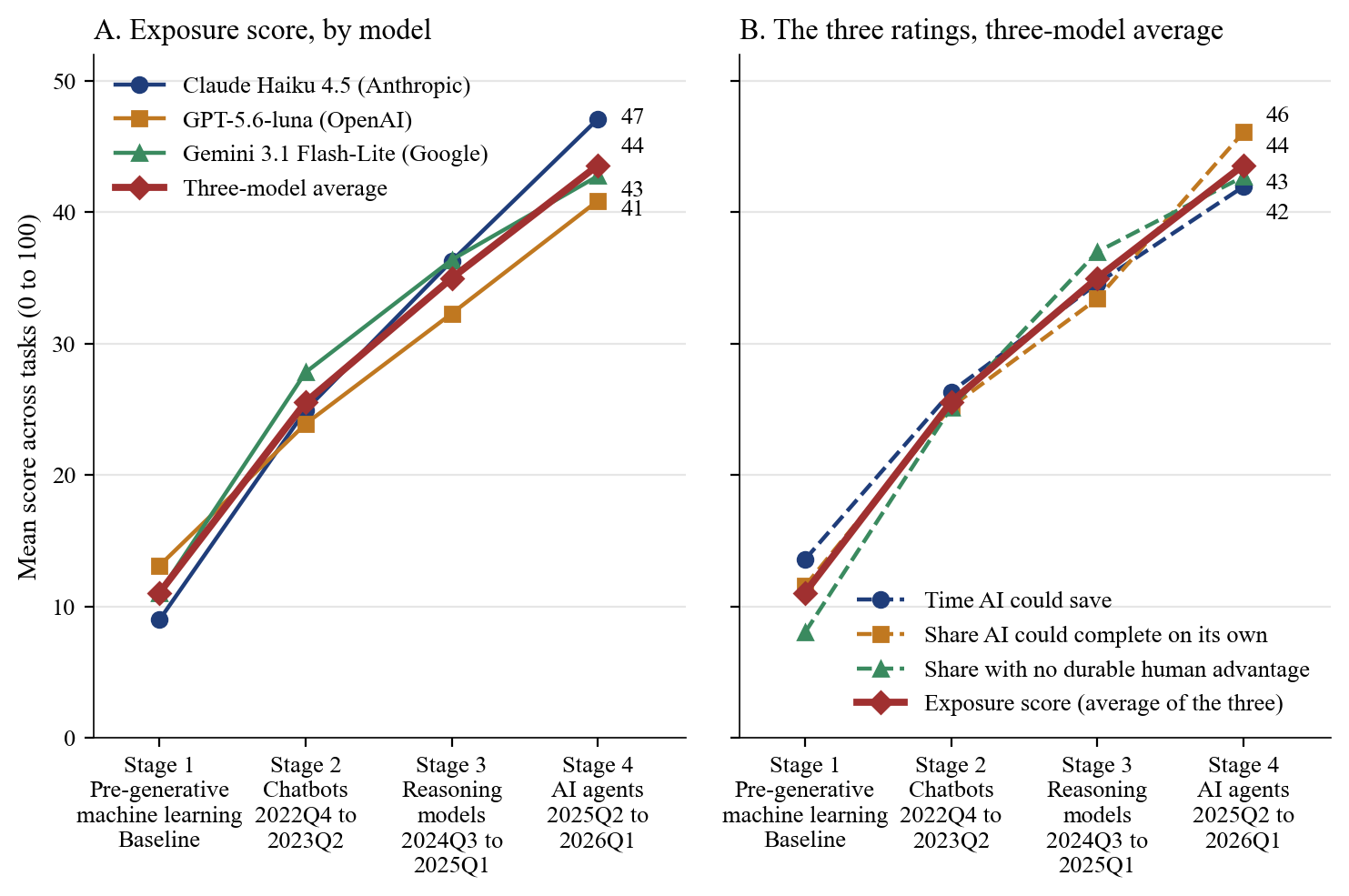}
\caption{The four capability stages: mean exposure by model and by rating, with the release and inflection dates}
\label{fig:ratermeans}
\begin{minipage}{0.92\textwidth}
\notesize
Notes: Panel A plots each model's composite exposure score and the three-model average, which is the score used in the analysis. Panel B plots, for the three-model average, the three ratings that make up the score. Under each stage are the quarters of its release and its inflection (Section~\ref{sec:construction}). The release months are November 2022, September 2024, and May 2025, and the inflection months March 2023, January 2025, and January 2026; the March 2023 inflection is assigned to 2023Q2 rather than to a quarter that was two-thirds pre-inflection.
\end{minipage}
\end{figure}

Figure~\ref{fig:ratermeans} plots each model's average exposure score at the four stages and, below it, the three ratings that make up the score, averaged over the three models.
In Panel A the three models' scores increase together across the stages but not in parallel: GPT is the highest at stage 1 and the lowest at stage 4, whereas Claude is the reverse, so the models disagree most about how much the agent generation adds.
In Panel B the human-advantage rating is the lowest at stage 1 and increases the most at stages 2 and 3, and the completion rating increases the most at stage 4.

Mean exposure across occupations rises from about 12 at stage 1 to 27 at stage 2, 36 at stage 3, and 45 at stage 4, and the standard deviation across occupations rise from 7 to 17 points. The share of tasks scoring 50 or more increases from 8 percent at stage 2 to 41 percent at stage 4.
Table~\ref{tab:topbottom} lists the ten occupations whose exposure increased the most from stage 1 to stage 4 and the ten whose exposure increased the least.
The occupations at the top are not limited to technology jobs.
Programmers and software developers appear alongside proofreaders, technical writers, medical transcriptionists, and office support workers, i.e., work that consists of reading, writing, and analysis, at wages that range from 14 dollars an hour for medical transcriptionists to 52 dollars for software developers.
The occupations at the bottom are physical work, from dishwashers and roofers to housekeepers and highway maintenance workers.

\begin{table}[htbp]
\centering
\begin{threeparttable}
\caption{The ten largest and ten smallest exposure changes}
\label{tab:topbottom}
\footnotesize\setlength{\tabcolsep}{4pt}
\begin{spacing}{1}
\begin{tabular}{@{}p{6.0cm}rrrrrr@{}}
\toprule
Occupation & $\Delta$S4$-$S1 & $\Delta$S3$-$S1 & Employment & Real wage & BA+ & AI adoption \\
\midrule
\multicolumn{7}{@{}l}{\emph{Ten largest changes}} \\
Computer programmers & 64.7 & 53.3 & 430k & \$38 & 75\% & 81\% \\
Proofreaders and copy markers & 59.2 & 51.1 & 6k & \$22 & 78\% & -- \\
Database administrators & 58.8 & 44.8 & 100k & \$43 & 77\% & 64\% \\
Miscellaneous office and administrative support workers, including desktop publishers & 58.5 & 39.4 & 578k & \$22 & 42\% & 44\% \\
Technical writers & 58.0 & 46.0 & 55k & \$38 & 87\% & -- \\
Medical transcriptionists & 57.9 & 45.2 & 39k & \$14 & 35\% & -- \\
Software developers, applications and systems software & 57.1 & 46.5 & 1.8M & \$52 & 88\% & 79\% \\
Web developers & 56.1 & 44.0 & 186k & \$35 & 68\% & 43\% \\
Miscellaneous media and communication workers & 55.7 & 47.1 & 108k & \$20 & 49\% & -- \\
Computer occupations, all other & 53.6 & 39.8 & 777k & \$34 & 61\% & 73\% \\
\addlinespace
\multicolumn{7}{@{}l}{\emph{Ten smallest changes}} \\
Miscellaneous food preparation and serving related workers, including dining room and cafeteria attendants and bartender helpers & 8.6 & 5.1 & 283k & \$11 & 8\% & 35\% \\
Dishwashers & 9.0 & 5.6 & 227k & \$11 & 5\% & -- \\
Roofers & 9.5 & 7.0 & 217k & \$18 & 6\% & -- \\
Cement masons, concrete finishers, and terrazzo workers & 9.5 & 6.5 & 63k & \$23 & 10\% & -- \\
Fence erectors & 10.7 & 7.9 & 41k & \$14 & 5\% & -- \\
Drywall installers, ceiling tile installers, and tapers & 10.8 & 7.5 & 171k & \$22 & 2\% & -- \\
Helpers, construction trades & 11.1 & 7.4 & 47k & \$17 & 7\% & 15\% \\
Plasterers and stucco masons & 11.2 & 7.4 & 23k & \$23 & 5\% & -- \\
Maids and housekeeping cleaners & 11.4 & 6.7 & 1.4M & \$12 & 8\% & 11\% \\
Highway maintenance workers & 12.0 & 7.7 & 106k & \$18 & 4\% & -- \\
\bottomrule
\end{tabular}
\end{spacing}
\begin{tablenotes}\notesize
\item Notes: Occupations ranked by the change in exposure from stage 1 to stage 4, with the change from stage 1 to stage 3 alongside. Employment and the share with a bachelor's degree or more are 2019 values; the real hourly wage is in January 2020 dollars; AI adoption is from the Real-Time Population Survey waves of 2025 to 2026 where the occupation is covered.
\end{tablenotes}
\end{threeparttable}
\end{table}

\subsection{Exposure, Adoption, and Why We Use the Change}
\label{sec:validation}

Adoption and exposure measure different things.
Whether a worker uses AI depends on the worker's perceptions, the firm's policies, and the labor market conditions, so adoption is endogenous.
Adoption is also a measure of the past, and is often available only for a part of the economy and occupations.
Exposure is built from task descriptions and from what the technology can do, so it is more removed from the hiring and pay decisions in economic data, and it can be constructed for every occupation whenever new model capabilities are released.

Though exposure and adoption are fundamentally different, an exposure measure would be of little value if it did not correlate with actual AI use.
We examine this with the Real-Time Population Survey of \citet{bick2026work}, which asks a representative sample of workers whether they use generative AI at work.
There are 143 occupations with at least 20 respondents, 141 of which match our exposure data, and adoption averages 45 percent and ranges from 5 to 87 percent.
Table~\ref{tab:adoption} reports the correlation between adoption and the levels and changes of our exposure score, as well as other exposure measures.

\begin{table}[htbp]
\centering
\begin{threeparttable}
\caption{Reported generative-AI adoption and each exposure measure}
\label{tab:adoption}
\begin{spacing}{1}
\begin{tabular}{@{}lccc@{}}
\toprule
Measure & Pearson & Spearman & $R^{2}$ \\
\midrule
\multicolumn{4}{@{}l}{\emph{Exposure levels}} \\
Stage 1 (pre-generative machine learning) & $-$0.01 & 0.08 & 0.00 \\
Stage 2 (chatbots) & 0.40 & 0.36 & 0.16 \\
Stage 3 (reasoning models) & 0.57 & 0.53 & 0.33 \\
Stage 4 (agents) & 0.57 & 0.54 & 0.32 \\
\addlinespace
\multicolumn{4}{@{}l}{\emph{Exposure changes}} \\
Stage 1 to 4 & 0.71 & 0.71 & 0.51 \\
Stage 1 to 3 & 0.74 & 0.74 & 0.55 \\
Stage 2 to 3 & 0.77 & 0.79 & 0.60 \\
\addlinespace
\multicolumn{4}{@{}l}{\emph{Existing measures}} \\
Eloundou et al.\ (2024), $\zeta$ & 0.73 & 0.71 & 0.54 \\
Felten et al.\ (2023), language-modeling exposure & 0.70 & 0.70 & 0.48 \\
\bottomrule
\end{tabular}
\end{spacing}
\begin{tablenotes}\notesize
\item Notes: Adoption is the share of workers in an occupation who report using generative AI at work, in percentage points, from four waves of the Real-Time Population Survey of \citet{bick2026work}, August 2025 to May 2026; the sample is the 141 occupations with at least 20 respondents and an exposure score. Pearson is the correlation weighted by respondent counts, Spearman the weighted correlation of ranks, and $R^{2}$ the squared Pearson correlation, i.e., the share of the variation in adoption that the measure explains.
\end{tablenotes}
\end{threeparttable}
\end{table}

Three patterns stand out.
First, exposure to pre-generative machine learning (stage 1 exposure) does not explain who uses generative AI.
Second, the correlation increases with each stage, from about 0.4 at stage 2 to about 0.57 at stages 3 and 4.
Third, changes in exposure predict adoption better than levels.
The change from stage 1 to stage 4 correlates 0.71 with adoption, the change from stage 1 to stage 3 correlates 0.74, and the step from chatbots to reasoning models alone reaches 0.77.
Widely used existing measures that track adoption closely are \citet{eloundou2024gpts}, at 0.73, and \citet{felten2023language}, at 0.70.
Appendix Figure~\ref{fig:adoptionall} plots adoption against the change in our score from stage 1 to stage 4 and against the Eloundou measure.

The explanatory variable in our regressions is the change in an occupation's exposure rather than its level.
The level and the change measure exposure to different technologies.
The stage-1 score captures exposure to pre-generative machine learning, which required workers trained in deep learning and diffused narrowly.
The adoption of generative AI, which has spread across occupations since the release of ChatGPT follows the capabilities added in stages 2 to 4.
Hence, the change in exposure, rather than its level, captures what generative AI can do, and Table~\ref{tab:adoption} confirms this.

Moreover, the level of exposure reflects persistent features of the work, such as how codifiable, digital, and analytical it is, which correlate with education and earnings and with the outcomes we examine.
The change nets out this common component and retains what a new generation of models added.
We use the change from stage 1 to stage 4 as the main explanatory variable and, since the agent era is recent, also report results for the change from stage 1 to stage 3.
The two have a correlation of 0.98.
We report results per standard deviation of the change in exposure, which is about 12 points for the stage 1 to 4 change and about 10 points for the stage 1 to 3 change, roughly the gap between the average office and administrative support occupation and the average computer occupation.

\section{Data and Empirical Design}
\label{sec:data}

\subsection{The CPS Occupation Panel}
\label{sec:cps}

Our labor market data come from the Current Population Survey (CPS), the monthly labor force survey of the U.S. Census Bureau and the Bureau of Labor Statistics, which samples about 60,000 households a month and underpins the official unemployment rate \citep{bls2026cps}.
The CPS is the nationally representative survey closest to real time, and it records, in addition to whether a person is working, usual weekly hours, part-time status, and, for the outgoing rotation groups, weekly earnings, so that we can examine the labor market at the extensive margin, at the intensive margin, and in pay.

We aggregate the monthly data to occupation-quarter cells, the level at which the exposure score is defined, from January 2017 through June 2026, so that the generative AI era, which we date from the first quarter of 2023, can be compared with the years before it period by period.\footnote{The CPS was not collected in October 2025 because of the federal government shutdown, and the Bureau of Labor Statistics did not collect it retroactively. Hence, our 2025Q4 cells average November and December 2025 only.}
We use the IPUMS harmonized OCC2010 occupation codes, which recode each year's census occupation codes to the 2010 census classification, so that occupations are consistent across the 2020 revision of the classification, and give 473 occupations over 38 quarters.
Twenty-two of them, residual categories and codes that end at the 2020 revision, have no exposure score and account for under 1 percent of employment, which leaves 451 occupations.
Appendix~B describes the crosswalk from O*NET to census occupations, and Section~\ref{sec:robustness} repeats the main results on the 2018 census codes as a robustness check.

For each occupation and quarter we measure employment in logs, the unemployment rate among those who report a last occupation, average weekly hours on the main job, average log real hourly earnings in the outgoing rotation groups, the part-time share, and the counts of part-time and full-time workers, and we build the same cells for each age group, each education group, and the six age-by-education groups.

Between the bottom and top quartiles of the change in exposure from stage 1 to stage 4, the share of workers with a bachelor's degree rises from 11 to 63 percent, the part-time share falls from 23 to 8 percent, the share of workers under 30 falls from about a third to under a fifth, and the real hourly wage rises from about 21 to 46 dollars (Appendix Table~\ref{tab:characteristics}).
Hence, any factor that affects educated office work differently from the rest, such as the pandemic, a tech-sector cycle, or monetary tightening, is a potential confounder.
The pandemic in particular hit the two groups very differently.
In the spring of 2020 the unemployment rate of the bottom quartile rose by about 16 percentage points and that of the top quartile by about 5, and the two recovered at different speeds through 2022.
This is why our main analyses do not use a base period in 2020 or 2021 (Appendix Figure~\ref{fig:outcomes}).
Three quarters of the employment-weighted variance of the change in exposure is between the 22 major occupation groups, and Section~\ref{sec:robustness} drops each major occupation group in turn.
Appendix~B presents the distribution of exposure across the workforce at each stage and the raw trends of the five outcomes by quartile from 2017 to 2026, overall and by age and education.

\subsection{The Empirical Specifications}
\label{sec:design}

Our main design compares changes in outcomes over a fixed interval across occupations whose exposure increased by different amounts.
For each group of workers $g$, which is all workers, an age group, an education group, or one of the six age-by-education cells, we estimate
\begin{equation}
\Delta_h Y_{ogt} \equiv Y_{ogt} - Y_{og,t-h} = \beta^{g}_{h}\,\Delta E_o + X_{o,t-h}'\gamma + \delta_t + \varepsilon_{ogt},\qquad t \in W ,
\label{eq:group}
\end{equation}
where $\Delta_h Y_{ogt}$ is the change in outcome $Y$ of group $g$ in occupation $o$ between period $t-h$ and period $t$, $h$ is the length of the difference in years, and $W$ is the set of end periods included in the regression.
$\Delta E_o$ is the change in occupation $o$'s exposure score from stage 1 to stage 4, standardized across occupations, so that every coefficient is read per standard deviation of the change in exposure, about 12 points, i.e., roughly the gap between the average office and administrative support occupation and the average computer occupation.
Since the agent era was two quarters old when the data end, we also report the change from stage 1 to stage 3, about 10 points per standard deviation, in Section~\ref{sec:robustness}.
$X_{o,t-h}$ is the vector of twelve workforce composition shares of the occupation in the base period, i.e., the shares female, under 30, 50 and over, white, Black, Asian, high school or less, bachelor's degree, post-bachelor's degree, South, Midwest, and Northeast, and $\delta_t$ denotes period fixed effects, which absorb any change common to all occupations in a period.
Hence, $\beta^{g}_{h}$ measures how much larger the $h$-year change in the outcome was in an occupation with a one standard deviation larger change in exposure, relative to occupations with a similar workforce in the same period.
Outcomes are in log points for employment and the part-time and full-time counts, in percentage points for the unemployment rate and the part-time share, and in hours for hours worked.
Regressions are weighted by the group's employment in the occupation in the base period, so that the estimates describe the average exposed worker rather than the average exposed occupation, and standard errors are clustered by occupation.

The headline estimate is the three-year difference that compares the first half of 2026 with the first half of 2023, i.e., the windows 2026Q1 versus 2023Q1 and 2026Q2 versus 2023Q2 stacked with a period effect for each.
It is the longest window whose base period lies after the release of ChatGPT and outside the pandemic.
Its base period straddles the chatbot inflection, so the estimate assumes that outcomes in early 2023 had not yet responded to the first generation, an assumption we relax in the appendix with a 2022 base.
Each horizon has a placebo window with both base and end periods before the release of ChatGPT and, where the data allow, with a base period before the pandemic, e.g., 2021Q4 versus 2018Q4 through 2022Q3 versus 2019Q3 for the three-year horizon.
If the design found an effect in the years before the technology existed, its estimates for the generative AI era could not be read as effects of AI.
To see when a change occurred, we also estimate Eq.~(\ref{eq:group}) one window at a time, shifting the end quarter from 2018 to 2026, so that the stacked estimates give the size of a change and the single-window estimates show the timing.

Our second design uses the timing of each capability jump within occupations.
We assign each occupation its stage-1 score until the beginning of the chatbot era, its stage-2 score from 2023Q2, its stage-3 score from 2025Q1, and its stage-4 score from 2026Q1, and estimate
\begin{equation}
Y_{ot} = \alpha_o + \delta_t + \beta_S\,E^{step}_o(t) + \varepsilon_{ot},\qquad t \in S ,
\label{eq:fe}
\end{equation}
where $E^{step}_o(t)$ is this step exposure score, standardized within the estimation window, $\alpha_o$ and $\delta_t$ are occupation and quarter fixed effects, and $S$ is the sample window, 2023Q1 to 2026Q2 in the main estimates.
$\beta_S$ measures how much higher the outcome is in quarters when the occupation's prevailing exposure is one standard deviation higher, relative to the occupation's own average and to the economy-wide level in that quarter, and a version that lets the coefficient shift at the reasoning model inflection shows when the change occurred.
These regressions are weighted by the occupation's average employment in 2019 and clustered by occupation, and we estimate them by group in the same way.
A single coefficient averages the relationship over the whole window, whereas the generations arrived at different dates.
Hence, we also let the coefficient change once, at the reasoning model inflection:
\begin{equation}
Y_{ot} = \alpha_o + \delta_t + \beta\,E^{step}_o(t) + \beta^{shift}\,E^{step}_o(t)\,\mathbf{1}[t \ge \text{2025Q1}] + \varepsilon_{ot},
\label{eq:feshift}
\end{equation}
where $\mathbf{1}[t \ge \text{2025Q1}]$ is an indicator for quarters from 2025Q1 onward.
$\beta$ is the coefficient before 2025Q1, $\beta^{shift}$ is the change in the coefficient at the inflection, and the coefficient from 2025Q1 onward is their sum.
We also conduct individual-level regressions of the fixed effects design, which controls for worker characteristics and allows us to separate changes for a given type of worker from changes in the composition of the workforce.

The estimates of Eqs.~(\ref{eq:group}) and~(\ref{eq:fe}) capture the differences in how the outcomes changed between occupations whose exposure rose more and occupations whose exposure rose less, conditional on the controls.
Interpreting those as effects of advancing AI capability requires that the change in exposure be unrelated to unobserved determinants of those changes, conditional on the controls.
Though we cannot be definitive, some features of the measure and research design support treating the exposure change as plausibly exogenous to the occupation's outcomes.
First, the stages describe general capabilities, to converse, to write, to reason, to act, etc., and not any occupation. 
So the variation in the exposure change across occupations comes from the interaction of a common advance in capability with each occupation's task content, which O*NET recorded before generative AI.
Second, the composition controls remove the part of the exposure change that is predicted by the age, education, sex, race, and regional make-up of an occupation's workforce, and the placebo windows show that occupations whose exposure would later rise more were not on different paths before generative AI.

Of course, we cannot fully rule out some relevant change that correlates with the change in exposure for reasons other than AI capability.
Often discussed candidates are the rise of remote work, the contraction of the technology sector, industry-specific trends, and a differential unwinding of the pandemic hiring boom in educated office work.
Section~\ref{sec:robustness} examines each, by controlling for each occupation's telework share and its change, by dropping the technology occupations and each major occupation group in turn, by adding industry shares and their interactions with time, and by the placebo windows.

\section{Results}
\label{sec:findings}

\subsection{Aggregate Labor Outcomes}
\label{sec:fd-results}

Table~\ref{tab:headline} reports the three-year first differences of Eq.~(\ref{eq:group}) for all workers, which measure the cumulative change from the start of the chatbot era to the agent era.
Panel A stacks the windows that compare the first and second quarters of 2026 with the same quarters of 2023, so that each window begins after the release of ChatGPT and outside the pandemic and ends after AI agents reached the workplace.
Standard errors, clustered by occupation, are in parentheses throughout the paper.
In these windows, each additional standard deviation of the increase in exposure is associated with a change in employment of $-1.89$ log points (1.45) and a change in the unemployment rate of 0.16 percentage points (0.12), neither of which is distinguishable from zero.\footnote{The most and least exposed quartiles of occupations differ by about 2.6 standard deviations of the increase in exposure, so the estimate implies about 5 log points less employment growth in the most exposed quartile than in the least exposed.}
The other three estimates do change.
The share of workers who usually work part-time falls by 0.74 percentage points (0.25), while the part-time share rose by 0.67 percentage points across all occupations over the same windows.
Hence, the more exposed occupations moved against the broader trend.
Average weekly hours rise by 0.20 hours (0.09), about 12 minutes, while hours fell slightly across all occupations, and average real hourly earnings rise by 1.94 log points (0.80), about 2 percent.\footnote{The earnings estimate is the least robust of the five. It is 0.49 log points (0.60) with six-month averages (Appendix Table~\ref{tab:cell_fd_main_1}), $-0.12$ (0.73) in the occupation fixed-effects design (Table~\ref{tab:fe-body}), and loses significance without weights, with industry controls, or with the CPS telework controls (Section~\ref{sec:robustness}).}
Panel B applies the same comparisons to windows that begin before the pandemic and end before the release of ChatGPT, and none of the five estimates is significant, which is reassuring for the design.

\begin{table}[htbp]
\centering
\caption{Three-year differences in the main outcomes: headline window and pre-AI placebo window}
\label{tab:headline}
\fontsize{8.3}{9.8}\selectfont
\setlength{\tabcolsep}{2.15pt}
\renewcommand{\arraystretch}{1.11}
\begin{adjustbox}{max width=\linewidth}
\begin{tabularx}{\linewidth}{@{}>{\raggedright\arraybackslash}p{0.225\linewidth}*{5}{>{\centering\arraybackslash}X}@{}}
\toprule
 & \shortstack{Log\\employment} & \shortstack{Unemployment\\(pp)} & Hours & \shortstack{Log real hourly\\earnings} & \shortstack{Part-time share\\(pp)} \\
\midrule
\addlinespace[0.16em]
\multicolumn{6}{@{}l}{\textbf{\textit{A. Headline window}}} \\
\multicolumn{6}{@{}l}{\parbox[t]{0.94\linewidth}{\fontsize{7.4}{8.8}\selectfont\itshape 2026Q1 vs. 2023Q1 and 2026Q2 vs. 2023Q2}} \\[0.14em]
Exposure change, stage 1 to 4 & -1.89 (1.45) & 0.16 (0.12) & 0.20\textsuperscript{**} (0.09) & 1.94\textsuperscript{**} (0.80) & -0.74\textsuperscript{***} (0.25) \\
\(R^{2}\) & 0.04 & 0.01 & 0.03 & 0.03 & 0.05 \\
\cmidrule(lr){1-6}
Mean outcome change & -1.21 & 0.60 & -0.23 & 1.61 & 0.67 \\
\(N\) & 900 & 900 & 900 & 846 & 900 \\
\specialrule{0.075em}{0.20em}{0.14em}
\multicolumn{6}{@{}l}{\textbf{\textit{B. Pre-AI placebo window}}} \\
\multicolumn{6}{@{}l}{\parbox[t]{0.94\linewidth}{\fontsize{7.4}{8.8}\selectfont\itshape 2021Q4 vs. 2018Q4 through 2022Q3 vs. 2019Q3}} \\[0.14em]
Exposure change, stage 1 to 4 & 0.49 (1.25) & -0.02 (0.11) & -0.08 (0.07) & 0.69 (0.61) & -0.00 (0.20) \\
\(R^{2}\) & 0.06 & 0.02 & 0.05 & 0.02 & 0.04 \\
\cmidrule(lr){1-6}
Mean outcome change & -2.27 & 0.25 & -0.29 & 3.08 & -0.45 \\
\(N\) & 1,794 & 1,794 & 1,793 & 1,733 & 1,794 \\
\specialrule{0.075em}{0.20em}{0.14em}
\bottomrule
\end{tabularx}\end{adjustbox}
\par\vspace{0.50em}\noindent
\begin{minipage}{\linewidth}
\notesize \textit{Notes:} Each entry is the coefficient on the standardized stage-1 to stage-4 change in exposure in the three-year first difference of Eq.~(\ref{eq:group}) for all workers, with standard errors clustered by occupation in parentheses. Panel A stacks 2026Q1 vs.\ 2023Q1 and 2026Q2 vs.\ 2023Q2; Panel B stacks the placebo windows 2021Q4 vs.\ 2018Q4 through 2022Q3 vs.\ 2019Q3. Regressions are weighted by base-period employment and include the twelve base-period composition controls and period effects. Log outcomes and shares are multiplied by 100; hours are in hours. The mean outcome change is the weighted mean of the first-differenced outcome in the sample; $N$ is the number of occupation-window observations. \mbox{\textsuperscript{*}\,$p<0.10$}, \mbox{\textsuperscript{**}\,$p<0.05$}, \mbox{\textsuperscript{***}\,$p<0.01$}. \end{minipage}

\end{table}

Hence, over three years, we find no significant change in employment or unemployment in occupations whose exposure rose more relative to occupations with similar workforces, but the types of positions, work hours, and the average pay changed.\footnote{The occupation fixed-effects design of Eq.~(\ref{eq:fe}), estimated over 2023Q1 to 2026Q2, points to the same conclusion on employment, at $-0.60$ log points (1.35) per standard deviation of the prevailing exposure (Appendix Table~\ref{tab:fe-main}).}
Sections~\ref{sec:who} and~\ref{sec:when} show that these changes reflect the composition of the workforce and the mix of part-time and full-time positions rather than higher pay or longer hours for a given worker.
That employment did not detectably change while the mix of positions and the average pay did is consistent with AI changing the composition of the workforce, which we examine next.
Earlier automation technologies also left total employment unchanged or higher, with the gains concentrated in services \citep{mann2023benign}.
It is worth noting that the employment estimates are negative at every horizon, which could reflect a modest decline that the estimates are too imprecise to detect.

\subsection{Heterogeneity by Age and Education}
\label{sec:who}

\begin{figure}[t]
\centering
\includegraphics[width=\linewidth]{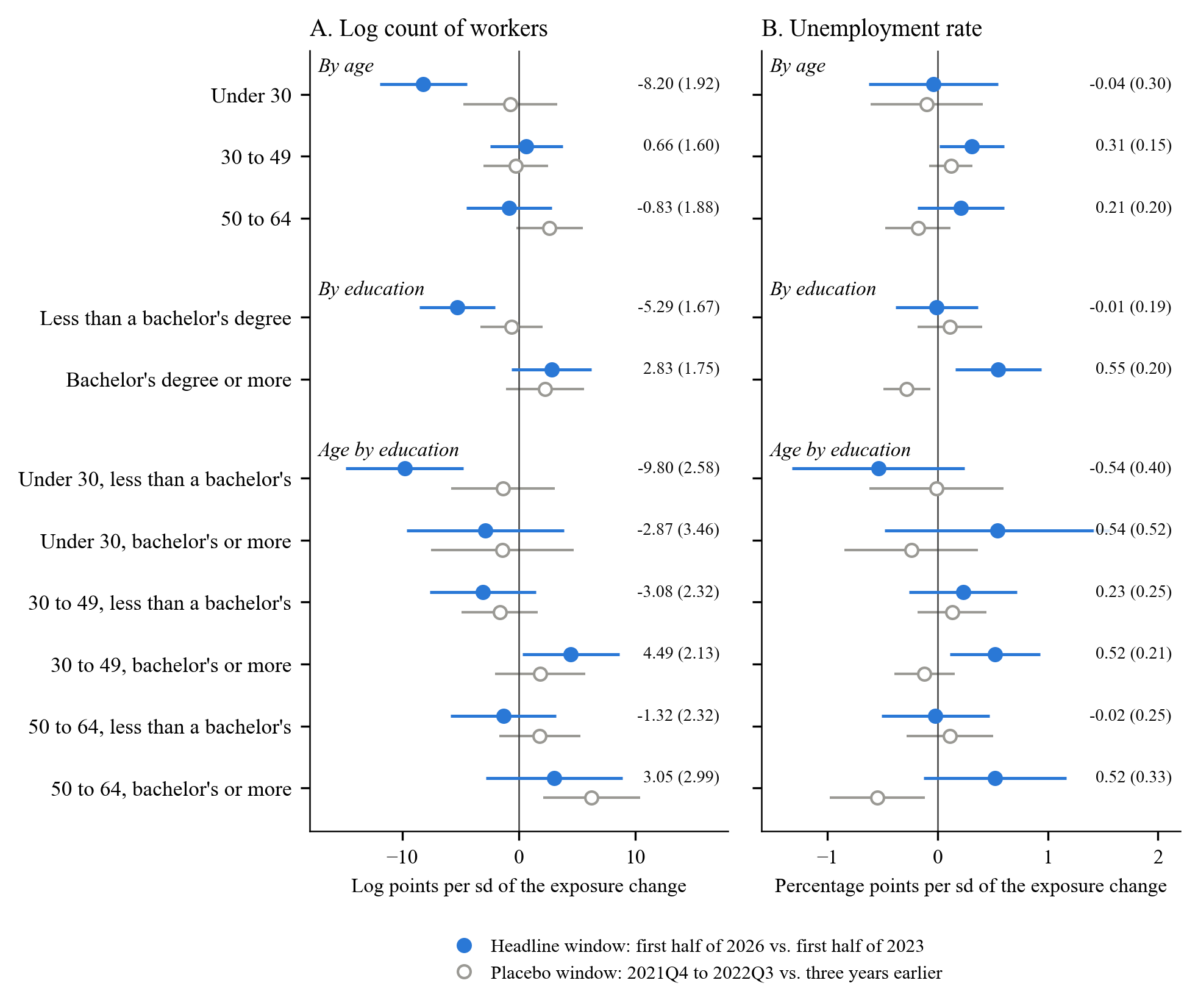}
\caption{Three-year changes in the number of workers and in the unemployment rate, by age, education, and age by education}
\label{fig:who}
\begin{minipage}{0.96\textwidth}
\notesize
Notes: Each marker is the coefficient of Eq.~(\ref{eq:group}) for the group's log count of workers (Panel A) or unemployment rate in percentage points (Panel B), per standard deviation of the stage-1 to stage-4 change in exposure, with its 95 percent interval. Filled markers are the headline window and hollow markers the pre-AI placebo window, 2021Q4 vs.\ 2018Q4 through 2022Q3 vs.\ 2019Q3; the headline estimates and standard errors are printed beside the markers. The specification is that of Table~\ref{tab:cell_fd_main_4}, which reports the headline estimates for all five outcomes; Appendix Tables~\ref{tab:cell_fd_main_5}, \ref{tab:cell_fd_main_11}, and~\ref{tab:cell_fd_main_7} report the six cells and the placebo window.
\end{minipage}
\end{figure}

The stable total hides changes for particular groups of workers.
Figure~\ref{fig:who} plots the three-year change in the number of workers by age, by education, and by age and education, with the placebo estimate beside each.
In the headline window, each additional standard deviation of the increase in exposure is associated with a decline of 8.20 log points (1.92) in the number of workers under 30, whereas the estimates for workers aged 30 to 49 and 50 to 64 are small and insignificant, at 0.66 (1.60) and $-0.83$ (1.88), and the placebo estimate for workers under 30 is $-0.70$ (2.06) (Table~\ref{tab:cell_fd_main_4} and Appendix Table~\ref{tab:cell_fd_main_11}).\footnote{The most and least exposed quartiles differ by about 2.6 standard deviations of the increase in exposure, so the estimate implies about 21 log points less growth in the number of workers under 30 in the most exposed quartile than in the least exposed.}
Workers under 30 are about a quarter of employment, so a decline of 8 log points among them corresponds to about 2 log points of total employment.
This is close to the aggregate estimate of $-1.89$ in Table~\ref{tab:headline}, so the flat aggregate and the decline among young workers are consistent with each other.

By education, the number of non-graduates, i.e., workers without a bachelor's degree, falls by 5.29 log points (1.67), whereas the estimate for graduates, i.e., workers with a bachelor's degree or more, is positive but insignificant, at 2.83 (1.75), and the placebo estimates for graduates are positive and nearly as large as the AI-era estimates, so the gains among graduates may partly continue a trend that predates generative AI.

Crossing age and education shows that the decline among young workers is concentrated among those without a degree: the number of workers under 30 without a degree falls by 9.80 log points (2.58), whereas the estimate for young graduates is $-2.87$ (3.46), too imprecise to distinguish from zero or from the estimate for non-graduates, and the placebo estimates are flat for both, at $-1.34$ (2.27) and $-1.39$ (3.13) (Appendix Tables~\ref{tab:cell_fd_main_5} and~\ref{tab:cell_fd_main_7}).
The gains are concentrated among college graduates aged 30 to 49, whose number rises by 4.49 log points (2.13), with an insignificant placebo estimate, and whose full-time count, weekly hours, and real hourly earnings also rise (Appendix Table~\ref{tab:cell_fd_secondary_5}).
The estimate for workers aged 30 to 49 without a degree is $-3.08$ (2.32), which is not statistically significant, and the age group as a whole changed little, at 0.66 (1.60).

The decline among young workers is not accompanied by higher unemployment among them, which suggests fewer young workers entering these occupations rather than more young workers losing their jobs.
The unemployment rate of workers under 30 in more exposed occupations changes by $-0.04$ percentage points (0.30), and by $-0.54$ (0.40) for those without a degree.
This is not a contradiction, since the CPS assigns unemployed workers to the occupation of their last job.
If exposed occupations hired fewer young workers, the young people who were never hired would not appear as unemployed in those occupations, whereas a rise in layoffs would show up as higher unemployment among young workers in these occupations.
The pattern is therefore consistent with evidence that the adjustment operates through reduced hiring rather than separations \citep{brynjolfsson2026canaries, atkinson2026young, hosseinimaasoum2025seniority}.
We note that young workers who left these occupations for jobs in other occupations could also produce this pattern.
However, such a transition would have to occur without a spell of unemployment, and given the reported difficulty and length of job search for young workers in recent years, this seems less likely.
We examine this directly by linking consecutive monthly interviews of the same respondent in the CPS rotating panel and counting the workers who enter and leave each occupation each month.
Given the noisiness of the linked sample, we report these estimates in Appendix Table~\ref{tab:flows} and regard them as informative but less precise than our main estimates.
Among workers under 30 in more exposed occupations, the number of entrants from all sources fell by 0.48 (0.45) per 100 young workers per month and the number of exits to all destinations fell by 0.21 (0.52), so that the net inflow fell by 0.37 (0.53).
The decline in entrants comes from fewer young workers entering from unemployment or from outside the labor force, at $-0.50$ (0.27), significant at the 10 percent level, whereas the number entering from other occupations did not change.
Among young workers without a degree, entrants fell by 1.45 (0.63) and exits by 0.77 (0.61).
Again, the decline in entrants came from non-employment, at $-1.04$ (0.42).
Hence, fewer young workers were hired into these occupations from outside employment, and those already in them were not more likely to leave.

However, unemployment did rise among college graduates.
The unemployment rate of graduates in more exposed occupations rose by 0.55 percentage points (0.20).
The increase is about half a percentage point in each age group, and the placebo estimates are zero or negative (Figure~\ref{fig:who}, Panel B).
Since the CPS assigns unemployed workers to the occupation of their last job, these are graduates who lost or left jobs in these occupations.
In the fixed-effects design, the unemployment rate of graduates rises by 0.42 percentage points (0.23) (Appendix Table~\ref{tab:cellfe-polished-primary-occ2010-groups}).
Hence, the number of graduates aged 30 to 49 in exposed occupations increased, but more graduates also lost their jobs in these occupations, which points to more churn among graduates rather than only more hiring.

That the decline falls on young workers without a degree may seem at odds with the widely publicized view that generative AI mainly threatens college-educated work.
However, exposed occupations employ many workers without a degree.
In the most exposed quartile, 37 percent of workers do not hold a bachelor's degree, and office and administrative support, which employs 11 percent of workers, saw its exposure rise by 41 points on average, which is close to that of business, legal, and engineering occupations.

The composition of the workforce in exposed occupations changed as a result.
Table~\ref{tab:cell_fd_main_4} reports the five outcomes of Table~\ref{tab:headline} by age and by education.
The number of workers under 30 fell and the number of non-graduates fell, whereas the number of graduates rose, so the workers who remained are older and more educated.
Among graduates, hours rose by 0.23 (0.13) and real hourly earnings by 2.60 log points (1.25), and among workers aged 30 to 49 by 0.29 (0.10) and 3.12 (1.16), whereas neither changed for workers under 30, at 0.18 (0.22) and 0.61 (1.76).
The part-time share fell for workers aged 30 to 49 and 50 to 64 and for graduates, by 1.05 (0.29), 0.80 (0.44), and 1.09 (0.37) percentage points, but not for workers under 30, at 0.13 (0.69), whose full-time and part-time counts fell together.
The placebo estimates are small for the number of workers in each age group and for the number of workers under 30 without a degree.
The shift is visible directly in the full-time workforce: the share of full-time workers under 30 fell by 1.59 percentage points (0.36) and the share with a bachelor's degree or more rose by 1.76 (0.35), with a flat placebo for the first and a smaller increase, 0.53 (0.30), for the second (Appendix Table~\ref{tab:cell_fd_composition_1}).
This shift matters for interpreting wages.
Average real hourly earnings in exposed occupations rose by 1.94 log points (0.80) in Table~\ref{tab:headline}, but the individual-level design, which controls for worker characteristics, shows no change in the pay of a given type of worker, at $-0.09$ log points (0.64) (Appendix Table~\ref{tab:fe-main}).
Hence, the rise in average wages reflects who works in these occupations rather than higher pay for the same workers.

\begin{table}[htbp]
\centering
\caption{Three-year differences in the main outcomes, by age and by education}
\label{tab:cell_fd_main_4}
\fontsize{8.3}{9.8}\selectfont
\setlength{\tabcolsep}{2.15pt}
\renewcommand{\arraystretch}{1.11}
\begin{adjustbox}{max width=\linewidth}
\begin{tabularx}{\linewidth}{@{}>{\raggedright\arraybackslash}p{0.225\linewidth}*{5}{>{\centering\arraybackslash}X}@{}}
\toprule
 & \shortstack{Log\\employment} & \shortstack{Unemployment\\(pp)} & Hours & \shortstack{Log real hourly\\earnings} & \shortstack{Part-time share\\(pp)} \\
\midrule
\addlinespace[0.16em]
\multicolumn{6}{@{}l}{\textbf{\textit{A. By age}}} \\[-0.06em]
Under 30 & -8.20\textsuperscript{***} (1.92) & -0.04 (0.30) & 0.18 (0.22) & 0.61 (1.76) & 0.13 (0.69) \\
\(R^{2}\) & 0.08 & 0.01 & 0.02 & 0.03 & 0.02 \\
\cmidrule(lr){1-6}
Mean outcome change & -3.42 & 0.83 & -0.49 & 2.94 & 1.76 \\
\(N\) & 831 & 840 & 829 & 591 & 831 \\
\cmidrule(lr){1-6}
30 to 49 & 0.66 (1.60) & 0.31\textsuperscript{**} (0.15) & 0.29\textsuperscript{***} (0.10) & 3.12\textsuperscript{***} (1.16) & -1.05\textsuperscript{***} (0.29) \\
\(R^{2}\) & 0.14 & 0.02 & 0.03 & 0.04 & 0.04 \\
\cmidrule(lr){1-6}
Mean outcome change & -1.15 & 0.65 & -0.17 & 0.99 & 0.39 \\
\(N\) & 891 & 891 & 891 & 756 & 891 \\
\cmidrule(lr){1-6}
50 to 64 & -0.83 (1.88) & 0.21 (0.20) & -0.04 (0.13) & -1.00 (1.68) & -0.80\textsuperscript{*} (0.44) \\
\(R^{2}\) & 0.14 & 0.02 & 0.02 & 0.04 & 0.02 \\
\cmidrule(lr){1-6}
Mean outcome change & -6.86 & 0.53 & -0.09 & 0.03 & 0.28 \\
\(N\) & 869 & 870 & 869 & 676 & 869 \\
\specialrule{0.075em}{0.20em}{0.14em}
\multicolumn{6}{@{}l}{\textbf{\textit{B. By education}}} \\[-0.06em]
Less than a bachelor's degree & -5.29\textsuperscript{***} (1.67) & -0.01 (0.19) & 0.16 (0.10) & 0.22 (1.06) & -0.22 (0.35) \\
\(R^{2}\) & 0.05 & 0.01 & 0.03 & 0.02 & 0.05 \\
\cmidrule(lr){1-6}
Mean outcome change & -4.09 & 0.54 & -0.23 & 2.46 & 1.04 \\
\(N\) & 844 & 844 & 844 & 734 & 844 \\
\cmidrule(lr){1-6}
Bachelor's degree or more & 2.83 (1.75) & 0.55\textsuperscript{***} (0.20) & 0.23\textsuperscript{*} (0.13) & 2.60\textsuperscript{**} (1.25) & -1.09\textsuperscript{***} (0.37) \\
\(R^{2}\) & 0.05 & 0.03 & 0.02 & 0.03 & 0.03 \\
\cmidrule(lr){1-6}
Mean outcome change & 0.61 & 0.74 & -0.25 & -0.32 & 0.29 \\
\(N\) & 813 & 815 & 813 & 643 & 813 \\
\bottomrule
\end{tabularx}\end{adjustbox}
\par\vspace{0.50em}\noindent
\begin{minipage}{\linewidth}
\notesize \textit{Notes:} Three-year differences stack 2026Q1 versus 2023Q1 and 2026Q2 versus 2023Q2. Each group coefficient is obtained from its own occupation-by-group-by-quarter regression using that group's base-period employment weights and the twelve full-occupation baseline workforce-composition controls. Regressions use S4--S1 exposure standardized at the aggregate occupation level. Log outcomes and shares are multiplied by 100; hours are in hours. $N$ is the group-outcome estimation sample. The ``Mean outcome change'' row is the within-group, base-period-employment-weighted mean of the first-differenced dependent variable (the later-period outcome minus the earlier-period outcome), not the mean outcome level. Logged outcomes and shares use the display scaling described above. \mbox{\textsuperscript{*}\,$p<0.10$}, \mbox{\textsuperscript{**}\,$p<0.05$}, \mbox{\textsuperscript{***}\,$p<0.01$}.
\end{minipage}

\end{table}

\subsection{Decomposing the Decline in Young Workers by Occupation}
\label{sec:decomposition}

\begin{figure}[tbp]
\centering
\includegraphics[width=\linewidth]{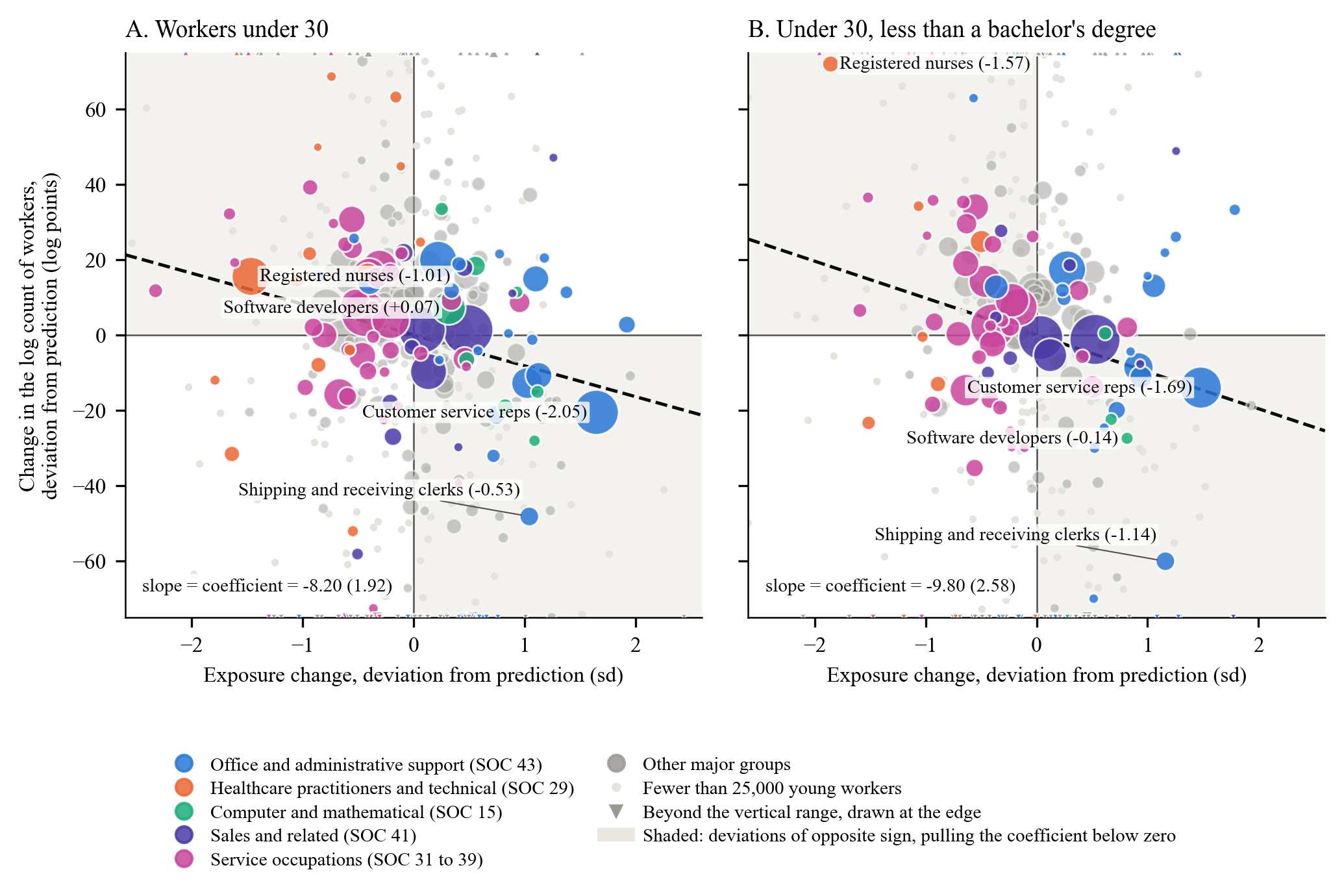}
\caption{Where the decline in young workers comes from: occupations' deviations in exposure change and in young employment}
\label{fig:decomposition}
\begin{minipage}{0.96\textwidth}
\notesize
Notes: Each bubble is an occupation, with area proportional to the group's base-period employment, the regression weight. The horizontal axis is the deviation of the occupation's stage-1 to stage-4 exposure change from the value predicted by the composition controls and the period effect of Eq.~(\ref{eq:group}), in standard deviations; the vertical axis is the deviation of its three-year change in the log count of the group's workers from the same prediction, in log points, 2026Q1 vs.\ 2023Q1 and 2026Q2 vs.\ 2023Q2 averaged. The dashed line through the origin has slope $\beta$ of Eq.~(\ref{eq:decomp}); occupations in the shaded quadrants, where the two deviations have opposite signs, pull the coefficient below zero. Panel A is workers under 30 and Panel B workers under 30 without a bachelor's degree (Table~\ref{tab:cell_fd_main_4} and Appendix Table~\ref{tab:cell_fd_main_5}). Labels give the contribution in log points for the three largest contributors in each panel, in absolute value, and for customer service representatives, registered nurses, and software developers in both panels; occupations with fewer than 25,000 young workers are small gray dots, and occupations beyond the vertical range are triangles at the edge. Colors mark SOC major groups 43, 29, 15, and 41 and the service occupations (31 to 39). Appendix Figure~\ref{fig:decomposition-ages} repeats Panel A for all workers and the two older age groups.
\end{minipage}
\end{figure}

\begin{figure}[tbp]
\centering
\includegraphics[width=\linewidth]{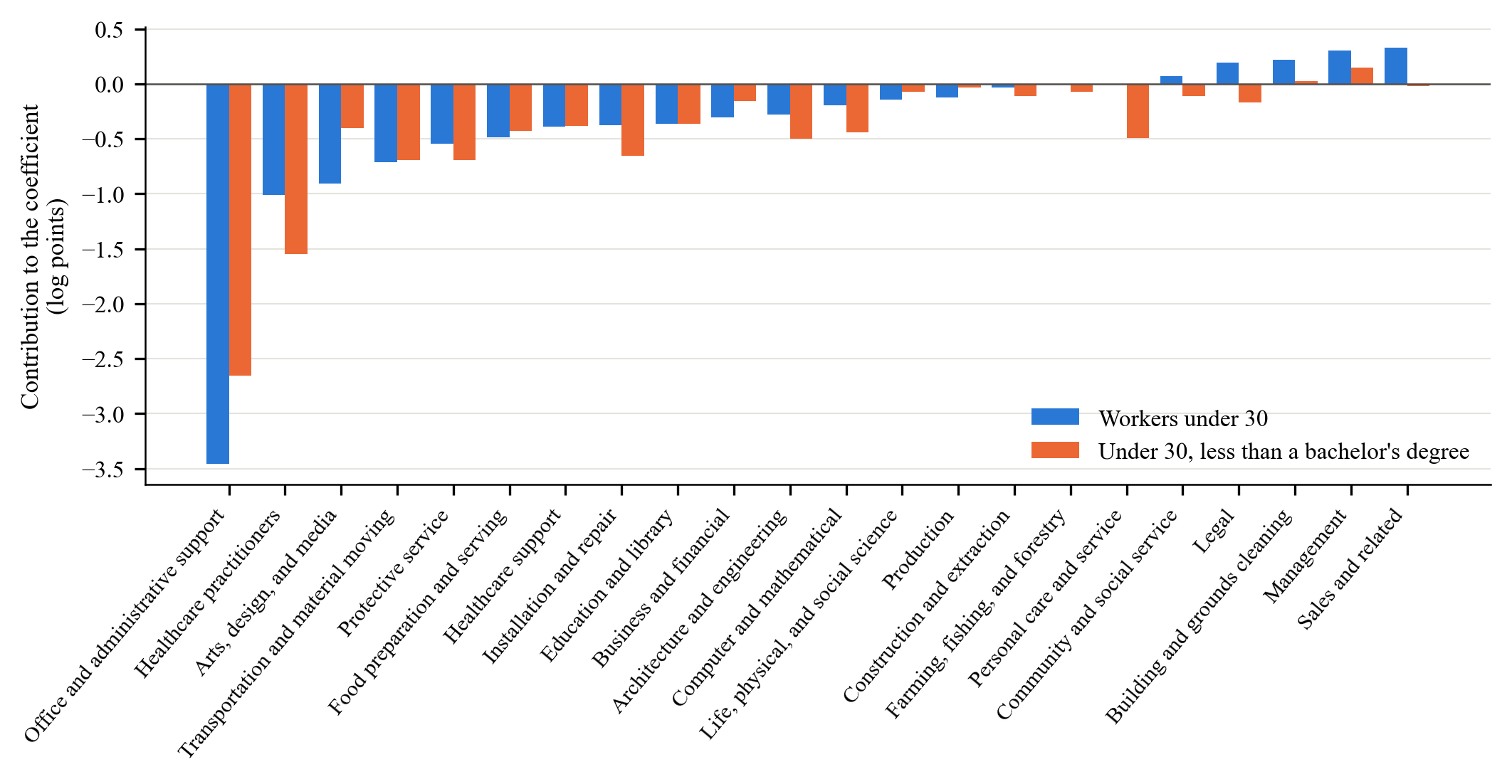}
\caption{Contributions to the young-worker coefficient by major occupation group}
\label{fig:decompgroups}
\begin{minipage}{0.96\textwidth}
\notesize
Notes: Each bar sums the contributions of the occupations in Figure~\ref{fig:decomposition} within one of the 22 major groups of the Standard Occupational Classification, for workers under 30 (Panel A of that figure) and for workers under 30 without a bachelor's degree (Panel B); the bars sum to the coefficients, $-8.20$ log points (1.92) and $-9.80$ (2.58). Appendix Table~\ref{tab:decomposition} reports the contributions with each group's share of the coefficient, of young employment, and its mean exposure change.
\end{minipage}
\end{figure}

A natural question is which occupations drive the decline in young workers.
The regression estimate is an average over occupations, so it can be decomposed to show how much each occupation, and each major occupation group, contributes to it.
After the composition controls and the period effect are partialled out of both variables, the coefficient of Eq.~(\ref{eq:group}) is
\begin{equation}
\beta = \frac{\sum_o w_o\, \tilde{e}_o\, \tilde{y}_o}{\sum_o w_o\, \tilde{e}_o^{\,2}},
\label{eq:decomp}
\end{equation}
where $\tilde{e}_o$ and $\tilde{y}_o$ are occupation $o$'s change in exposure and change in the log count of workers net of the controls, and $w_o$ is its base-period employment, the regression weight.
Each occupation's contribution to $\beta$ is its term in the numerator divided by the denominator, and the contributions sum to $\beta$.
Figure~\ref{fig:decomposition} presents this decomposition for workers under 30 (Panel A) and for young workers without a bachelor's degree (Panel B), Appendix Figure~\ref{fig:decomposition-ages} repeats it for all workers and for workers aged 30 to 49 and 50 to 64, and Figure~\ref{fig:decompgroups} plots the contributions by major occupation group, which Appendix Table~\ref{tab:decomposition} reports.

Office and administrative support occupations account for 42 percent of the under-30 coefficient while employing 13 percent of young workers, and customer service representatives alone account for about a quarter of it.\footnote{
Appendix Table~\ref{tab:topoccs} ranks the occupations in the most exposed quartile by their share of the quartile's loss of young workers between the first half of 2023 and the first half of 2026.
Customer service representatives account for 35 percent of the loss in both samples, followed by secretaries and administrative assistants, human resources workers, management analysts, and bookkeeping clerks, each at 5 to 8 percent.
Across the quartile, the number of young workers decreased by 9.3 log points on average, compared with an increase of 3.0 log points in the least exposed quartile, and by 16.2 log points for young workers without a degree, compared with an increase of 1.9.
}
Healthcare practitioners contribute 12 percent, arts and media 11 percent, and transportation 9 percent.
Computer and mathematical occupations, whose exposure increased the most, by 53 points on average, contribute only 2 percent, and business, financial, architecture, and engineering occupations together contribute 7 percent.
Sales, management, and legal occupations pull the coefficient toward zero, since their exposure rose more than average but their young employment did not fall relative to prediction.
For young workers without a bachelor's degree the pattern is similar, with office and administrative support at 27 percent and healthcare practitioners at 16 percent.

The large contribution of healthcare practitioners may seem puzzling, since their exposure increased the least.
However, a contribution to the coefficient estimate does not require a loss of workers.
Occupations whose exposure rose less than average but whose employment rose relative to prediction also contribute to the estimate, since they serve as the comparison group.
A below-average increase in exposure paired with above-average employment growth pushes the coefficient in the same direction as high exposure paired with falling employment.
Registered nurses, for example, contribute as much to the non-graduate estimate as customer service representatives, even though the number of young nurses without a degree increased.
Occupation groups whose exposure change is below the mean account for about two fifths of the under-30 coefficient and about half of the non-graduate coefficient.
In short, the decline in young workers in exposed occupations is concentrated in office and administrative support, and in customer service work in particular.
The estimate compares these occupations with healthcare, protective service, transportation, and food service, where the number of young workers held up or increased, and these comparison occupations account for about two fifths of the negative estimate.
The most exposed technical occupations, in which AI capabilities advanced the most and which have drawn the most public attention, contribute little to either estimate.

\FloatBarrier

\subsection{Timing and the Types of Positions}
\label{sec:when}

\begin{figure}[t]
\centering
\includegraphics[width=0.92\linewidth]{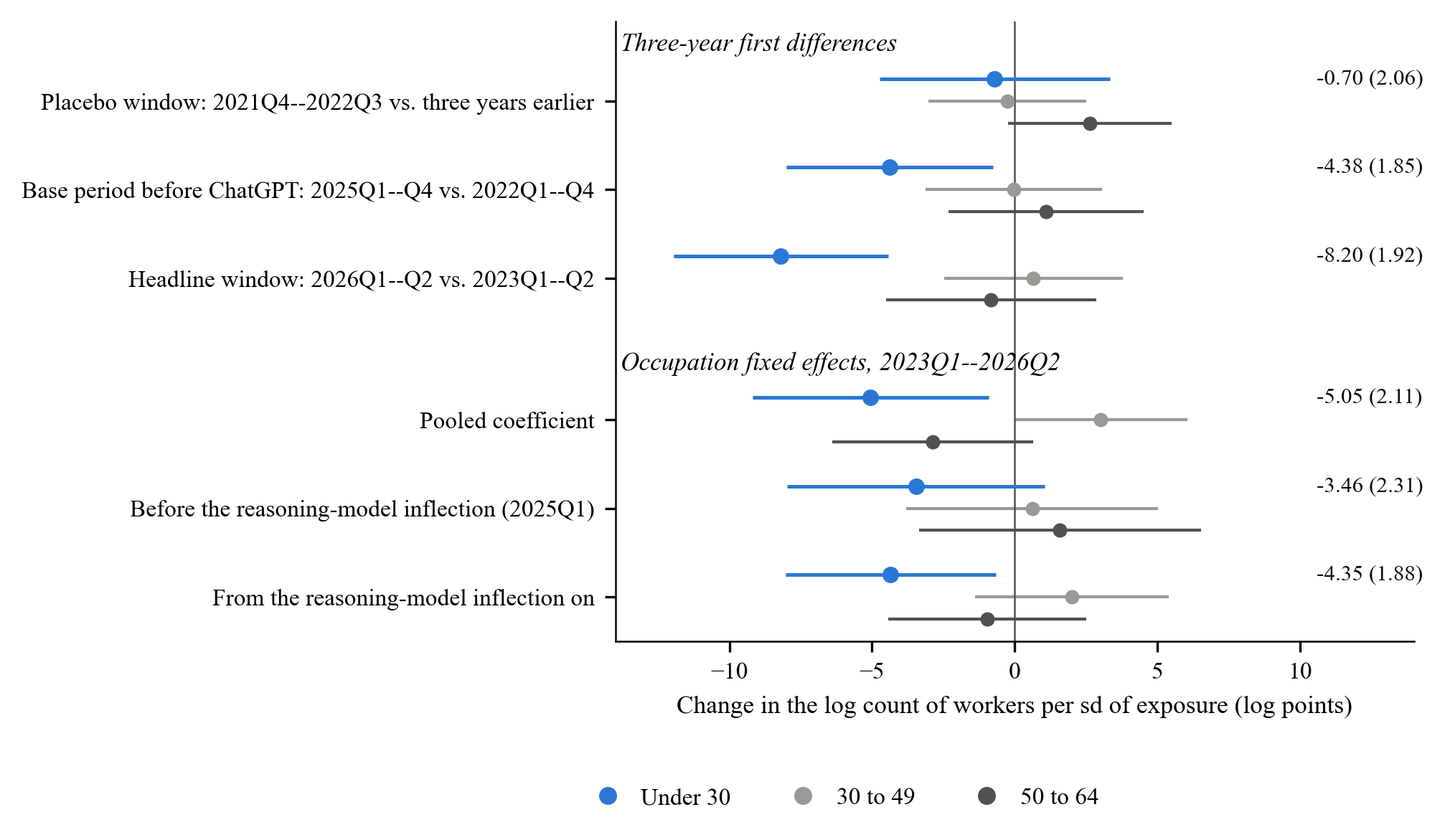}
\caption{The decline in workers under 30 across windows and designs, with the older age groups}
\label{fig:timing}
\begin{minipage}{0.96\textwidth}
\notesize
Notes: Each marker is a coefficient for the log count of workers in the age group, with its 95 percent interval; the under-30 estimates are printed at the right. The first three rows are three-year first differences of Eq.~(\ref{eq:group}) per standard deviation of the stage-1 to stage-4 change in exposure, for the placebo window (2021Q4 vs.\ 2018Q4 through 2022Q3 vs.\ 2019Q3), the four windows 2025Q1 vs.\ 2022Q1 through 2025Q4 vs.\ 2022Q4 stacked, and the headline window (2026Q1 vs.\ 2023Q1 and 2026Q2 vs.\ 2023Q2). The last three rows are the occupation fixed-effects coefficients over 2023Q1 to 2026Q2 per standard deviation of the prevailing step exposure, pooled (Eq.~\ref{eq:fe}) and before and from 2025Q1 (Eq.~\ref{eq:feshift}). Sources: Table~\ref{tab:cell_fd_main_4} and Appendix Tables~\ref{tab:cell_fd_main_11} and~\ref{tab:cellfe-polished-primary-occ2010-groups}; the 2022-base window and the coefficient from 2025Q1 are estimated from the same data with the same specifications (replication files).
\end{minipage}
\end{figure}

\begin{figure}[t]
\centering
\includegraphics[width=0.92\linewidth]{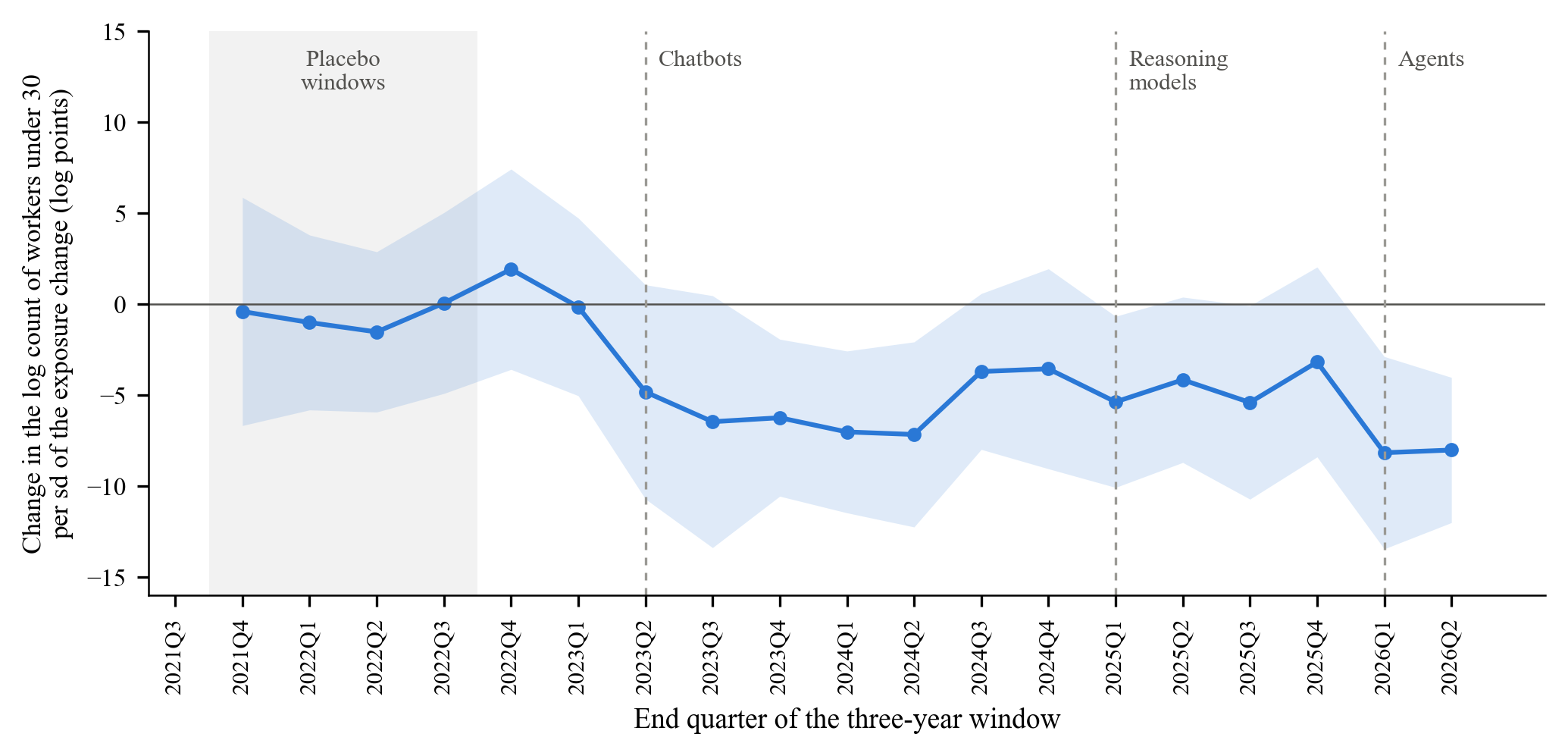}
\caption{The three-year difference in the number of workers under 30, by end quarter}
\label{fig:byquarter}
\begin{minipage}{0.96\textwidth}
\notesize
Notes: Each point is the coefficient of Eq.~(\ref{eq:group}) for the log count of workers under 30 from a separate regression for the three-year window ending in the quarter shown, per standard deviation of the stage-1 to stage-4 change in exposure, with its 95 percent interval. The placebo windows, shaded, end between 2021Q4 and 2022Q3 and compare each quarter with the same quarter three years earlier; the dashed lines mark the inflection quarters. Appendix Figure~\ref{fig:fd-groups-byquarter} shows all three age groups and five outcomes.
\end{minipage}
\end{figure}

Figure~\ref{fig:timing} places the headline three-year estimate for workers under 30 beside two other three-year windows and beside the occupation fixed-effects estimates, with the two older age groups for comparison.
In the placebo window, which compares 2021Q4 through 2022Q3 with the same quarters three years earlier, the estimate for workers under 30 is $-0.70$ log points (2.06).
Hence, we find no evidence that exposed occupations were already shedding young workers before generative AI, or that the pandemic left a lasting difference in young-worker employment between more and less exposed occupations on the eve of the ChatGPT release.
Moving the base period back to 2022, before the release of ChatGPT, gives $-4.38$ (1.85) for the four windows ending in 2025, about half the headline estimate, which is consistent with the decline continuing through 2025 and 2026.
The fixed-effects design of Eq.~(\ref{eq:fe}) uses every quarter of the generative AI era and the exposure prevailing in each quarter, and Table~\ref{tab:fe-body} reports its estimates for the five main outcomes.
Its pooled estimate for workers under 30 is $-5.05$ log points (2.11).
The estimate is $-3.46$ (2.31) before the reasoning model inflection and $-4.35$ (1.88) after (Appendix Table~\ref{tab:cellfe-polished-primary-occ2010-groups} reports the estimate before the inflection and the shift at it, and Figure~\ref{fig:timing} the estimate after it).
For workers aged 30 to 49 and 50 to 64, no window and neither design shows a decline.
Figure~\ref{fig:byquarter} plots the three-year difference for workers under 30 for every end quarter from 2021Q4 to 2026Q2.
Once generative AI reaches the workplace the estimate turns negative and is largest in the windows ending in 2026.\footnote{The estimates for workers under 30 are also negative in the windows ending between 2023Q2 and 2024Q4. These windows have base periods in 2020 and 2021. In those years young workers in less exposed occupations had lost far more jobs than those in more exposed occupations and had not yet recovered. A three-year change from such a base period therefore picks up the recovery of young employment in less exposed occupations, which appears as a relative decline in more exposed ones. We do not read these windows as effects of AI. The placebo windows compare periods before the pandemic with periods after the recovery, so they are not affected.}
The estimate is $-3.18$ log points (2.66) in the window ending in 2025Q4 and $-8.16$ (2.70) in the window ending in 2026Q1.
Hence, the headline estimate of about 8 percent rests on the two windows ending in 2026, whereas the windows ending in 2025 give 3 to 5 percent.
The fixed-effects design averages the relationship over every quarter from 2023Q1 to 2026Q2, and its estimate of about 5 percent is in line with the windows ending in 2025 (Table~\ref{tab:fe-body}, Panel B).
We note that the first half of 2026 is also the AI agent stage, and two quarters of data cannot separate the agent stage from other changes in early 2026.

\begin{table}[htbp]
\centering
\begin{threeparttable}
\caption{Occupation fixed-effects estimates, 2023Q1 to 2026Q2}
\label{tab:fe-body}
\begin{spacing}{1}\small\setlength{\tabcolsep}{3pt}
\begin{tabular*}{\linewidth}{@{\extracolsep{\fill}}>{\raggedright\arraybackslash}p{3.1cm}*{5}{c}@{}}
\toprule
 & \shortstack{Log\\employment} & \shortstack{Unemployment\\rate (pp)} & \shortstack{Main-job\\hours} & \shortstack{Log real\\hourly earnings} & \shortstack{Part-time\\share (pp)} \\
\midrule
\multicolumn{6}{@{}l}{\emph{A. All workers}} \\
Pooled, Eq.~(\ref{eq:fe}) & -0.60 (1.35) & 0.09 (0.13) & 0.01 (0.08) & -0.12 (0.73) & -0.78\textsuperscript{***} (0.28) \\
\addlinespace
\multicolumn{6}{@{}l}{\emph{Era split, Eq.~(\ref{eq:feshift})}} \\
Before 2025Q1 & -0.51 (1.74) & 0.27 (0.22) & -0.25\textsuperscript{**} (0.12) & -0.59 (1.09) & -0.14 (0.42) \\
Shift at 2025Q1 & -0.06 (0.96) & -0.10 (0.11) & 0.14\textsuperscript{***} (0.05) & 0.27 (0.46) & -0.37\textsuperscript{*} (0.19) \\
From 2025Q1 & -0.56 (1.28) & 0.16 (0.15) & -0.10 (0.09) & -0.32 (0.80) & -0.50 (0.31) \\
\midrule
$N$ & 6,295 & 6,295 & 6,295 & 6,108 & 6,295 \\
\midrule
\multicolumn{6}{@{}l}{\emph{B. Pooled coefficient by group, Eq.~(\ref{eq:fe}) estimated by age group and by education}} \\
Under 30 & -5.05\textsuperscript{**} (2.11) & -0.26 (0.31) & 0.08 (0.19) & 0.32 (1.39) & -0.17 (0.64) \\
30 to 49 & 3.03\textsuperscript{*} (1.55) & 0.09 (0.16) & 0.02 (0.09) & -0.31 (1.08) & -0.85\textsuperscript{***} (0.28) \\
50 to 64 & -2.87 (1.80) & 0.49\textsuperscript{***} (0.17) & -0.36\textsuperscript{***} (0.12) & -1.54 (1.32) & -0.14 (0.42) \\
Less than a bachelor's degree & -3.53\textsuperscript{*} (1.81) & -0.25 (0.19) & -0.02 (0.12) & 0.87 (1.03) & -0.59 (0.39) \\
Bachelor's degree or more & -0.79 (2.06) & 0.42\textsuperscript{*} (0.23) & 0.16 (0.10) & -0.05 (1.28) & -0.91\textsuperscript{**} (0.38) \\
\bottomrule
\end{tabular*}
\end{spacing}
\begin{tablenotes}\notesize
\item Notes: Each coefficient is for a one-standard-deviation increase in the step exposure score, the score of the capability stage prevailing in the quarter, standardized within the window, in occupation-quarter regressions with occupation and quarter fixed effects, weighted by the occupation's 2019 employment, with standard errors clustered by occupation in parentheses. The era-split rows report the coefficient before 2025Q1, the change at the reasoning model inflection, and the coefficient from 2025Q1 on. Log outcomes and shares are multiplied by 100; hours are in hours. Panel B reports the pooled coefficient of the same regression estimated separately by age group and by education, weighted by the group's 2019 employment in the occupation, with occupation-by-group and quarter-by-group fixed effects (Appendix Table~\ref{tab:cellfe-polished-primary-occ2010-groups} gives the era split by group). Appendix Table~\ref{tab:fe-main} adds the window from 2021Q3 and the individual-level estimates. \textsuperscript{*}$p<0.10$, \textsuperscript{**}$p<0.05$, \textsuperscript{***}$p<0.01$.
\end{tablenotes}
\end{threeparttable}
\end{table}

That the decline falls on the youngest workers is what the theories of expertise predict \citep{autor2025expertise,ide2025knowledge,garicano2025training}, and the same pattern appears in payroll data \citep{brynjolfsson2026canaries}.
The estimates here show it in nationally representative public data and with an exposure measure that changes as each generation of the technology arrives.

Exposed occupations also changed the types of positions.
In the three-year comparison, each additional standard deviation of the increase in exposure is associated with a decline of 0.74 percentage points (0.25) in the share of workers who usually work part-time, the most precisely estimated coefficient in Table~\ref{tab:headline}, and the placebo estimate is $-0.00$ (0.20).
The number of part-time workers fell by 7.34 log points (2.51) while the number of full-time workers did not detectably change, at $-0.87$ (1.43) (Appendix Table~\ref{tab:cell_fd_composition_1}).
The part-time share fell among workers aged 30 to 49 and 50 to 64 and among graduates, but not among workers under 30, whose full-time and part-time counts fell together (Table~\ref{tab:cell_fd_main_4}).
The fixed-effects design gives a pooled estimate of $-0.78$ percentage points (0.28) for the part-time share.
Split at the reasoning model inflection, the estimates are $-0.14$ (0.42) before and $-0.50$ (0.31) after, neither significant on its own (Table~\ref{tab:fe-body}).
Average weekly hours rose by 0.20 hours (0.09) in the three-year comparison, with a placebo of $-0.08$ (0.07), but the fixed-effects design shows no rise, at 0.01 (0.08) pooled (Table~\ref{tab:fe-body}).
Because part-time workers work far fewer hours than full-time workers, having fewer of them accounts for most of the rise in average hours in the three-year comparison.
\FloatBarrier

\section{Robustness}
\label{sec:robustness}

\begin{figure}[t]
\centering
\includegraphics[width=\linewidth]{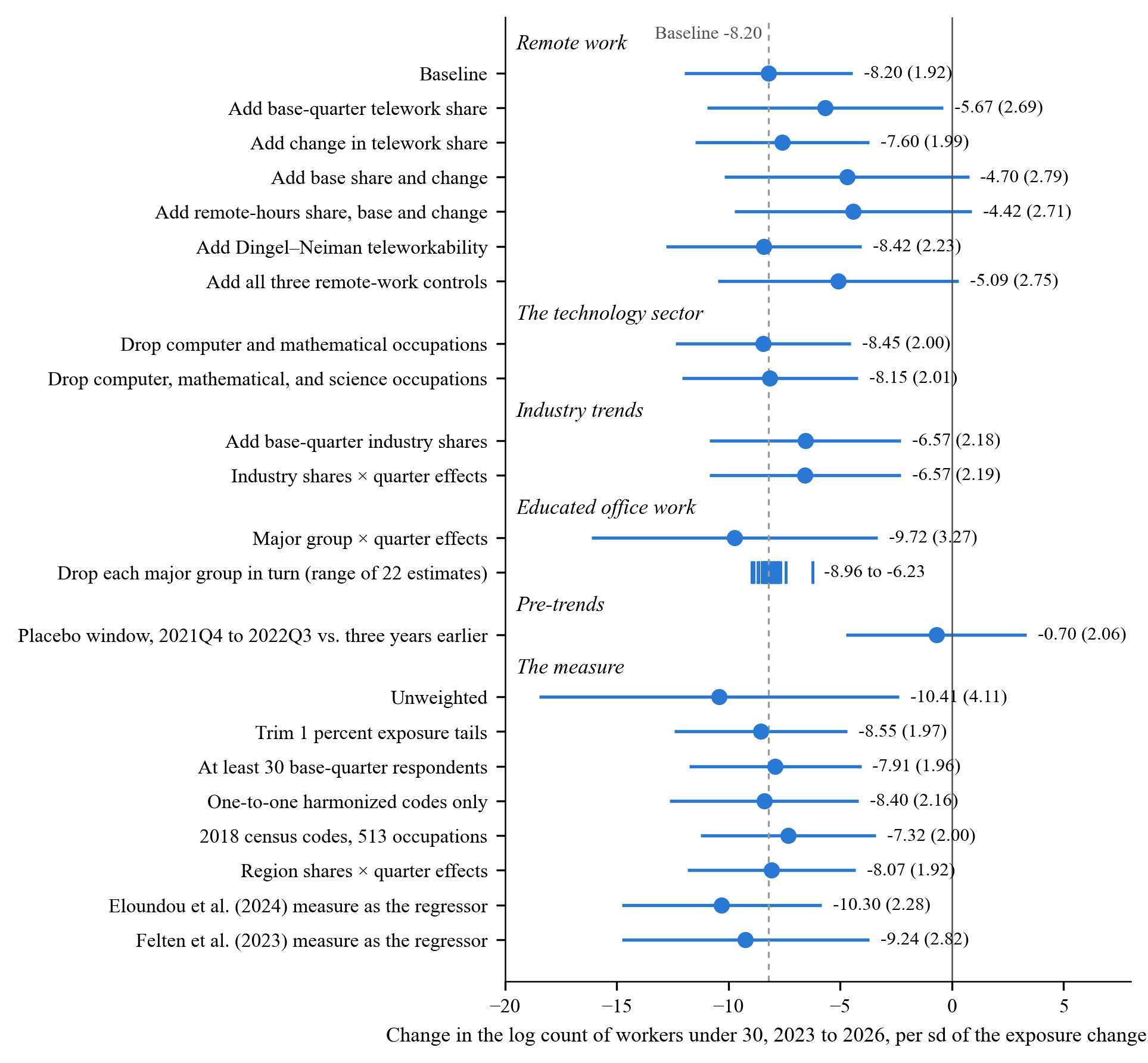}
\caption{The decline in workers under 30 under the specifications that address each alternative explanation}
\label{fig:alternatives}
\begin{minipage}{0.96\textwidth}
\notesize
Notes: Each marker is the coefficient on the standardized change in exposure from stage 1 to stage 4 in the three-year first difference of Eq.~(\ref{eq:group}) for the log count of workers under 30, 2026Q1 vs.\ 2023Q1 and 2026Q2 vs.\ 2023Q2 stacked, with its 95 percent interval; the dashed line is the baseline. Regressions are weighted by base-period young employment, except the unweighted row, and include the twelve composition controls and period effects, with standard errors clustered by occupation. The remote-work rows add the base-quarter CPS telework share, its change over the window, the remote-hours share, or the \citet{dingel2020home} teleworkability index (Appendix Table~\ref{tab:cell_fd_telework_age_18_29_1}); the leave-one-out row marks the 22 estimates from dropping each major group in turn; the other rows are the specification checks of Appendix Table~\ref{tab:cell_fd_spec_secondary_1}, the placebo window of Appendix Table~\ref{tab:cell_fd_main_11}, and the 2018 census codes of Appendix Table~\ref{tab:cell_fd_main_17}. The last two rows replace our change in exposure with the standardized measure of \citet{eloundou2024gpts} or \citet{felten2023language}.
\end{minipage}
\end{figure}

The estimates of Section~\ref{sec:findings} are differences between occupations whose exposure rose more and occupations whose exposure rose less, controlling for composition. 
The question is whether something other than advancing AI capability could have produced those estimates.
This section examines alternative explanations that could explain a decline in young workers, and Figure~\ref{fig:alternatives} reports the under-30 employment estimate under the specification that addresses each. Appendix~F reports the same checks for the other outcomes and specifications.
The first is remote work.
\citet{emanuel2026remote} estimate that remote work explains about two-thirds of the increase in unemployment among young college graduates, and \citet{lambert2026ladder} find that the coefficient on AI attenuates and often loses significance once remote-work exposure is added.
The reasoning is that remote work makes training inexperienced workers more costly, which reduces employers' incentives to hire them.
This matters for our results because occupations with larger increases in AI exposure also tend to offer greater opportunities for remote work.
We use the CPS telework responses to construct, for each occupation and quarter, the share of workers who teleworked and the share of hours worked remotely, and the \citet{dingel2020home} index of work-from-home feasibility as a measure of potential rather than realized remote work.\footnote{Since October 2022, the CPS has asked respondents who worked during the reference week whether they teleworked or worked at home for pay and, if so, for how many hours.}
Controlling for the change in the telework share over the window leaves the under-30 estimate at $-7.60$ log points (1.99), close to the baseline of $-8.20$ and significant at the 1 percent level.
The base-quarter telework share reduces the estimate more, to $-5.67$ (2.69).
Controlling for both the share and its change gives $-4.70$ (2.79), significant at the 10 percent level, and the remote-hours measures give $-4.42$ (2.71), which is not statistically significant.
Some further reduction when several telework variables enter together is to be expected, since they are correlated with each other and with exposure, so that each additional control removes variation and raises the standard error.
With all three controls the estimate is $-5.09$ (2.75).
Moreover, telework changes are measured over the same period as employment and may themselves respond to AI, so controlling for them could attenuate the estimated decline.
In short, holding fixed the change in remote work over the window leaves the decline in young workers almost unchanged.
Holding fixed the level of remote work as well reduces the estimate by a third to a half.
The estimate is then significant at the 5 percent level in one of the four specifications and at the 10 percent level in two.
However, the decline keeps its sign and at least half of its size in every specification.
This differs from \citet{lambert2026ladder}, in whose data the coefficient on AI exposure falls to about zero once remote work is held fixed.
In ours, the estimate is unchanged when the share of an occupation's work that can be done from home is held fixed, and at least half of it remains under every remote-work control.

The second explanation is the contraction of the technology sector, which began in late 2022, just as ChatGPT was released.
Dropping computer and mathematical occupations, alone or with the life, physical, and social science occupations, leaves the under-30 estimate at $-8.45$ (2.00) and $-8.15$ (2.01) and every other estimate close to baseline, and Section~\ref{sec:decomposition} showed that computer and mathematical occupations contribute only 2 percent of the under-30 coefficient.

The third is industry-specific trends, since exposed occupations are concentrated in some industries.
Adding base-quarter industry shares, alone or interacted with quarter effects, reduces the under-30 estimate to $-6.57$ (2.18), still significant at the 1 percent level, so about a fifth of the baseline estimate can be attributed to differences across industries.

The fourth is that educated office work as a whole fared differently after 2022 for reasons unrelated to AI, such as a differential unwinding of the pandemic hiring boom.
Since three quarters of the variance in exposure changes lies between the 22 major occupation groups (Section~\ref{sec:cps}), replacing the quarter effects with major-group-by-quarter effects removes most of the identifying variation and compares occupations only with others in the same group.
Under this specification the under-30 decline is $-9.72$ log points (3.27), significant at the 1 percent level.
Dropping each major group in turn gives estimates between $-8.96$ and $-6.23$, significant in every case, with the smallest when office and administrative support occupations are excluded.

The fifth is a pre-existing trend.
The placebo window, which compares 2021Q4 through 2022Q3 with the same quarters three years earlier, gives $-0.70$ (2.06) for workers under 30.
The single-window estimates of Figure~\ref{fig:byquarter} are close to zero before generative AI.
The differential unemployment trend of the pandemic recovery does not appear in the employment of young workers.

The last concerns the measure and the sample.
Without weights the estimate is $-10.41$ (4.11), significant at the 5 percent level.
It is between $-7.91$ and $-8.55$ when the top and bottom 1 percent of exposure changes are trimmed, when occupations with fewer than 30 base-quarter respondents are dropped, and when the sample is restricted to occupations that map one-to-one across the 2020 classification revision.
It is $-8.07$ (1.92) with region shares interacted with quarter effects.
On the data built with the 2018 census codes, which cover 513 occupations from 2020 onward, it is $-7.32$ (2.00), so the findings do not depend on the occupation-code crosswalk.
The change in exposure from stage 1 to stage 3 ranks occupations almost identically to the change from stage 1 to stage 4.
It gives nearly identical estimates, at $-10.65$ (2.78) against $-9.80$ (2.58) for young workers without a degree (Appendix Table~\ref{tab:cell_fd_main_6}).
The models that rate our score ran in 2026, after the outcomes we study, so their ratings of the earlier stages could in principle reflect what had happened in the labor market.
The measures of \citet{eloundou2024gpts}, rated by GPT-4 in March 2023, and of \citet{felten2023language} predate the outcomes.
With each of them in place of our change in exposure, standardized in the same way, the under-30 estimate is $-10.30$ log points (2.28) and $-9.24$ (2.82).
Hence, the decline among young workers does not hinge on our measure or on hindsight in the ratings.

Appendix~F applies the same checks to the other outcomes and to the fixed-effects designs.
The decline in the part-time share keeps its sign in every specification but loses statistical significance without weights or with major-group-by-quarter effects, the hours increase at the reasoning model inflection remains significant under most specifications but not all, and the three-year increase in average hourly earnings is the least robust, losing significance without weights, with industry controls, or with the CPS telework-share controls.
\FloatBarrier

\section{Conclusion}
\label{sec:conclusion}

Whether generative AI is taking away jobs continues to be hotly debated.
This paper examines the question using the CPS and an occupational AI exposure score that is updated at each generation of the technology, from chatbots to reasoning models and agents, so that exposure varies within occupations over time rather than being fixed at one date.
We compare occupations of similar workforce composition that differ in how much their exposure increased.
The number of workers under 30 in exposed occupations decreased by about 8 percent per standard deviation of the change in exposure between the first half of 2023 and the first half of 2026, and by about 10 percent among those without a college degree.
The decline is largest in the most recent windows.
The windows ending in 2025 and the fixed-effects design, which averages over every quarter from 2023 to 2026, put it at 3 to 5 percent.
The same comparison shows no such change in the years before the technology existed.
The decline is unchanged when the change in remote work over the period is held fixed.
At least half of it remains, with less precision, when the level of remote work is held fixed as well.

A decomposition of the estimate by occupation shows that the decline is in office and administrative support, and in customer service jobs in particular, rather than in the technical occupations in which AI capabilities advanced the most.
Part of the estimate reflects the comparison group, since young employment in healthcare and other in-person services held up whereas their exposure changed little.
The jobs that decreased for young workers are the clerical and customer-facing ones, in which much of the work is routine information handling.
The unemployment rate of young workers in these occupations did not increase, which is consistent with less entry rather than more job loss.
College graduates aged 30 to 49 in the same occupations gained full-time positions and hours.
Total employment in exposed occupations changed by $-1.89$ log points (1.45), which is not significantly different from zero.
This estimate is about the size implied by the decline among young workers alone, since workers under 30 are about a quarter of employment in these occupations.
The overall unemployment rate did not change either, at 0.16 percentage points (0.12).
However, it rose by about half a percentage point among college graduates, whose number also increased, which points to more churn among graduates in these occupations.
Exposed occupations also offered fewer part-time positions, and their full-time workforce became older and more educated.
This shift in composition accounts for the rise in their average wage, since workers of given characteristics were not paid more.

The flat aggregate and the changes in the composition of the workforce documented here are an early snapshot of the labor market's response to generative AI and will likely evolve.

Our evidence has limitations and should be read with them in mind.
The CPS is small at the occupation level.
It tells us who works in which occupation but not which tasks they perform or whether their employer has adopted AI.
The exposure score measures what AI could do in each occupation and, though validated against adoption, is not a measure of adoption.
The AI agent era is only two quarters old in our data.

That aggregate employment has not changed does not mean that the labor market has been unaffected.
The divergences across groups may matter as much for understanding AI's impact as the aggregate does, and understanding them in finer detail will require data on tasks, on workers' use of AI, and on firm adoption.
Whether the changes we document are the beginning of a larger adjustment remains to be seen, and continued examination as new data arrive would help shed light on how AI and human labor will interact in the years ahead.

\clearpage
\bibliographystyle{aer}
\bibliography{references}

\clearpage
\setcounter{section}{0}
\setcounter{table}{0}
\setcounter{figure}{0}
\setcounter{equation}{0}
\renewcommand{\thesection}{\Alph{section}}
\counterwithin{table}{section}
\counterwithin{figure}{section}
\counterwithin{equation}{section}
\begin{center}
{\Large\bfseries Online Appendix\par}
\vspace{0.6em}
{\large Advancing AI Capabilities and Evolving Labor Outcomes\par}
\vspace{0.8em}
{Jacob Dominski, Tyler Boes, and Yong Suk Lee\par}
\vspace{0.4em}
{November 2026\par}
\end{center}
\vspace{1em}

\noindent\textbf{Contents.}
\begin{itemize}\setlength{\itemsep}{0pt}
\item Appendix A --- The AI exposure score: construction, prompts, reliability, and validation
\item Appendix B --- Data: the occupation crosswalk and descriptive figures
\item Appendix C --- Aggregate results in full
\item Appendix D --- Results by age and education in full
\item Appendix E --- The decomposition by occupation
\item Appendix F --- Robustness in full
\item Appendix G --- Placebo windows
\end{itemize}

\clearpage

\section{The AI Exposure Score: Construction, Prompts, Reliability, and Validation}
\label{app:measurement}

This appendix documents how the exposure score of Section~\ref{sec:measurement} is built, so that a reader can reproduce it or adapt it.
Appendices~\ref{app:recipe} to~\ref{app:adapting} are the manual: the steps in order, the stage descriptions and prompts verbatim, the models and their settings and cost, the checks applied to the ratings, and how to extend the score to a new stage, a new model, or another occupation classification.
Appendices~\ref{app:subtasks} to~\ref{app:validation} present supporting evidence on the subtasks, the two versions of the ratings, the agreement across models and across runs, and the validation against adoption.

\subsection{The Recipe}
\label{app:recipe}

The score is built in five steps, which Figure~\ref{fig:pipeline} summarizes.
The inputs are the 19,281 task statements of the O*NET 2019 database, the O*NET task ratings, from which we take the relevance weights, the four stage descriptions of Appendix~\ref{app:stages}, and the four prompts of Appendix~\ref{app:prompts}.
First, one model, Claude Haiku 4.5, splits each task into three to seven subtasks, each with an integer share of the task's time and one of seven activity types (Prompt 0).
This step is run once, and the subtasks are held fixed for every model, so that differences between models reflect their judgments about the same subtasks and not different ways of splitting the task.
Second, each of the three models rates every subtask at each of the four stages on three questions (Prompts A, B, and C): the percent of the subtask's time that AI could save, whether AI could complete the subtask alone to a professional standard, and whether people keep a durable advantage on it.
Ratings A and B are given twice, once for the technology alone and once after the economic, regulatory, and organizational constraints on using it, and the paper uses the constraint-adjusted version.
Third, each response is checked against its output schema, the task-level ratings are recomputed from the subtask ratings and their time shares, and the ratings are made monotone across stages (Appendix~\ref{app:checks}).
Fourth, the three task-level ratings are combined into the task's exposure score at each stage, $E = [A + B + (100 - C)]/3$, for each model, and the three models' scores are averaged.
Fifth, an occupation's score at each stage is the relevance-weighted mean of its tasks' scores, and the O*NET occupation scores are carried to the census occupation codes of the CPS through the crosswalk of Appendix~\ref{app:data}.
The outputs are a task-level file with every rating of every model at every stage and an occupation-level file with the scores at the four stages, from which the changes between stages used in the paper are differences.

\begin{figure}[htbp]
\centering
\begin{tikzpicture}[
  node distance=4mm and 6mm,
  box/.style={draw, rounded corners=2pt, align=left, text width=7.2cm, inner sep=5pt, font=\small},
  side/.style={align=left, text width=5.6cm, font=\footnotesize},
  arr/.style={-{Latex[length=2mm]}, thick}
]
\node[box] (s1) {\textbf{1. Occupations to tasks}\\ O*NET: 19,281 task statements across 879 occupations};
\node[box, below=of s1] (s2) {\textbf{2. Tasks to subtasks}\\ Each task split once into 3 to 6 subtasks with time shares $\omega_{it}$; the split is frozen};
\node[box, below=of s2] (s3) {\textbf{3. Three ratings per subtask, at each of four stages}\\ A: percent of time AI could save\\ B: could AI complete it alone (0/1)\\ C: does a person keep a durable advantage (0/1)};
\node[box, below=of s3] (s4) {\textbf{4. Subtasks to a task score}\\ Time-weighted averages $A_t, B_t, C_t$; composite $E_t = \tfrac{1}{3}[A_t + B_t + (100 - C_t)]$};
\node[box, below=of s4] (s5) {\textbf{5. Three raters}\\ Language models from Anthropic, OpenAI, and Google each rate every task; the score averages them at the task level};
\node[box, below=of s5] (s6) {\textbf{6. Tasks to occupations}\\ Relevance-weighted task average $E_o(s)$; crosswalk to 513 census codes and 473 harmonized codes};
\draw[arr] (s1) -- (s2);
\draw[arr] (s2) -- (s3);
\draw[arr] (s3) -- (s4);
\draw[arr] (s4) -- (s5);
\draw[arr] (s5) -- (s6);
\node[side, right=of s3] {Inputs to step 3: the four stage descriptions (Appendix~\ref{app:stages}), with example systems. Raters see only the task text, the subtasks, and the stage text.};
\node[side, right=of s6] {Output: one score per occupation at each stage, $E_o(1)$ to $E_o(4)$. Stage differences are simple subtractions.};
\end{tikzpicture}
\caption{From tasks to occupation scores}
\label{fig:pipeline}
\begin{minipage}{0.92\textwidth}
\notesize
Notes: The steps that produce an occupation's exposure score at each capability stage. The subsections below give the prompts, output formats, checks, and weighting variants for each step.
\end{minipage}
\end{figure}

\subsection{Stage Definitions, Verbatim}
\label{app:stages}

The release and inflection dates of Section~\ref{sec:construction} of the paper rest on the following sources. ChatGPT was released in November 2022 \citep{openai2022chatgpt} and reached 100 million monthly users within two months \citep{hu2023chatgpt}; GPT-4 followed in March 2023 \citep{openai2023gpt4}. OpenAI's o1 was released in September 2024 \citep{openai2024o1}; in January 2025 DeepSeek released R1 with open weights \citep{deepseek2025r1}, its app became the most downloaded in the United States within a week \citep{reuters2025deepseek}, and OpenAI made o3-mini available to all ChatGPT users \citep{openai2025o3mini}. Claude Code was released in May 2025 \citep{anthropic2025claude4}; in January 2026 Anthropic released Claude Cowork \citep{anthropic2026cowork}, and the number of weekly active users of Claude Code doubled between January 1 and mid-February 2026 \citep{anthropic2026seriesg}.
The four capability-stage descriptions below were spliced into every scoring prompt exactly as shown; the file calls stage 1 ``Traditional Machine Learning'', which the paper calls pre-generative machine learning.
Table~\ref{tab:stages-cando} condenses them.

\begin{table}[htbp]
\centering
\begin{threeparttable}
\caption{What AI can and cannot do at each capability stage, condensed}
\label{tab:stages-cando}
\footnotesize\setlength{\tabcolsep}{4pt}
\begin{spacing}{1}
\newcolumntype{L}[1]{>{\raggedright\arraybackslash}p{#1}}
\begin{tabular}{@{}L{3.0cm}L{5.6cm}L{5.6cm}@{}}
\toprule
Stage & Can do & Cannot do \\
\midrule
1. Pre-generative machine learning & Classification, forecasting, optical character recognition, rule-based process automation & Generate coherent text, reason across domains, handle ambiguity \\
\addlinespace
2. Chatbots & Fluent text, summarization, conversational answers, code drafting, multimodal understanding & Reason reliably beyond a few steps, avoid fabrication, use tools or memory \\
\addlinespace
3. Reasoning models & Multi-step analytical reasoning at professional-exam level, self-critique, far less fabrication & Work beyond one session or without a person framing each task \\
\addlinespace
4. AI agents & Autonomous multi-step execution with tools, persistence over hours or days, self-correction & Run long chains reliably, act in the physical world, carry legal authority or client trust \\
\bottomrule
\end{tabular}
\end{spacing}
\begin{tablenotes}\notesize
\item Notes: A condensation of the stage descriptions reproduced verbatim below, which are the only text the rating models see. Example systems and dates are in Table 2.1 of the paper.
\end{tablenotes}
\end{threeparttable}
\end{table}

\begin{spacing}{1}
\begin{Verbatim}
### Stage 1: Traditional Machine Learning

CORE CAPABILITIES:
- Supervised learning: classification, regression, ranking on structured/tabular data
- Unsupervised learning: clustering, anomaly detection, dimensionality reduction
- Time-series forecasting and demand prediction
- Rule-based and statistical NLP: named entity recognition, sentiment classification, keyword extraction, spam filtering, topic modeling (LDA), TF-IDF search
- Computer vision (CNNs): image classification, object detection, facial recognition, OCR (optical character recognition)
- Robotic process automation (RPA): rule-based automation of repetitive digital workflows (form filling, data entry, file transfers between systems)
- Recommendation engines and personalization algorithms
- Optimization algorithms (scheduling, routing, resource allocation)

KEY LIMITATIONS:
- Cannot generate coherent free-form text, creative content, or natural dialogue
- Cannot reason across domains or handle novel, unstructured problems
- Cannot understand context, nuance, or ambiguity in language
- Requires labeled training data and task-specific model development
- Brittle when inputs deviate from training distribution
- No ability to explain reasoning or engage in multi-step logical analysis

EXAMPLE SYSTEMS: XGBoost, random forests, logistic regression, BERT (classification fine-tuned), spaCy NER pipelines, Tesseract OCR, UiPath/Automation Anywhere (RPA), TensorFlow object detection, scikit-learn pipelines.

### Stage 2: Foundation Models — LLMs and Multimodal AI

CORE CAPABILITIES:
- Coherent text generation: drafting, summarization, paraphrasing, translation across dozens of languages
- Conversational question-answering and instruction-following
- Zero-shot and few-shot task performance without task-specific training
- Code generation, debugging, and explanation in major programming languages
- Multimodal understanding: process images, audio, video, and documents as inputs
- Multimodal generation: create images, audio, and video from text descriptions
- Basic reasoning and analysis on short, self-contained problems
- Information synthesis across documents provided in context
- Style adaptation: adjust tone, register, reading level for audience

KEY LIMITATIONS:
- Prone to hallucination: confidently generates plausible but incorrect facts
- Unreliable on tasks requiring precise factual recall or numerical computation
- Limited multi-step reasoning — performance degrades on chains longer than 3-4 steps
- No persistent memory across sessions; cannot learn or update from experience
- Cannot access external tools, databases, or live information independently
- Context windows of 4K-128K tokens constrain input length
- Cannot verify its own outputs or detect its own errors
- Struggles with tasks requiring spatial reasoning, physical intuition, or causal inference
- No ability to take actions in the world — generates text/media only

EXAMPLE SYSTEMS: GPT-3.5, GPT-4, Claude 2/3, Gemini 1.0/1.5, Llama 2/3, DALL-E 3, Midjourney, Stable Diffusion, Whisper (transcription).

### Stage 3: Reasoning Models

CORE CAPABILITIES (in addition to all Stage 2 capabilities):
- Extended chain-of-thought reasoning: reliably solves multi-step logical, mathematical, and analytical problems (10+ reasoning steps)
- High factual consistency: significantly reduced hallucination rates compared to Stage 2
- Strong performance on expert-level tasks in law, medicine, science, finance, and software engineering (e.g., passes professional licensing exams at high percentiles)
- Can decompose complex problems into sub-problems and solve them sequentially
- Improved calibration: better at expressing uncertainty and flagging knowledge gaps
- Capable of reviewing, critiquing, and iterating on its own outputs when prompted

KEY LIMITATIONS:
- Still operates within a single session — no persistent learning or memory
- Cannot autonomously access tools, browse the web, or execute code unless a human initiates the interaction and provides the interface
- Requires human-in-the-loop: a person must frame the task, provide inputs, and evaluate outputs
- Cannot manage multi-step workflows that span multiple systems or time periods
- Cannot monitor ongoing processes, respond to real-time changes, or take initiative
- Performance still degrades on highly novel problems outside training distribution
- Does not replace the need for domain expertise to evaluate output quality

EXAMPLE SYSTEMS: o1, o3, Claude Sonnet 4.5 (extended thinking), Gemini 2.0 Flash Thinking, DeepSeek-R1.

KEY DISTINCTION FROM STAGE 2: Stage 3 models are dramatically better at complex analytical tasks but still require a human operator to drive the workflow. They make workers more productive; they do not replace the worker's role in the process.

### Stage 4: AI Agents

CORE CAPABILITIES (in addition to all Stage 3 capabilities):
- Autonomous multi-step task execution: plans, executes, monitors, and adjusts workflows with minimal human oversight
- Tool use: independently calls APIs, queries databases, executes code, manages files, sends communications, and navigates software interfaces
- Error detection and self-correction: identifies when outputs are wrong or when a plan is failing, and revises approach
- Persistent operation: can work on tasks over extended periods (hours to days), maintaining state and context
- Coordination: can orchestrate multiple sub-agents or tools to accomplish complex goals
- Learns from feedback within a task: adjusts strategy based on intermediate results

KEY LIMITATIONS:
- Reliability is imperfect: autonomous operation introduces compounding error risk across long task chains
- Cannot handle situations requiring physical presence, embodied action, or real-time physical manipulation
- Cannot navigate truly novel situations that require human judgment, ethical reasoning, or stakeholder negotiation
- Cannot build trust, empathy, or interpersonal rapport
- Accountability and oversight gaps: when errors occur, attribution and correction are harder than with human workers
- Current systems are domain-specific — general-purpose autonomous agents remain limited
- Cannot replace tasks that require professional licensure, legal authority, or fiduciary responsibility (e.g., signing contracts, making medical decisions of record)

EXAMPLE SYSTEMS: Claude with computer use, Devin (software engineering), multi-agent orchestration frameworks (CrewAI, AutoGen), Operator-style browser agents.

KEY DISTINCTION FROM STAGE 3: Stage 4 systems can independently execute workflows that Stage 3 models can only assist with. The human shifts from operator to supervisor.
\end{Verbatim}
\end{spacing}

\subsection{Prompts 0, A, B, and C, Verbatim}
\label{app:prompts}
Prompt 0 decomposes each task into subtasks with time shares, and Prompts A, B, and C produce ratings A, B, and C for each subtask at each stage; each prompt ends with the output schema the model was required to follow.

\subsubsection{Prompt 0: subtask decomposition}
\begin{spacing}{1}
\begin{Verbatim}
You are an expert occupational analyst. Your task is to decompose a specific
occupational task from the O*NET database into its constituent subtasks — the
distinct activities a worker performs to carry out the full task.

This decomposition will be used as a fixed input to downstream AI exposure
assessments. It must be NEUTRAL with respect to AI capability — describe what a
human worker does, not what AI could do.

---

## DECOMPOSITION INSTRUCTIONS

For the given occupation and task:

Step 1: Identify the distinct activities a worker performs to complete this task
from start to finish. Think about the task as a sequence of actions: what does the
worker do first, then next, then next?

Step 2: For each subtask, estimate its share of the total time required to complete
the full task. Shares must sum to 100.

Step 3: Classify each subtask by its PRIMARY activity type from the typology below.

---

## SUBTASK ACTIVITY TYPOLOGY

Classify each subtask into exactly one primary type:

### INFORMATION GATHERING
Collecting, retrieving, reading, or receiving data, documents, or verbal input
needed to perform the task. Includes research, interviewing, observation, and
monitoring.
Examples: reading patient charts, reviewing financial statements, conducting site
inspections, monitoring dashboards, interviewing witnesses.

### INFORMATION PROCESSING
Analyzing, synthesizing, calculating, interpreting, or transforming information
into a different form. Includes quantitative analysis, pattern recognition,
diagnosis, and evaluation.
Examples: analyzing lab results, calculating structural loads, interpreting legal
precedent, evaluating vendor proposals, coding qualitative data.

### CONTENT GENERATION
Producing written, visual, or multimedia output — reports, plans, designs,
correspondence, presentations, or creative work.
Examples: drafting reports, writing code, creating designs, composing emails,
preparing slide decks.

### COMMUNICATION AND COORDINATION
Conveying information to others, negotiating, persuading, teaching, managing
expectations, or coordinating activities across people. Includes meetings,
presentations, and interpersonal exchanges.
Examples: presenting findings to a client, negotiating contract terms, leading a
team meeting, explaining a diagnosis to a patient, delegating subtasks.

### DECISION AND JUDGMENT
Making a choice, rendering a verdict, approving or rejecting, setting strategy,
or exercising professional discretion where the act of deciding is itself the
work product.
Examples: approving a loan application, selecting a treatment plan, making a
hiring decision, setting project priorities, issuing a ruling.

### PHYSICAL EXECUTION
Performing manual, bodily, or sensorimotor activities — operating equipment,
manipulating objects, physically moving through space, or providing hands-on care.
Examples: performing surgery, operating machinery, assembling components, physically
inspecting a site, transporting materials.

### ADMINISTRATIVE AND PROCEDURAL
Routine recordkeeping, filing, scheduling, compliance documentation, data entry,
or following established procedural checklists.
Examples: updating patient records, filing regulatory paperwork, scheduling
appointments, entering data into systems, maintaining logs.

---

## GUIDANCE

- Decompose into 3-7 subtasks. If a task seems to require more, combine closely
  related activities. If fewer than 3, check whether you are describing the task
  at too high a level.
- Time shares must reflect realistic time allocation for a typical worker. The
  most time-consuming activity should have the largest share.
- Order subtasks in the sequence a worker would typically perform them.
- Each subtask should be specific enough that a downstream assessor can judge
  whether AI could perform it, but general enough to avoid listing micro-steps.
- Physical components must be identified even if they are a small share of time.
  Do not fold physical activities into other categories.
- If the task description is ambiguous, decompose the most common interpretation
  for the given occupation.

CRITICAL CONSISTENCY RULES:
- "num_subtasks" MUST exactly equal the number of objects in the "task_decomposition"
  array. Count your subtasks before setting this field.
- "sequence_position" values MUST be sequential integers from 1 to N (where N is the
  number of subtasks), with no gaps, no duplicates, and no skipped values.
- Double-check both of these before returning your response.

---

## OUTPUT FORMAT

Return a JSON object with the following structure. Do not include any text outside
the JSON object.

{
  "occupation": "<O*NET occupation title>",
  "onet_code": "<O*NET-SOC code>",
  "task_id": <integer>,
  "task_text": "<exact O*NET task text>",
  "num_subtasks": <integer, 3-7>,
  "task_decomposition": [
    {
      "subtask_id": <1-indexed integer>,
      "subtask": "<description of the subtask — one concise sentence>",
      "activity_type": "<one of: information_gathering, information_processing,
        content_generation, communication_coordination, decision_judgment,
        physical_execution, administrative_procedural>",
      "pct_of_total_time": <integer, 1-100>,
      "sequence_position": <1-indexed, reflecting typical execution order>
    }
  ],
  "time_share_sum_check": 100
}
\end{Verbatim}
\end{spacing}

\subsubsection{Prompt A: feasible time savings}
\begin{spacing}{1}
\begin{Verbatim}
You are an expert labor economist and AI capability analyst. Your task is to assess how
much AI at each of four capability stages could reduce the time required for a worker to
complete a specific occupational task from the O*NET database.

For each task and stage, provide TWO scores:

1. TECHNICAL CAPABILITY (time_savings_pct_technical): The maximum percentage time
   reduction assuming the AI system is optimally deployed, the worker is fully trained,
   and there are no cost, regulatory, or organizational barriers. This is the ceiling —
   what is technically possible with this class of AI.

2. FEASIBILITY-ADJUSTED (time_savings_pct_feasible): The realistic percentage time
   reduction accounting for the constraints below, as faced by the median employer in
   this occupation.

The feasible score should work DOWN from the technical score. Start from what is
technically possible, then discount for the binding constraints that would prevent
full realization in practice.

FEASIBILITY CONSTRAINTS (apply to feasible score only):
- Economic viability: Implementation costs and expected productivity gains must justify
  deployment for the median employer. Consider hardware, software licensing, integration,
  training, and ongoing maintenance costs relative to labor cost savings.
- Regulatory and legal barriers: Deployment must comply with professional licensing
  requirements, liability rules, data privacy regulations, and industry-specific
  compliance standards applicable to this occupation.
- Organizational readiness: Consider whether the median employer in this occupation has
  the technical infrastructure, data systems, and IT capacity to support deployment.

You will receive a pre-specified subtask decomposition with time shares. Use these
subtasks and time shares as given. Do not modify, add, or remove subtasks. Assess AI
time savings on each subtask, then compute the task-level score as the time-weighted
sum.

---

## AI CAPABILITY STAGES

[INSERT SHARED STAGE DEFINITIONS HERE]

---

## INPUT FORMAT

You will receive:

{
  "occupation": "<O*NET occupation title>",
  "onet_code": "<O*NET-SOC code>",
  "task_id": <integer>,
  "task_text": "<exact O*NET task text>",
  "task_decomposition": [
    {
      "subtask_id": <integer>,
      "subtask": "<description>",
      "activity_type": "<type>",
      "pct_of_total_time": <integer>
    }
  ]
}

---

## ASSESSMENT INSTRUCTIONS

For each subtask provided in the input, estimate how much AI at each stage could
reduce the time a worker spends on that subtask.

SCORING GUIDANCE:
- Both scores use a 0-100% scale representing percentage of total task time saved.
- Both scores must be MONOTONICALLY NON-DECREASING across stages (Stage 4 >= Stage 3
  >= Stage 2 >= Stage 1). If a later stage adds no relevant capability, carry forward
  the prior score.
- The feasible score must be LESS THAN OR EQUAL TO the technical score at every stage.
- Use the full 0-100 range. Do not default to moderate values when uncertain.
- Consider the ENTIRE task via its subtasks, not just the most AI-amenable subtask.
- For the feasible score, identify the binding constraint(s) — economic, regulatory,
  organizational, or a combination.

---

## OUTPUT FORMAT — PRODUCTION

Return a JSON object with the following structure. Do not include any text outside
the JSON object.

{
  "occupation": "<O*NET occupation title>",
  "onet_code": "<O*NET-SOC code>",
  "task_id": <integer>,
  "task_text": "<exact task text>",
  "subtask_assessments": [
    {
      "subtask_id": <integer>,
      "subtask": "<description from input>",
      "pct_of_total_time": <from input>,
      "time_savings_by_stage": {
        "stage_1": {"technical": <0-100>, "feasible": <0-100>},
        "stage_2": {"technical": <0-100>, "feasible": <0-100>},
        "stage_3": {"technical": <0-100>, "feasible": <0-100>},
        "stage_4": {"technical": <0-100>, "feasible": <0-100>}
      }
    }
  ],
  "stage_1": {
    "time_savings_pct_technical": <0-100>,
    "time_savings_pct_feasible": <0-100>,
    "primary_constraint": "<comma-separated list from: none, economic, regulatory, organizational>"
  },
  "stage_2": { ... },
  "stage_3": { ... },
  "stage_4": { ... }
}

## OUTPUT FORMAT — PILOT (add these fields)

Add to each stage block:
  "confidence": "high/medium/low",
  "rationale_technical": "<1-2 sentences: which subtasks can this stage address and how>",
  "rationale_feasible": "<1 sentence: what binding constraint(s) reduce the feasible score, or 'No significant gap' if technical ~ feasible>"

Add to each subtask_assessments entry:
  "rationale": "<1 sentence: what AI capability applies to this subtask at which stage>"
\end{Verbatim}
\end{spacing}

\subsubsection{Prompt B: independent completion to a professional standard}
\begin{spacing}{1}
\begin{Verbatim}
You are an expert labor economist and AI capability analyst. Your task is to assess what
percentage of an occupational task from the O*NET database could be independently completed
by AI at each of four capability stages.

This is a TASK COMPLETION assessment — not a time savings assessment. The question is:
"What share of this task can AI perform to an acceptable professional standard without
human intervention?"

For each task and stage, provide TWO scores:

1. TECHNICAL CAPABILITY (completion_pct_technical): The percentage of the task that AI
   could independently complete assuming optimal deployment, full data access, and no
   cost, regulatory, or organizational barriers.

2. FEASIBILITY-ADJUSTED (completion_pct_feasible): The realistic percentage of the task
   AI could independently complete, accounting for the constraints below, as faced by the
   median employer in this occupation.

The feasible score should work DOWN from the technical score.

FEASIBILITY CONSTRAINTS (apply to feasible score only):
- Economic viability: Implementation costs and expected productivity gains must justify
  deployment for the median employer.
- Regulatory and legal barriers: Deployment must comply with professional licensing
  requirements, liability rules, data privacy regulations, and industry-specific
  compliance standards.
- Organizational readiness: Consider whether the median employer has the technical
  infrastructure, data systems, and IT capacity to support deployment.

You will receive a pre-specified subtask decomposition with time shares. Use these
subtasks and time shares as given. Do not modify, add, or remove subtasks.

HOW TO ASSESS COMPLETION:

For each subtask, make a binary judgment at each stage: could AI at this stage complete
this subtask to an ACCEPTABLE PROFESSIONAL STANDARD without a human performing or
redoing it?

"Acceptable professional standard" means the output would be considered satisfactory by
a supervisor, client, or regulatory body — not necessarily perfect, but meeting the
quality bar expected of a competent human worker.

The task-level completion percentage is the TIME-WEIGHTED share of subtasks that AI can
independently complete.

IMPORTANT DISTINCTIONS:
- "AI completes the subtask" means AI does the work END-TO-END and the output is usable
  as-is. A human may review it, but does not need to redo or substantially revise it.
- If AI can draft something but a human must substantially revise it to reach professional
  standard, that subtask is NOT independently completed by AI.
- If AI produces the final output and a human only performs a brief quality check (similar
  to how a supervisor would review any worker's output), that subtask IS independently
  completed by AI.
- Physical subtasks that require bodily presence, manual dexterity, or real-time physical
  interaction cannot be completed by AI at any current stage.

---

## AI CAPABILITY STAGES

[INSERT SHARED STAGE DEFINITIONS HERE]

---

## INPUT FORMAT

You will receive:

{
  "occupation": "<O*NET occupation title>",
  "onet_code": "<O*NET-SOC code>",
  "task_id": <integer>,
  "task_text": "<exact O*NET task text>",
  "task_decomposition": [
    {
      "subtask_id": <integer>,
      "subtask": "<description>",
      "activity_type": "<type>",
      "pct_of_total_time": <integer>
    }
  ]
}

---

## ASSESSMENT INSTRUCTIONS

SCORING GUIDANCE:
- Both scores use a 0-100% scale representing the time-weighted share of subtasks AI
  can independently complete to professional standard.
- Both scores must be MONOTONICALLY NON-DECREASING across stages.
- The feasible score must be LESS THAN OR EQUAL TO the technical score at every stage.
- Use the full 0-100 range. Do not default to moderate values when uncertain.
- The binary judgment at the subtask level is critical. Avoid partial credit — either
  AI can do the subtask to professional standard independently, or it cannot.

CRITICAL MONOTONICITY RULES FOR SUBTASK BINARY JUDGMENTS:
- If technical=true at Stage N, then technical MUST be true at Stages N+1, N+2, etc.
  Later stages have ALL capabilities of earlier stages plus more. A subtask that AI can
  complete at an earlier stage can ALWAYS be completed at a later stage.
- If feasible=true at Stage N, then feasible MUST be true at Stages N+1, N+2, etc.
  Feasibility constraints do not get WORSE at later stages.
- If feasible=true at any stage, then technical MUST also be true at that stage.
- Before returning your response, verify each subtask's judgments across all four stages
  to confirm these rules are satisfied. If you find a violation, correct it.

---

## OUTPUT FORMAT — PRODUCTION

Return a JSON object with the following structure. Do not include any text outside
the JSON object.

{
  "occupation": "<O*NET occupation title>",
  "onet_code": "<O*NET-SOC code>",
  "task_id": <integer>,
  "task_text": "<exact task text>",
  "subtask_assessments": [
    {
      "subtask_id": <integer>,
      "subtask": "<description from input>",
      "pct_of_total_time": <from input>,
      "ai_can_complete_independently": {
        "stage_1": {"technical": true/false, "feasible": true/false},
        "stage_2": {"technical": true/false, "feasible": true/false},
        "stage_3": {"technical": true/false, "feasible": true/false},
        "stage_4": {"technical": true/false, "feasible": true/false}
      }
    }
  ],
  "stage_1": {
    "completion_pct_technical": <0-100>,
    "completion_pct_feasible": <0-100>,
    "primary_constraint": "<comma-separated list from: none, economic, regulatory, organizational>"
  },
  "stage_2": { ... },
  "stage_3": { ... },
  "stage_4": { ... }
}

## OUTPUT FORMAT — PILOT (add these fields)

Add to each stage block:
  "confidence": "high/medium/low",
  "rationale_technical": "<1-2 sentences: which subtasks can this stage complete independently and why>",
  "rationale_feasible": "<1 sentence: what binding constraint(s) reduce the feasible score>"

Add to each subtask_assessments entry:
  "rationale": "<1 sentence: why AI can or cannot complete this subtask independently at each stage>"
\end{Verbatim}
\end{spacing}

\subsubsection{Prompt C: human comparative advantage}
\begin{spacing}{1}
\begin{Verbatim}
You are an expert labor economist and AI capability analyst. Your task is to assess where
humans retain a COMPARATIVE ADVANTAGE over AI at each of four capability stages for a
specific occupational task from the O*NET database.

Comparative advantage is relative — it depends on what the AI at a given stage can and
cannot do. A human may have strong comparative advantage over Stage 1 ML for a writing
task but lose that advantage at Stage 2 when foundation models can draft competently.
Conversely, a human's advantage in building patient trust may persist across all stages.

For each subtask, the question is: at this AI capability stage, is the subtask performed
MORE EFFECTIVELY by a human, REQUIRED to be done by a human for legal or institutional
reasons, or does it LOSE ITS VALUE if performed by AI?

You will receive a pre-specified subtask decomposition with time shares. Use these
subtasks and time shares as given. Do not modify, add, or remove subtasks.

---

## HUMAN COMPARATIVE ADVANTAGE TYPOLOGY

When identifying human advantages, classify each into one or more of the following types:

### PHYSICAL AND SENSORIMOTOR
The subtask requires bodily presence, manual dexterity, physical manipulation of objects,
real-time sensorimotor coordination, navigation of unstructured physical environments, or
direct physical interaction with people or surroundings. Includes tasks where the physical
co-location of the worker is essential to performing or verifying work.

This advantage is grounded in the persistent difficulty of embodied AI: low-level
sensorimotor skills that are trivial for humans require enormous computational resources
and remain beyond current robotic capability in unstructured environments.

Examples: performing surgery, operating heavy machinery, physically inspecting a
construction site, hands-on patient care, assembling delicate components, navigating
a disaster site.

### INTERPERSONAL AND RELATIONAL
The subtask's effectiveness depends on a genuine human relationship, emotional connection,
perceived authenticity, or the ability to read and respond to social and emotional cues
in real time. This includes not only dyadic trust and rapport but also persuasion,
negotiation, coordination across people, and social-emotional interaction where the human
presence is constitutive of — not incidental to — the value of the work.

Examples: therapy and counseling, negotiating a sensitive business deal, delivering
difficult medical news, mentoring a junior employee, de-escalating a crisis, leading
a team through a difficult transition, building a coalition among stakeholders.

### INSTITUTIONAL AUTHORITY
Legal, regulatory, professional, or social frameworks require or strongly expect a human
in this role for the output to be accepted as valid, legitimate, or authoritative. This
encompasses both hard constraints (a human must serve as the responsible decision-maker,
signatory, or fiduciary by law or regulation) and soft constraints (clients, patients,
the public, or institutional norms expect a human in this role for the output to carry
credibility and legitimacy, even absent legal mandate).

Even if AI produces the analysis or performs the underlying work, a human must own the
decision and bear the consequences, or a human must be present for the output to be
accepted by its intended audience.

Examples:
- Hard: a physician signing a diagnosis, a lawyer advising a client, a CPA certifying
  financial statements, a judge issuing a ruling, an engineer stamping structural plans.
- Soft: a CEO presenting earnings to shareholders, a teacher leading a classroom
  discussion, a clergy member officiating a ceremony, an elected official casting a vote.

### TACIT JUDGMENT AND EXPERTISE
The subtask involves reasoning under genuine novelty or ambiguity, integration of tacit
knowledge and contextual cues, expert improvisation that cannot be reduced to explicit
rules, or creative problem-solving that goes beyond recombination of existing patterns.
This encompasses both situations where historical patterns and training data are
insufficient (novelty-driven) and situations where the practitioner draws on embodied,
experiential, or context-dependent knowledge that is difficult to articulate or capture
in data (expertise-driven).

The core of this advantage is non-codifiable human cognition: the expert knows more than
they can tell, and their performance depends on holistic pattern recognition, flexible
knowledge transfer, and situational judgment that resists decomposition into rules.

Examples: setting strategy during an organizational crisis, adjudicating an ethically
novel case, making triage decisions with incomplete information, an experienced nurse
reading a patient's non-verbal distress signals, a chef adjusting a recipe based on
ingredient quality, a field geologist interpreting rock formations in person, an
experienced trader reading market sentiment.

---

## AI CAPABILITY STAGES

[INSERT SHARED STAGE DEFINITIONS HERE]

---

## INPUT FORMAT

You will receive:

{
  "occupation": "<O*NET occupation title>",
  "onet_code": "<O*NET-SOC code>",
  "task_id": <integer>,
  "task_text": "<exact O*NET task text>",
  "task_decomposition": [
    {
      "subtask_id": <integer>,
      "subtask": "<description>",
      "activity_type": "<type>",
      "pct_of_total_time": <integer>
    }
  ]
}

---

## ASSESSMENT INSTRUCTIONS

For each subtask AND each AI capability stage, make a judgment: does a human retain
comparative advantage over AI at this stage? Apply the typology above. A subtask has
human comparative advantage at a given stage if at least one advantage type applies
relative to the capabilities and limitations of AI at that stage.

Then compute the HUMAN ADVANTAGE SHARE at each stage — the time-weighted percentage
of the task where humans retain comparative advantage.

IMPORTANT GUIDANCE:
- Human advantage shares must be MONOTONICALLY NON-INCREASING across stages
  (Stage 1 >= Stage 2 >= Stage 3 >= Stage 4).
- Multiple advantage types can apply to the same subtask. List all that apply.
- Some subtasks will have NO human comparative advantage at any stage.
- Physical and sensorimotor advantage is typically the most persistent.
- Institutional authority advantages can shift if legal frameworks change, but assess
  based on current institutional structures.
- Tacit judgment and expertise advantages erode as AI reasoning improves across stages,
  but may persist for genuinely novel or unprecedented situations even at Stage 4.
- Use the full 0-100 range.

---

## OUTPUT FORMAT — PRODUCTION

Return a JSON object with the following structure. Do not include any text outside
the JSON object.

{
  "occupation": "<O*NET occupation title>",
  "onet_code": "<O*NET-SOC code>",
  "task_id": <integer>,
  "task_text": "<exact task text>",
  "subtask_assessments": [
    {
      "subtask_id": <integer>,
      "subtask": "<description from input>",
      "pct_of_total_time": <from input>,
      "human_has_advantage": {
        "stage_1": true/false,
        "stage_2": true/false,
        "stage_3": true/false,
        "stage_4": true/false
      },
      "advantage_types": {
        "stage_1": ["<from: physical_sensorimotor, interpersonal_relational, institutional_authority, tacit_judgment_expertise>"],
        "stage_2": ["..."],
        "stage_3": ["..."],
        "stage_4": ["..."]
      }
    }
  ],
  "stage_1": {
    "human_advantage_share": <0-100>,
    "dominant_advantage_type": "<most prevalent type at this stage, by time-weighted share>"
  },
  "stage_2": { ... },
  "stage_3": { ... },
  "stage_4": { ... },
  "residual_human_advantage": "<2-3 sentences: characterization of the advantages that persist at Stage 4>"
}

## OUTPUT FORMAT — PILOT (add these fields)

Add to each stage block:
  "confidence": "high/medium/low",
  "rationale": "<1-2 sentences: where and why humans retain or lose advantage at this stage>"

Add to each subtask_assessments entry:
  "rationale": "<1 sentence: why humans do or do not retain advantage, and how it changes across stages>"
\end{Verbatim}
\end{spacing}

\subsection{Models, Settings, Cost, and Runtime}
\label{app:models}

Table~\ref{tab:models} lists the models, their exact identifiers, and the settings of each run.
Claude Haiku 4.5 ran through Anthropic's batch interface on April 10, 2026, and the other models ran through OpenRouter on August 1 and 2, 2026, with the same prompt texts and the same subtasks.
All runs used temperature 0 where the model accepts the setting, a fixed seed, and JSON output.
The GPT-5 family does not accept a temperature setting, so the stability of its ratings was measured by the retest of Appendix~\ref{app:reliability} rather than assumed.
The three models were chosen after the pilot described in Appendix~\ref{app:reliability}, and DeepSeek V4 Flash was run as a fourth model for comparison but does not enter the score.

Table~\ref{tab:cost} reports the cost of the runs.
The three-model score cost 520 dollars in API charges: 304 for Claude Haiku 4.5, which also ran the subtask decomposition, 141 for Gemini 3.1 Flash-Lite, and 75 for GPT-5.6-luna.
The fourth model cost 51 dollars.
These figures are estimates from token counts at list prices and exclude the pilot, the retests, and the whole-task check of Appendix~\ref{app:subtasks}.
Running times differed considerably across the OpenRouter models, from about two hours for Gemini to about 27 hours for DeepSeek, because the models differ in how much reasoning text they emit before answering (Table~\ref{tab:models}).
Every task has a score from all three models except three, which OpenAI's content filter refused and which average the Claude and Gemini scores.

\begin{table}[htbp]
\centering
\begin{threeparttable}
\caption{The models and their settings}
\label{tab:models}
\scriptsize\setlength{\tabcolsep}{4pt}
\begin{spacing}{1}
\newcolumntype{L}[1]{>{\raggedright\arraybackslash}p{#1}}
\begin{tabular}{@{}L{2.4cm}L{3.4cm}L{2.3cm}L{2.0cm}L{2.8cm}L{2.2cm}@{}}
\toprule
Model & Identifier and interface & Prompts & Run & Settings & Wall-clock time \\
\midrule
Claude Haiku 4.5 (Anthropic) & \texttt{claude-haiku-\allowbreak 4-5-20251001}, Anthropic Message Batches API & 0 (decomposition), A, B, C & April 10, 2026 & temperature 0; maximum output 2,048 tokens for Prompt 0 and 4,096 for A, B, C; shared system prompt cached & batches, completed within the day \\
\addlinespace
GPT-5.6-luna (OpenAI) & \texttt{openai/gpt-5.6-luna}, OpenRouter, provider OpenAI & A, B, C & August 1 to 2, 2026 & no temperature setting (the GPT-5 family does not accept one); seed 12345; JSON output; maximum output 10,000 tokens, 16,000 on re-queries & about 15 hours \\
\addlinespace
Gemini 3.1 Flash-Lite (Google) & \texttt{google/\allowbreak gemini-3.1-\allowbreak flash-lite}, OpenRouter, providers Google AI Studio and Google & A, B, C & August 1 to 2, 2026 & temperature 0; seed 12345; JSON output; maximum output 10,000 tokens & about 2 hours \\
\addlinespace
DeepSeek V4 Flash (DeepSeek), fourth model & \texttt{deepseek/\allowbreak deepseek-v4-flash}, OpenRouter, providers GMICloud, DeepInfra, and CoreWeave & A, B, C & August 1 to 2, 2026 & temperature 0; seed 12345; JSON output; maximum output 10,000 tokens & about 27 hours \\
\bottomrule
\end{tabular}
\end{spacing}
\begin{tablenotes}\notesize
\item Notes: Every model received the same prompt texts (Appendix~\ref{app:prompts}) and the same subtasks from Claude's Prompt 0 run, and every response went through the same parser and checks (Appendix~\ref{app:checks}). The three OpenRouter models were pinned to the providers listed so that a run was not spread across serving hosts; the maximum output allows for the reasoning tokens these models emit before answering. Wall-clock time is for ratings A, B, and C on all 19,281 tasks, run one prompt at a time. Re-queries for the few responses that failed were completed on August 6, 2026.
\end{tablenotes}
\end{threeparttable}
\end{table}

\begin{table}[htbp]
\centering
\begin{threeparttable}
\caption{API cost of the scoring runs, by model and prompt (US dollars)}
\label{tab:cost}
\footnotesize
\begin{spacing}{1}
\begin{tabular}{@{}l@{\hspace{8pt}}r@{\hspace{8pt}}r@{\hspace{8pt}}r@{\hspace{8pt}}r@{}}
\toprule
 & Claude Haiku 4.5 & GPT-5.6-luna & Gemini 3.1 Flash-Lite & DeepSeek V4 Flash \\
\midrule
Prompt 0, subtask decomposition & 38.8 &  &  &  \\
Rating A, time savings & 83.9 & 20.2 & 45.5 & 15.2 \\
Rating B, independent completion & 84.1 & 28.5 & 44.5 & 18.4 \\
Rating C, human advantage & 97.5 & 25.9 & 51.1 & 17.3 \\
\midrule
Total & 304.3 & 74.6 & 141.1 & 50.9 \\
Tokens, millions & 304 & 297 & 251 & 425 \\
\bottomrule
\end{tabular}
\end{spacing}
\begin{tablenotes}\notesize
\item Notes: Each cell is the cost of running the prompt on all 19,281 tasks at all four stages. Claude Haiku 4.5 ran through the Anthropic Message Batches API on April 10, 2026, and its costs are token counts from the batch results priced at the batch list prices (0.50 and 2.50 dollars per million input and output tokens, 0.05 for cached input). The other three models ran through OpenRouter on August 1 and 2, 2026, and their costs are the runner's estimates from token counts at list prices. Prompt 0 was run once, by Claude, and its subtasks were used for every model. The three-model score (Claude, GPT, and Gemini) cost 520 dollars; DeepSeek V4 Flash is the fourth model of Appendix~\ref{app:reliability} and is not part of the score. The totals exclude the pilot of Appendix Table~\ref{tab:pilot}, the retests, and the whole-task check of Appendix~\ref{app:subtasks}; the OpenRouter account as a whole recorded 652 dollars, 265 thousand requests, and 1.41 billion tokens over the project.
\end{tablenotes}
\end{threeparttable}
\end{table}

\subsection{Checks and Post-Processing}
\label{app:checks}

Every response went through the same three steps, whichever model produced it.
First, the response was checked against the output schema of its prompt: the time shares of a decomposition sum to 100, the constraint-adjusted rating does not exceed the capability-only rating, and categorical fields take one of the listed values.
A response that could not be parsed, was cut off before the schema was complete, or did not fit the schema was re-queried, with up to six attempts and a larger output limit on the later attempts.
Fewer than one percent of requests needed a re-query, and after the re-queries every task has ratings from every model except the three tasks noted above.
Second, the task-level ratings are recomputed from the subtask ratings and their time shares rather than taken from the model's own arithmetic, since the models make arithmetic errors.
Third, the ratings are made monotone across stages, since the stages are cumulative: ratings A and B may not fall and rating C may not rise from one stage to the next, and the constraint-adjusted rating may not exceed the capability-only rating.
A rating that violates one of these conditions is set equal to the value at the previous stage, and the task-level ratings are then recomputed from the corrected subtask ratings.

\subsection{Adapting the Score}
\label{app:adapting}

The pipeline keeps three things separate: the tasks and their subtasks, the stage descriptions, and the models.
Hence, each can be changed without the others.
To add a stage when a new generation of the technology arrives, one appends its description to the stage text of Appendix~\ref{app:stages} and reruns ratings A, B, and C, since the prompts take the stage descriptions as input and contain no dates; the decomposition does not change.
To replace or add a model, one runs the same prompts on the same fixed subtasks through the model's interface and applies the same checks, as we did for three models through one interface, at a cost of 50 to 150 dollars per model (Table~\ref{tab:cost}).
To apply the score to another country's occupations, one needs either task statements for its occupations, which would be decomposed and rated in the same way, or a crosswalk from the O*NET occupations to its classification, as Appendix~\ref{app:data} builds for the census codes.

\subsection{Subtasks: Descriptives and a Worked Example}
\label{app:subtasks}

Table~\ref{tab:subtasks} describes the subtasks that Prompt 0 produced.
The 19,281 tasks were split into 99,157 subtasks, i.e., 5.1 per task.
Most tasks have five subtasks (65 percent) or six (25 percent), three tasks have three, and no task received the seven that the prompt allowed.
The median subtask accounts for 15 percent of its task's time, and the middle 80 percent of subtasks account for between 10 and 35 percent.
Physical execution accounts for about a quarter of all task time, gathering and processing information together for 38 percent, communication and coordination for 13 percent, and generating content and decision and judgment for 8 and 6 percent.
Table~\ref{tab:worked} shows the six subtasks of one task, the preparation of a tax return, with one model's ratings at each stage and the three-model means.

\begin{table}[htbp]
\centering
\begin{threeparttable}
\caption{The subtask decomposition: counts, time shares, and activity types}
\label{tab:subtasks}
\footnotesize
\begin{spacing}{1}
\begin{tabular}{@{}ll@{}}
\toprule
 &  \\
\midrule
\multicolumn{2}{@{}l}{\emph{A. Subtasks per task}} \\
3 subtasks & 3 tasks (0.0 percent) \\
4 subtasks & 2,001 tasks (10.4 percent) \\
5 subtasks & 12,518 tasks (64.9 percent) \\
6 subtasks & 4,759 tasks (24.7 percent) \\
Total & 99,157 subtasks in 19,281 tasks, 5.14 per task \\
Time share of a subtask & median 15 percent, 10th to 90th percentile 10 to 35 \\
\addlinespace
\multicolumn{2}{@{}l}{\emph{B. Share of task time by activity type (percent)}} \\
Physical execution & 24.6 \\
Gathering information & 20.1 \\
Processing information & 18.3 \\
Communication and coordination & 12.9 \\
Administrative and procedural work & 10.0 \\
Generating content & 8.0 \\
Decision and judgment & 6.1 \\
\bottomrule
\end{tabular}
\end{spacing}
\begin{tablenotes}\notesize
\item Notes: Prompt 0 (Appendix~\ref{app:prompts}) split each of the 19,281 O*NET task statements into three to seven subtasks with integer time shares summing to 100 and one of seven activity types; no task received seven. Panel B weights each subtask by its time share.
\end{tablenotes}
\end{threeparttable}
\end{table}

\begin{table}[htbp]
\centering
\begin{threeparttable}
\caption{A worked example: one task through every step}
\label{tab:worked}
\footnotesize\setlength{\tabcolsep}{3pt}
\begin{spacing}{1}
\begin{tabular}{@{}>{\raggedright\arraybackslash}p{4.6cm}r*{12}{c}@{}}
\toprule
 & & \multicolumn{4}{c}{A: time saved (\%)} & \multicolumn{4}{c}{B: completes alone} & \multicolumn{4}{c}{C: human advantage} \\
\cmidrule(lr){3-6}\cmidrule(lr){7-10}\cmidrule(lr){11-14}
Subtask & Share & S1 & S2 & S3 & S4 & S1 & S2 & S3 & S4 & S1 & S2 & S3 & S4 \\
\midrule
1. Gather and review the client's documents & 20 & 15 & 35 & 40 & 50 & 0 & 0 & 0 & 1 & 1 & 1 & 0 & 0 \\
2. Compute taxable income, deductions, and credits & 25 & 20 & 40 & 55 & 60 & 0 & 0 & 0 & 1 & 1 & 1 & 0 & 0 \\
3. Enter the data into tax software & 20 & 45 & 55 & 60 & 70 & 0 & 0 & 0 & 1 & 0 & 0 & 0 & 0 \\
4. Review the completed return & 15 & 25 & 45 & 60 & 65 & 0 & 0 & 0 & 1 & 1 & 1 & 0 & 0 \\
5. Communicate with the client & 12 & 0 & 15 & 25 & 30 & 0 & 0 & 0 & 0 & 1 & 1 & 1 & 1 \\
6. Authorize the return for submission & 8 & 5 & 15 & 30 & 35 & 0 & 0 & 0 & 0 & 1 & 1 & 1 & 1 \\
\midrule
Task rating, Eq.~(\ref{eq:taskrating}) & 100 & 21.1 & 37.8 & 48.1 & 55.1 & 0 & 0 & 0 & 80 & 80 & 80 & 20 & 20 \\
\addlinespace
 & & \multicolumn{12}{l}{Composite $E_t(s)=[A_t(s)+B_t(s)+(100-C_t(s))]/3$, Eq.~(\ref{eq:composite})} \\
 & & \multicolumn{3}{l}{\quad Claude Haiku 4.5} & & \multicolumn{2}{c}{13.7} & \multicolumn{2}{c}{19.2} & \multicolumn{2}{c}{42.7} & \multicolumn{2}{c}{71.7} \\
 & & \multicolumn{3}{l}{\quad Three-model mean} & & \multicolumn{2}{c}{16.5} & \multicolumn{2}{c}{24.3} & \multicolumn{2}{c}{51.4} & \multicolumn{2}{c}{69.2} \\
\bottomrule
\end{tabular}
\end{spacing}
\begin{tablenotes}[flushleft]\notesize
\item Notes: The task is ``prepare or assist in preparing simple to complex tax returns for individuals or small businesses'' (O*NET 13-2082.00, tax preparers), which Prompt 0 split into the six subtasks shown, with their shares of the task's time in percent. The subtask ratings are those of Claude Haiku 4.5 at each of the four stages, in the constraint-adjusted version that the paper uses: A is the percent of the subtask's time that AI could save, B is 1 if AI could complete the subtask alone to a professional standard and 0 otherwise, and C is 1 if people keep a durable advantage on the subtask. The task rating is the time-share-weighted mean of the subtask ratings, with B and C in percent of the task's time. The last two rows give the composite score at each stage for Claude Haiku 4.5 and averaged over the three models (Claude Haiku 4.5, Gemini 3.1 Flash-Lite, and GPT-5.6-luna), which rated the same six subtasks; the three-model composite is the task's exposure score. The ratings are made monotone across stages as described in Appendix~\ref{app:checks}.
\end{tablenotes}
\end{threeparttable}
\end{table}

Table~\ref{tab:wholetask} compares the subtask-based score with a score obtained by asking the same three models to judge each task as a whole.
Scored as a whole, tasks come out less exposed for every model, and the whole-task ratings take far fewer distinct values, since a model handed a whole task answers in round numbers.
The two methods agree on the ranking of tasks to about 0.85 at stage 4 and about 0.78 on the change from stage 1 to stage 4.
The whole-task score is not noisier by the usual tests: the three models agree with one another at least as well under it, and a second run reproduces it more exactly than the subtask-based score.
The whole-task score does predict adoption a little less well: for the three-model average, the change from stage 1 to stage 3 correlates 0.74 with adoption under the subtask method and 0.70 under the whole-task method, and when both standardized changes enter one regression the subtask-based change keeps its coefficient while the whole-task change goes to zero.
The case for scoring subtasks therefore rests on what the method lets the model answer and the reader inspect, as Section 2 argues, on resolution, and on a modest edge in predicting use, not on agreement or stability.

\begin{table}[htbp]
\centering
\begin{threeparttable}
\caption{Scoring the task as a whole versus scoring its subtasks}
\label{tab:wholetask}
\footnotesize\setlength{\tabcolsep}{5pt}
\begin{spacing}{1}
\begin{tabular}{@{}lrrrr@{}}
\toprule
 & Claude Haiku 4.5 & GPT-5.6-luna & Gemini 3.1 Flash-Lite & Three-model average \\
\midrule
\multicolumn{5}{@{}l}{\emph{Mean exposure score at stage 4}} \\
Subtask-based & 47.1 & 40.9 & 42.7 & -- \\
Whole task & 32.6 & 38.9 & 33.7 & -- \\
\addlinespace
\multicolumn{5}{@{}l}{\emph{Rank correlation between the two methods, same model}} \\
Stage-4 level & 0.84 & 0.86 & 0.87 & -- \\
Change, stage 1 to 4 & 0.77 & 0.78 & 0.79 & -- \\
\addlinespace
\multicolumn{5}{@{}l}{\emph{Distinct values of the time-saved rating at stage 4}} \\
Subtask-based & 726 & 863 & 682 & -- \\
Whole task & 44 & 100 & 26 & -- \\
\addlinespace
\multicolumn{5}{@{}l}{\emph{Rank correlation of the stage-1 to stage-4 change with the other two models, average}} \\
Subtask-based & 0.76 & 0.78 & 0.80 & -- \\
Whole task & 0.84 & 0.84 & 0.86 & -- \\
\addlinespace
\multicolumn{5}{@{}l}{\emph{Correlation with adoption across 141 occupations, change from stage 1 to 3}} \\
Subtask-based & 0.74 & 0.72 & 0.72 & 0.74 \\
Whole task & 0.69 & 0.75 & 0.64 & 0.70 \\
\addlinespace
\multicolumn{5}{@{}l}{\emph{Correlation with adoption across 141 occupations, change from stage 1 to 4}} \\
Subtask-based & 0.67 & 0.75 & 0.71 & 0.71 \\
Whole task & 0.66 & 0.76 & 0.65 & 0.70 \\
\addlinespace
\multicolumn{5}{@{}l}{\emph{Test-retest rank correlation of the time-saved rating at stage 4, 1,000-task pilot}} \\
Subtask-based & -- & -- & 0.97 & -- \\
Whole task & -- & -- & 0.99 & -- \\
\bottomrule
\end{tabular}
\end{spacing}
\begin{tablenotes}\notesize
\item Notes: Each model scored all 19,281 O*NET tasks twice: once subtask by subtask with the fixed decomposition and the three prompts of Section 2, and once as a whole, with the task statement alone and the same three questions asked as percentages of the task, using the same stage definitions, typology, and monotonicity rules. Whole-task scores were obtained through OpenRouter on October 8, 2026, at temperature zero. The composite is the same average of the three ratings in both cases. Correlations are across tasks. The adoption rows aggregate each score to census occupations as in Section 2 and correlate the change with the share of workers using generative AI at work in the Real-Time Population Survey, weighted by respondents, for the 141 occupations with at least 20 respondents. The test-retest rows repeat the 1,000-task pilot sample a second time with the same settings for one model.
\end{tablenotes}
\end{threeparttable}
\end{table}

\subsubsection{The whole-task prompt, verbatim}
\begin{spacing}{1}
\begin{Verbatim}
You are an expert labor economist and AI capability analyst. Your task is to assess the
exposure of a specific occupational task from the O*NET database to AI at each of four
capability stages.

You will receive the task statement ONLY. You will NOT receive a breakdown of the task
into subtasks, and you must NOT produce one. Judge the task as a whole, in one pass,
and report your assessments as percentages of the task.

For each stage, provide three assessments of the task as a whole.

1. TIME SAVINGS: the percentage of the time a worker now spends on this task that AI at
   this stage could save. Provide two scores:
   - time_savings_pct_technical: the maximum reduction assuming optimal deployment, a
     fully trained worker, and no cost, regulatory, or organizational barriers.
   - time_savings_pct_feasible: the realistic reduction for the median employer in this
     occupation, accounting for the feasibility constraints below. Work DOWN from the
     technical score.

2. INDEPENDENT COMPLETION: the percentage of this task that AI at this stage could
   complete END-TO-END to an acceptable professional standard without a human performing
   or redoing the work. "Acceptable professional standard" means the output would be
   considered satisfactory by a supervisor, client, or regulatory body. If AI can draft
   something but a human must substantially revise it, that part of the task is NOT
   independently completed. If a human only performs a brief quality check, it IS.
   Physical work that requires bodily presence, manual dexterity, or real-time physical
   interaction cannot be completed by AI at any current stage. Provide two scores:
   - completion_pct_technical: assuming optimal deployment and no barriers.
   - completion_pct_feasible: for the median employer, accounting for the constraints
     below. Work DOWN from the technical score.

3. HUMAN COMPARATIVE ADVANTAGE: the percentage of this task on which a human retains a
   comparative advantage over AI at this stage, meaning the work is performed MORE
   EFFECTIVELY by a human, is REQUIRED to be done by a human for legal or institutional
   reasons, or LOSES ITS VALUE if performed by AI. Comparative advantage is relative to
   what AI at the given stage can and cannot do. Classify the advantage using the
   typology below and report the dominant type at each stage.

FEASIBILITY CONSTRAINTS (apply to the feasible scores only):
- Economic viability: implementation costs and expected productivity gains must justify
  deployment for the median employer.
- Regulatory and legal barriers: deployment must comply with professional licensing
  requirements, liability rules, data privacy regulations, and industry-specific
  compliance standards.
- Organizational readiness: whether the median employer has the technical
  infrastructure, data systems, and IT capacity to support deployment.

---

## HUMAN COMPARATIVE ADVANTAGE TYPOLOGY

### PHYSICAL AND SENSORIMOTOR
The work requires bodily presence, manual dexterity, physical manipulation of objects,
real-time sensorimotor coordination, navigation of unstructured physical environments, or
direct physical interaction with people or surroundings. Examples: performing surgery,
operating heavy machinery, physically inspecting a construction site, hands-on patient
care, assembling delicate components.

### INTERPERSONAL AND RELATIONAL
The work's effectiveness depends on a genuine human relationship, emotional connection,
perceived authenticity, or the ability to read and respond to social and emotional cues
in real time, including persuasion, negotiation, and coordination across people.
Examples: therapy and counseling, negotiating a sensitive deal, delivering difficult
medical news, mentoring, de-escalating a crisis.

### INSTITUTIONAL AUTHORITY
Legal, regulatory, professional, or social frameworks require or strongly expect a human
in this role for the output to be accepted as valid, legitimate, or authoritative, whether
as a hard constraint (a human must be the responsible decision-maker, signatory, or
fiduciary) or a soft one (clients, patients, or the public expect a human for the output
to carry credibility). Examples: a physician signing a diagnosis, a CPA certifying
statements, a teacher leading a classroom discussion.

### TACIT JUDGMENT AND EXPERTISE
The work involves reasoning under genuine novelty or ambiguity, integration of tacit
knowledge and contextual cues, expert improvisation that cannot be reduced to explicit
rules, or creative problem-solving beyond recombination of existing patterns. Examples:
setting strategy in a crisis, triage with incomplete information, an experienced nurse
reading non-verbal distress, a chef adjusting a recipe to ingredient quality.

---

## AI CAPABILITY STAGES

[INSERT SHARED STAGE DEFINITIONS HERE]

---

## INPUT FORMAT

You will receive:

{
  "occupation": "<O*NET occupation title>",
  "onet_code": "<O*NET-SOC code>",
  "task_id": <integer>,
  "task_text": "<exact O*NET task text>"
}

---

## ASSESSMENT INSTRUCTIONS

Judge the task as a whole. Consider everything the task statement covers, including any
physical, interpersonal, and administrative parts, not only the part AI handles best.

SCORING GUIDANCE:
- All scores use a 0-100 scale and refer to percentages of the task.
- Time savings and completion scores must be MONOTONICALLY NON-DECREASING across stages
  (Stage 4 >= Stage 3 >= Stage 2 >= Stage 1). If a later stage adds no relevant
  capability, carry forward the prior score.
- Human advantage shares must be MONOTONICALLY NON-INCREASING across stages
  (Stage 1 >= Stage 2 >= Stage 3 >= Stage 4).
- Each feasible score must be LESS THAN OR EQUAL TO its technical score at every stage.
- Use the full 0-100 range. Do not default to moderate values when uncertain.
- Report integers. Do not restrict yourself to multiples of 5 or 10; use the precision
  your judgment supports.

---

## OUTPUT FORMAT — PRODUCTION

Return a JSON object with the following structure. Do not include any text outside
the JSON object.

{
  "occupation": "<O*NET occupation title>",
  "onet_code": "<O*NET-SOC code>",
  "task_id": <integer>,
  "task_text": "<exact task text>",
  "stage_1": {
    "time_savings_pct_technical": <0-100>,
    "time_savings_pct_feasible": <0-100>,
    "completion_pct_technical": <0-100>,
    "completion_pct_feasible": <0-100>,
    "human_advantage_share": <0-100>,
    "dominant_advantage_type": "<one of: physical_sensorimotor, interpersonal_relational, institutional_authority, tacit_judgment_expertise, none>",
    "primary_constraint": "<comma-separated list from: none, economic, regulatory, organizational>"
  },
  "stage_2": { ... },
  "stage_3": { ... },
  "stage_4": { ... }
}

## OUTPUT FORMAT — PILOT (add these fields)

Add to each stage block:
  "confidence": "high/medium/low",
  "rationale": "<1-2 sentences: what AI at this stage can and cannot do for this task, and why>"
\end{Verbatim}
\end{spacing}

\subsection{Capability-Only and Constraint-Adjusted Ratings}
\label{app:technical}

Ratings A and B were each asked in two versions, for the capability of the technology alone and after the economic, regulatory, and organizational constraints on using AI for the subtask, and the paper uses the constraint-adjusted version.
Table~\ref{tab:techfeas} compares the two.
The constraint-adjusted score is lower at every stage, at 43.6 against 55.4 at stage 4 for the three-model mean, and its change from stage 1 to stage 4 is 32.5 points against 39.3.
However, the two versions rank tasks almost identically: the correlation across tasks is above 0.96 at every stage and 0.93 for the change from stage 1 to stage 4.
The gap between the versions differs somewhat by model, at about 13 points for Claude and GPT and 10 points for Gemini at stage 4.
Rating C, human comparative advantage, was asked once, so it enters both composites unchanged.
Hence, the choice between the two versions matters for the level of the score but little for the cross-section of changes that the regressions use.
We have not rerun the regressions with the capability-only score.

\begin{table}[htbp]
\centering
\begin{threeparttable}
\caption{Capability-only and constraint-adjusted scores across tasks}
\label{tab:techfeas}
\footnotesize
\begin{spacing}{1}
\begin{tabular}{@{}lcccc@{}}
\toprule
 \\
\midrule
\multicolumn{5}{@{}l}{\emph{A. Three-model score, by stage}} \\
 & Capability only & Constraint adjusted & Pearson & Spearman \\
Stage 1 & 16.1 & 11.0 & 0.975 & 0.962 \\
Stage 2 & 36.1 & 25.5 & 0.964 & 0.965 \\
Stage 3 & 46.6 & 35.0 & 0.961 & 0.962 \\
Stage 4 & 55.4 & 43.6 & 0.969 & 0.972 \\
Change, stage 1 to 4 & 39.3 & 32.5 & 0.933 & 0.928 \\
\addlinespace
\multicolumn{5}{@{}l}{\emph{B. Each model, stage 4}} \\
Claude Haiku 4.5 & 59.8 & 47.1 & 0.928 & 0.930 \\
GPT-5.6-luna & 53.9 & 40.9 & 0.948 & 0.951 \\
Gemini 3.1 Flash-Lite & 52.6 & 42.7 & 0.971 & 0.973 \\
\bottomrule
\end{tabular}
\end{spacing}
\begin{tablenotes}\notesize
\item Notes: Ratings A and B were each given in two versions: the capability of the technology alone, and the rating after the economic, regulatory, and organizational constraints on using AI for the subtask, which the paper uses. Rating C has one version and enters both composites. Means are across the 19,278 tasks with all three models; correlations are across tasks. The change rows compare each version's change from stage 1 to stage 4.
\end{tablenotes}
\end{threeparttable}
\end{table}

\subsection{Reliability}
\label{app:reliability}

We chose the models in a pilot on a stratified sample of 1,000 tasks, in which seven candidate models from five developers rated the same fixed subtasks (Table~\ref{tab:pilot}).
Within each developer the cheaper model agreed with the original Claude scores at least as well as the dearer one, so we selected Gemini 3.1 Flash-Lite and GPT-5.6-luna for the score and DeepSeek V4 Flash as a fourth, open-weights model.
Table~\ref{tab:agreement} reports the agreement across models on all tasks and across runs of the same model.
The three models in the score rank tasks similarly, with Spearman correlations of 0.87 to 0.91 between their stage-4 scores, and DeepSeek correlates 0.80 to 0.83 with each of them.
The models differ in levels rather than in ranking: the mean stage-4 score is 47.1 for Claude, 42.7 for Gemini, 40.9 for GPT, and 27.3 for DeepSeek, which is why we report the DeepSeek scores but do not include them in the score.
A second run of the Claude pipeline on the 1,000 pilot tasks reproduces the ratings closely at the task level, with correlations of 0.97 to 0.98 for rating A, 0.83 to 0.93 for rating B, and 0.89 to 0.96 for rating C across the four stages.
Since an occupation's score averages about 22 tasks, the reliability of the occupation scores is higher still, and the Spearman--Brown formula puts it above 0.98.
At the occupation level, each model's stage-4 score correlates 0.97 to 0.99 with the three-model score and each model's change from stage 1 to stage 4 correlates 0.95 to 0.99 with it, and adding DeepSeek as a fourth model leaves the ranking of occupations essentially unchanged, at 0.999 for the level and 0.997 for the change.

Table~\ref{tab:weights} shows that the occupation scores do not depend on how an occupation's tasks are weighted.
Weighting tasks equally or by O*NET importance or frequency instead of relevance gives scores that correlate 0.995 or more with the relevance-weighted score, for the stage-4 level and for the change from stage 1 to stage 4.
The three ratings also move together across occupations: the first principal component of ratings A, B, and 100 minus C at stage 4 accounts for 97 percent of their variance with equal loadings, and it correlates 0.9996 with the equal-weight score.

\begin{table}[htbp]
\centering
\begin{threeparttable}
\caption{The pilot: seven candidate models on 1,000 tasks, August 1, 2026}
\label{tab:pilot}
\scriptsize
\begin{spacing}{1}
\begin{tabular}{@{}llccccl@{}}
\toprule
Model & Developer & \shortstack{Agreement\\with Claude} & \shortstack{Retest,\\rating A} & \shortstack{Coverage,\\A / B / C (percent)} & \shortstack{Cost\\(dollars)} & Decision \\
\midrule
Gemini 3.1 Flash-Lite & Google & 0.898 & 0.973 & 100 / 100 / 100 & 7.4 & selected \\
GPT-5.6-luna & OpenAI & 0.866 & 0.899 & 100 / 99.6 / 100 & 3.8 & selected \\
GPT-5 mini & OpenAI & 0.798 & 0.860 & 100 / 100 / 100 & 24.8 & not selected \\
DeepSeek V4 Flash & DeepSeek & 0.830 & 0.845 & 100 / 100 / 100 & 2.7 & selected as the fourth model \\
DeepSeek V4 Pro & DeepSeek & 0.812 & 0.811 & 100 / 100 / 100 & 26.9 & not selected \\
Kimi K2.5 & Moonshot &  &  & stopped after A and B & 32.2 & not selected \\
Kimi K2.6 & Moonshot &  &  & stopped after A & 22.7 & not selected \\
\midrule
Claude Haiku 4.5, second run & Anthropic & 0.975 & & 100 / 100 / 100 & about 10 & the original model \\
\bottomrule
\end{tabular}
\end{spacing}
\begin{tablenotes}\notesize
\item Notes: The 1,000 tasks are a stratified sample of the 19,281, with every major occupation group represented. Every model rated the same fixed subtasks with ratings A, B, and C at all four stages through OpenRouter, with temperature 0, seed 12345, and JSON output. Agreement is the Spearman correlation across tasks between the model's stage-4 composite score and Claude Haiku 4.5's original score; retest is the Spearman correlation between two runs of rating A, stage 4. Cost covers ratings A, B, and C on the 1,000 tasks. The two Kimi models were stopped because their long reasoning output made them slow (about 400 requests an hour); Gemini 3.5 Flash was dropped before the pilot on its projected cost. Within each developer the cheaper model agreed with Claude at least as well as the dearer one. The last row re-ran the original Claude pipeline on the same 1,000 tasks through the Anthropic batch API; its agreement column is the correlation between the two Claude runs.
\end{tablenotes}
\end{threeparttable}
\end{table}

\begin{table}[htbp]
\centering
\begin{threeparttable}
\caption{Agreement across models and across runs}
\label{tab:agreement}
\footnotesize
\begin{spacing}{1}
\begin{tabular}{@{}lcccc@{}}
\toprule
 \\
\midrule
\multicolumn{5}{@{}l}{\emph{A. Task level: Spearman correlations of the stage-4 score across models (19,276 tasks)}} \\
 & Claude & Gemini & GPT & DeepSeek \\
Claude & 1 &  &  &  \\
Gemini & 0.900 & 1 &  &  \\
GPT & 0.865 & 0.905 & 1 &  \\
DeepSeek & 0.804 & 0.829 & 0.823 & 1 \\
Mean stage-4 score & 47.1 & 42.7 & 40.9 & 27.3 \\
\addlinespace
\multicolumn{5}{@{}l}{\emph{B. Task level: Claude Haiku 4.5 rated twice, Spearman correlation between runs on 1,000 tasks}} \\
 & Stage 1 & Stage 2 & Stage 3 & Stage 4 \\
Rating A, feasible time savings & 0.974 & 0.976 & 0.976 & 0.967 \\
Rating B, feasible completion & 0.834 & 0.860 & 0.903 & 0.931 \\
Rating C, human advantage & 0.885 & 0.958 & 0.962 & 0.947 \\
\addlinespace
\multicolumn{5}{@{}l}{\emph{C. Occupation level: Spearman correlation with the three-model score (879 occupations)}} \\
 & Gemini & GPT & DeepSeek & Four-model score \\
Stage-4 score & 0.993 & 0.987 & 0.973 & 0.999 \\
Change from stage 1 to stage 4 & 0.986 & 0.973 & 0.953 & 0.997 \\
\bottomrule
\end{tabular}
\end{spacing}
\begin{tablenotes}\notesize
\item Notes: Panel A uses the production runs on all tasks; each model's score is the composite of its own three ratings. Panel B compares the original April 2026 Claude run with a second run of the same pipeline in August 2026 on the stratified 1,000-task pilot sample. Panel C aggregates each model's task scores to occupations with the relevance weights of Section~\ref{sec:construction} and correlates them with the three-model score of the paper; the four-model score adds DeepSeek V4 Flash. The Claude column of Panel C is omitted because Claude's occupation scores enter the three-model score.
\end{tablenotes}
\end{threeparttable}
\end{table}

\begin{table}[htbp]
\centering
\begin{threeparttable}
\caption{Occupation scores under alternative task weights}
\label{tab:weights}
\footnotesize
\begin{spacing}{1}
\begin{tabular}{@{}lcccc@{}}
\toprule
 & \multicolumn{2}{c}{Stage-4 score} & \multicolumn{2}{c}{Change, stage 1 to 4} \\
\midrule
 & Pearson & Spearman & Pearson & Spearman \\
Simple average & 0.995 & 0.995 & 0.996 & 0.995 \\
Importance weights & 0.999 & 0.998 & 0.999 & 0.998 \\
Frequency weights & 0.998 & 0.998 & 0.998 & 0.998 \\
\bottomrule
\end{tabular}
\end{spacing}
\begin{tablenotes}\notesize
\item Notes: Three-model score across the 879 O*NET occupations. The paper weights an occupation's tasks by O*NET relevance, the share of incumbents who report the task as part of their job; the alternatives weight tasks equally, by O*NET importance, or by O*NET frequency. Each cell is the correlation across occupations between the alternative score and the relevance-weighted score. A principal-components analysis of the three relevance-weighted ratings (A, B, and 100 minus C) at stage 4 across occupations gives a first component that accounts for 97.3 percent of their variance with loadings of 0.58, 0.58, and 0.58, and that correlates 0.9996 with the equal-weight score.
\end{tablenotes}
\end{threeparttable}
\end{table}

\subsection{Validation Extras}
\label{app:validation}

\begin{figure}[htbp]
\centering
\includegraphics[width=\linewidth]{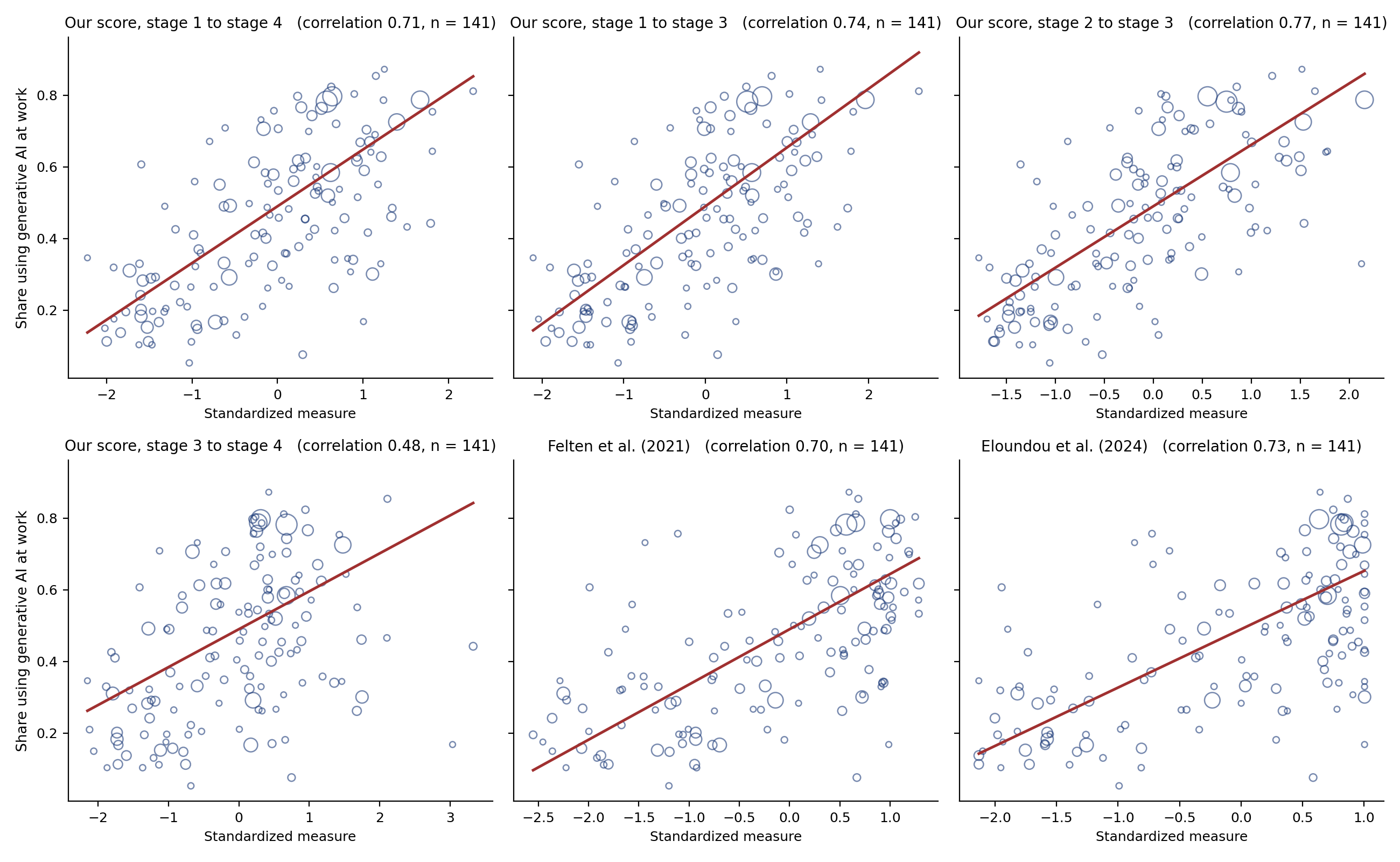}
\caption{Adoption against each stage change in our score, and against the Felten and Eloundou measures}
\label{fig:adoptionall}
\begin{minipage}{0.92\textwidth}
\notesize
Notes: Each dot is an occupation, sized by its number of survey respondents. Adoption is the share of workers who use generative AI at work in the Real-Time Population Survey of \citet{bick2026work}, pooled over four waves from August 2025 to May 2026, for the 141 occupations with at least 20 respondents. The first four panels plot changes in our score between stages; the last two plot the \citet{felten2023language} index and the \citet{eloundou2024gpts} measure. Each measure is standardized. Lines are weighted least-squares fits, and the correlations are weighted by respondents. Section~\ref{sec:validation} of the paper discusses the first and last panels.
\end{minipage}
\end{figure}

Figure~\ref{fig:adoptionall} plots reported adoption against each stage change in our score and against the Felten and Eloundou measures, and Table~\ref{tab:adoptionreg} reports the corresponding regressions.
A standard deviation of our change from stage 1 to stage 4 is associated with 15.9 percentage points more reported adoption, and the change alone explains half of the variance in adoption across occupations.
The stage-4 level alone explains less, and when the stage-4 and stage-1 levels enter together the stage-4 level enters positively and the stage-1 level negatively, i.e., what predicts adoption is how much exposure grew since pre-generative machine learning, not how exposed an occupation is.
The \citet{eloundou2024gpts} measure predicts adoption about as well as our change, and since the two correlate 0.90 across occupations, neither is estimated precisely when both enter; with our change from stage 1 to stage 3, both remain significant.
The \citet{felten2023language} measure explains somewhat less, and our change keeps a coefficient of 9.8 (1.8) when it is added.

\begin{table}[htbp]
\centering
\begin{threeparttable}
\caption{Reported adoption regressed on exposure measures}
\label{tab:adoptionreg}
\footnotesize
\begin{spacing}{1}
\begin{tabular}{@{}l*{8}{c}@{}}
\toprule
 \\
\midrule
 & (1) & (2) & (3) & (4) & (5) & (6) & (7) & (8) \\
\midrule
Our change, stage 1 to 4 & 15.85 &  &  &  & 6.18 &  &  & 9.75 \\
 & (1.43) &  &  &  & (3.32) &  &  & (1.84) \\
Our change, stage 1 to 3 &  &  &  &  &  & 9.13 &  &  \\
 &  &  &  &  &  & (2.45) &  &  \\
Our stage-4 level &  & 12.61 & 21.84 &  &  &  &  &  \\
 &  & (1.84) & (1.72) &  &  &  &  &  \\
Our stage-1 level &  &  & $-$14.32 &  &  &  &  &  \\
 &  &  & (2.10) &  &  &  &  &  \\
Eloundou et al.\ (2024), $\zeta$ &  &  &  & 16.29 & 10.71 & 8.25 &  &  \\
 &  &  &  & (1.48) & (3.34) & (2.86) &  &  \\
Felten et al.\ (2023) &  &  &  &  &  &  & 15.43 & 7.98 \\
 &  &  &  &  &  &  & (1.60) & (1.81) \\
\midrule
$R^{2}$ & 0.51 & 0.32 & 0.57 & 0.54 & 0.55 & 0.58 & 0.48 & 0.56 \\
Occupations & 141 & 141 & 141 & 141 & 141 & 141 & 141 & 141 \\
\bottomrule
\end{tabular}
\end{spacing}
\begin{tablenotes}\notesize
\item Notes: The dependent variable is the share of workers in the occupation who report using generative AI at work, in percentage points, from four waves of the Real-Time Population Survey of \citet{bick2026work}, August 2025 to May 2026, for the occupations with at least 20 respondents. Each regressor is standardized to mean zero and standard deviation one, weighted by respondents, so a coefficient is the change in adoption, in percentage points, per standard deviation of the measure. Regressions are weighted by respondents, with heteroskedasticity-robust standard errors in parentheses. Columns (5) and (6) add the Eloundou et al.\ measure to our change, and column (8) adds the Felten et al.\ measure. Across the 511 census occupations with all measures, our change from stage 1 to stage 4 correlates 0.90 with the Eloundou et al.\ measure and 0.79 with the Felten et al.\ measure, and our stage-1 level correlates 0.50 with the Eloundou et al.\ measure.
\end{tablenotes}
\end{threeparttable}
\end{table}

\section{Data: The Occupation Crosswalk and Descriptive Figures}
\label{app:data}

This appendix documents how the exposure score, which is built on O*NET task statements, is carried over to the occupation codes of the CPS, and it presents the descriptive exhibits that Section~\ref{sec:cps} of the paper discusses.
Appendix~\ref{app:crosswalk} describes the mapping from O*NET occupations to the 2018 census occupation codes and then to the harmonized codes used throughout the paper.
Appendix~\ref{app:unscored} lists the harmonized codes that carry no score.
Appendix~\ref{app:descriptives} presents the distribution of exposure across the workforce at each stage, the characteristics of the workforce by quartile of the change in exposure, and the raw trends of the five outcomes by quartile.

\subsection{From O*NET Occupations to Harmonized Census Codes}
\label{app:crosswalk}

Table~\ref{tab:crosswalk} Panel A counts the units at each step of the mapping.
O*NET attaches its 19,281 task statements to 923 detailed occupations in its 2019 taxonomy, whose codes are the 2018 Standard Occupational Classification (SOC) codes with a two-digit suffix that distinguishes specializations within a SOC code, and the occupation score is the relevance-weighted average of the occupation's task scores, as described in Section~\ref{sec:measurement}.
The CPS, however, records occupation in the 2018 census occupation classification, which has 526 codes and is coarser than the 832 SOC codes, so that a census code typically combines several SOC codes.
We first average the scores of the O*NET occupations that share a SOC code, and we then average the SOC codes that feed each census code, weighting each SOC code by its national employment in the May 2023 Occupational Employment and Wage Statistics of the Bureau of Labor Statistics.
This yields a score for 513 of the 526 census codes.
The remaining 13 are, with two exceptions, residual ``all other'' categories, i.e., codes such as ``sales and related workers, all other'', for which O*NET has no occupation with task statements.

The second step carries the score from the 2018 census codes to the IPUMS harmonized OCC2010 codes, which we use because the census classification changed in January 2020, as Section~\ref{sec:data} explains.
Following the guidance that IPUMS provides for building custom crosswalks between an unharmonized and a harmonized variable, we take every person-month in the CPS from January 2020, when the 2018 codes came into use, to June 2026, in which both the 2018 code and the harmonized code are recorded, and we count the pairs.
Every 2018 code maps into exactly one harmonized code, and a harmonized code collects one or more 2018 codes, so the mapping is one-to-one or many-to-one.
The score of a harmonized code is the average of the scores of the 2018 codes that feed it, weighted by the number of person-months in each pair.
Panel B of Table~\ref{tab:crosswalk} classifies the 451 harmonized codes of the estimation sample by the number of 2018 codes that feed them.
In other words, 409 harmonized codes collect a single 2018 code and cover 82 percent of person-months, 28 collect two 2018 codes and cover 7 percent, and 14 collect three or more and cover 11 percent.
For the one-to-one codes the change in exposure from stage 1 to stage 4 is identical in the two files, as it should be.
For the many-to-one codes the 2018 codes that are combined do not always have the same exposure, and the median range of the stage 1 to stage 4 change within a harmonized code is 4.7 points, with a 90th percentile of 15.7 points, against a standard deviation of about 12 points across occupations.
Hence, the harmonized file blurs the exposure of a few dozen occupations, and Section~\ref{sec:robustness} repeats the main estimates on the 2018-code file and on the one-to-one codes only.

\begin{table}[htbp]
\centering
\caption{From O*NET occupations to harmonized census codes}
\label{tab:crosswalk}
\begin{threeparttable}
\begin{tabular}{@{}>{\raggedright\arraybackslash}p{9.6cm}rr@{}}
\toprule
\multicolumn{3}{@{}l}{\textit{A. The mapping, step by step}} \\
\midrule
& Units & \\
\midrule
O*NET task statements & 19,281 & \\
O*NET-SOC 2019 occupations with task statements & 923 & \\
SOC 2018 codes & 832 & \\
2018 census occupation codes & 526 & \\
\quad with an exposure score & 513 & \\
Harmonized (OCC2010) codes in the 2017--2026 panel & 473 & \\
\quad with an exposure score (estimation sample) & 451 & \\
\midrule
\multicolumn{3}{@{}l}{\textit{B. Harmonized codes by the number of 2018 codes that feed them}} \\
\midrule
& Codes & Share of person-months \\
\midrule
One 2018 code & 409 & 82.1 \\
Two 2018 codes & 28 & 6.9 \\
Three or more 2018 codes & 14 & 11.0 \\
\midrule
Correlation of the S4--S1 change across the two files, one-to-one codes & 1.000 & \\
Median within-code range of the S4--S1 change, many-to-one codes & 4.7 & \\
90th percentile of that range & 15.7 & \\
\bottomrule
\end{tabular}
\begin{tablenotes}[flushleft]\notesize
\item Notes: Panel A counts the units at each step of the mapping from O*NET task statements to the harmonized census codes of the estimation sample. Panel B classifies the harmonized codes of the estimation sample by how many 2018 census codes map into them; shares are of CPS person-months from January 2020 to June 2026, the period in which the 2018 codes are in use. The S4--S1 change is the change in the exposure score from stage 1 to stage 4, in points; the within-code range is the difference between the largest and the smallest change among the 2018 codes that feed one harmonized code.
\end{tablenotes}
\end{threeparttable}
\end{table}

\subsection{Harmonized Codes without an Exposure Score}
\label{app:unscored}

Table~\ref{tab:unscored} lists the 22 harmonized codes that appear in the 2017Q1 to 2026Q2 panel but carry no exposure score.
Five of them are residual ``all other'' categories, which are present in every quarter but have no counterpart in O*NET, and the remaining 17 are codes that the 2018 census classification discontinued, so that no 2018 code maps into them and they appear in the panel only through 2019Q4.
Together, the 22 codes account for 0.66 percent of employment over the 38 quarters, and the largest of them, ``sales and related workers, all other'', accounts for 0.22 percent.
Hence, the estimation sample consists of the remaining 451 harmonized codes.

\begin{table}[htbp]
\centering
\caption{Harmonized codes without an exposure score}
\label{tab:unscored}
\begin{threeparttable}
\footnotesize
\begin{tabular}{@{}l>{\raggedright\arraybackslash}p{6.2cm}cr@{}}
\toprule
Code & Title & Quarters in the panel & Share of employment (percent) \\
\midrule
\multicolumn{4}{@{}l}{\textit{A. Residual categories with no O*NET counterpart}} \\
2060 & Religious workers, all other & 2017Q1--2026Q2 & 0.04 \\
4650 & Personal care and service workers, all other & 2017Q1--2026Q2 & 0.09 \\
4965 & Sales and related workers, all other & 2017Q1--2026Q2 & 0.22 \\
7855 & Food processing workers, all other & 2017Q1--2026Q2 & 0.09 \\
9150 & Motor vehicle operators, all other & 2017Q1--2026Q2 & 0.08 \\
\midrule
\multicolumn{4}{@{}l}{\textit{B. Codes that end with the 2020 reclassification}} \\
330 & Gaming managers & 2017Q1--2019Q4 & 0.00 \\
1930 & Geological and petroleum technicians, and nuclear technicians & 2017Q1--2019Q4 & 0.00 \\
4060 & Counter attendants, cafeteria, food concession, and coffee shop & 2017Q1--2019Q4 & 0.03 \\
4300 & First-line supervisors of gaming workers & 2017Q1--2019Q4 & 0.04 \\
4410 & Motion picture projectionists & 2017Q1--2019Q4 & 0.00 \\
5030 & Communications equipment operators, all other & 2017Q1--2019Q4 & 0.00 \\
5130 & Gaming cage workers & 2017Q1--2019Q4 & 0.00 \\
5200 & Brokerage clerks & 2017Q1--2019Q4 & 0.00 \\
5800 & Computer operators & 2017Q1--2019Q4 & 0.01 \\
6300 & Paving, surfacing, and tamping equipment operators & 2017Q1--2019Q4 & 0.00 \\
7110 & Electronic equipment installers and repairers, motor vehicles & 2017Q1--2019Q4 & 0.00 \\
7710 & Aircraft structure, surfaces, rigging, and systems assemblers & 2017Q1--2019Q4 & 0.00 \\
7940 & Rolling machine setters, operators, and tenders, metal and plastic & 2017Q1--2019Q4 & 0.00 \\
8400 & Textile bleaching and dyeing, and cutting machine setters, operators, and tenders & 2017Q1--2019Q4 & 0.00 \\
8410 & Textile knitting and weaving machine setters, operators, and tenders & 2017Q1--2019Q4 & 0.00 \\
9240 & Railroad conductors and yardmasters & 2017Q1--2019Q4 & 0.01 \\
9520 & Dredge, excavating, and loading machine operators & 2017Q1--2019Q4 & 0.01 \\
\midrule
All 22 codes & & & 0.66 \\
\bottomrule
\end{tabular}
\begin{tablenotes}[flushleft]\notesize
\item Notes: The harmonized (IPUMS OCC2010) codes that appear in the 2017Q1--2026Q2 occupation-by-quarter panel but carry no exposure score. Titles are the IPUMS labels. The share of employment is the code's CPS-weighted employment summed over the quarters it appears in, as a share of total employment over all 38 quarters. Panel A lists residual ``all other'' categories, for which O*NET has no occupation; Panel B lists codes that the 2018 census classification discontinued, so that no 2018 code maps into them and they do not appear after 2019Q4.
\end{tablenotes}
\end{threeparttable}
\end{table}

\clearpage
\subsection{Descriptive Figures and Tables}
\label{app:descriptives}

\begin{figure}[htbp]
\centering
\includegraphics[width=0.8\linewidth]{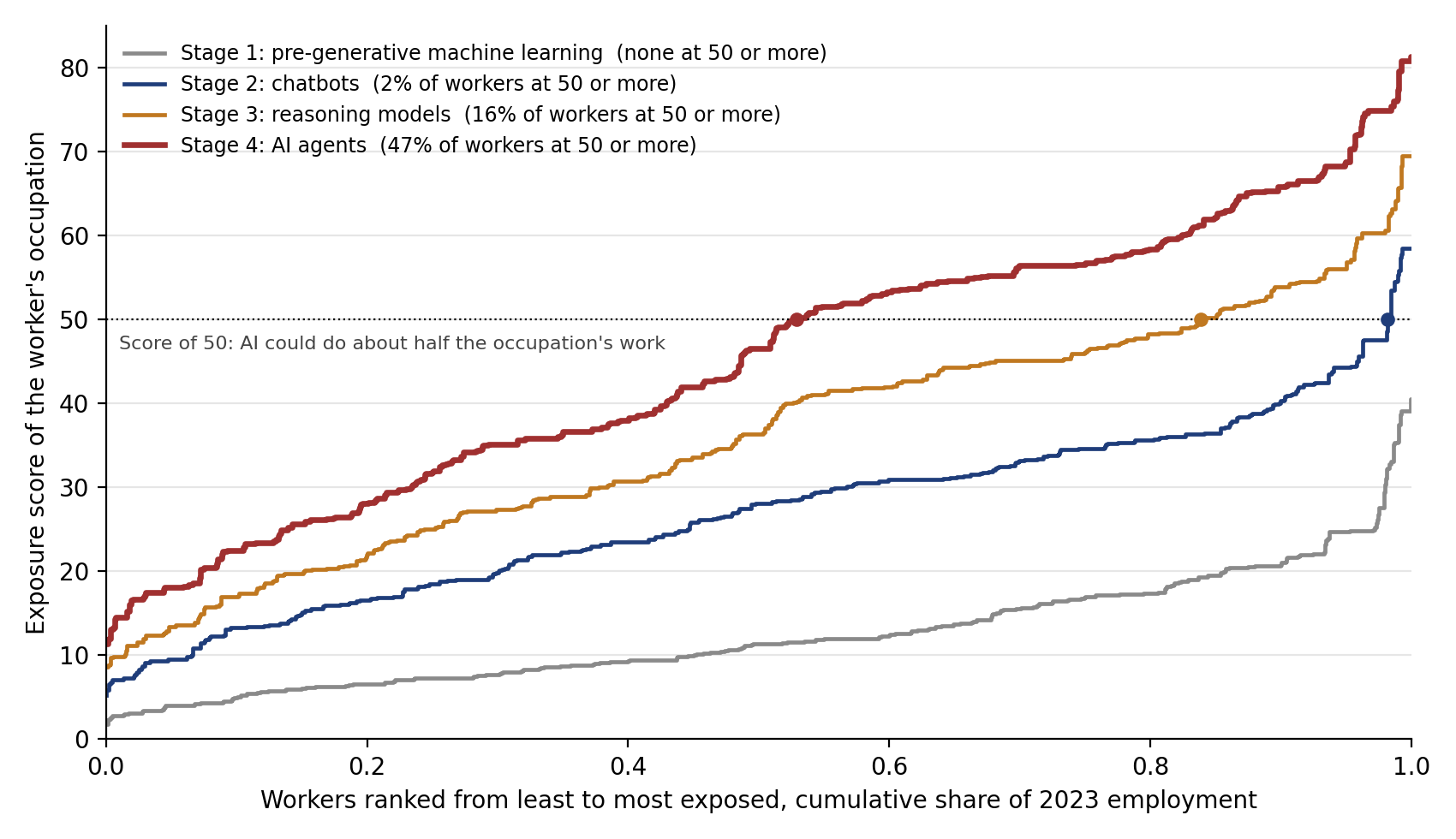}
\caption{The expanding frontier: exposure across the workforce at each capability stage}
\label{fig:frontier}
\begin{minipage}{0.92\textwidth}
\notesize
Notes: Workers are ranked from the least to the most exposed occupation, and each curve plots the exposure score of the worker's occupation at one capability stage, so that a curve is the employment-weighted distribution of the score at that stage. Where a curve crosses 50, the point at which AI could do about half the occupation's work, the share of workers to the right is marked. Occupations are the harmonized census occupations with scores at all four stages, weighted by average employment in 2023, the first year generative AI was in use at work.
\end{minipage}
\end{figure}

\begin{figure}[htbp]
\centering
\includegraphics[width=\linewidth]{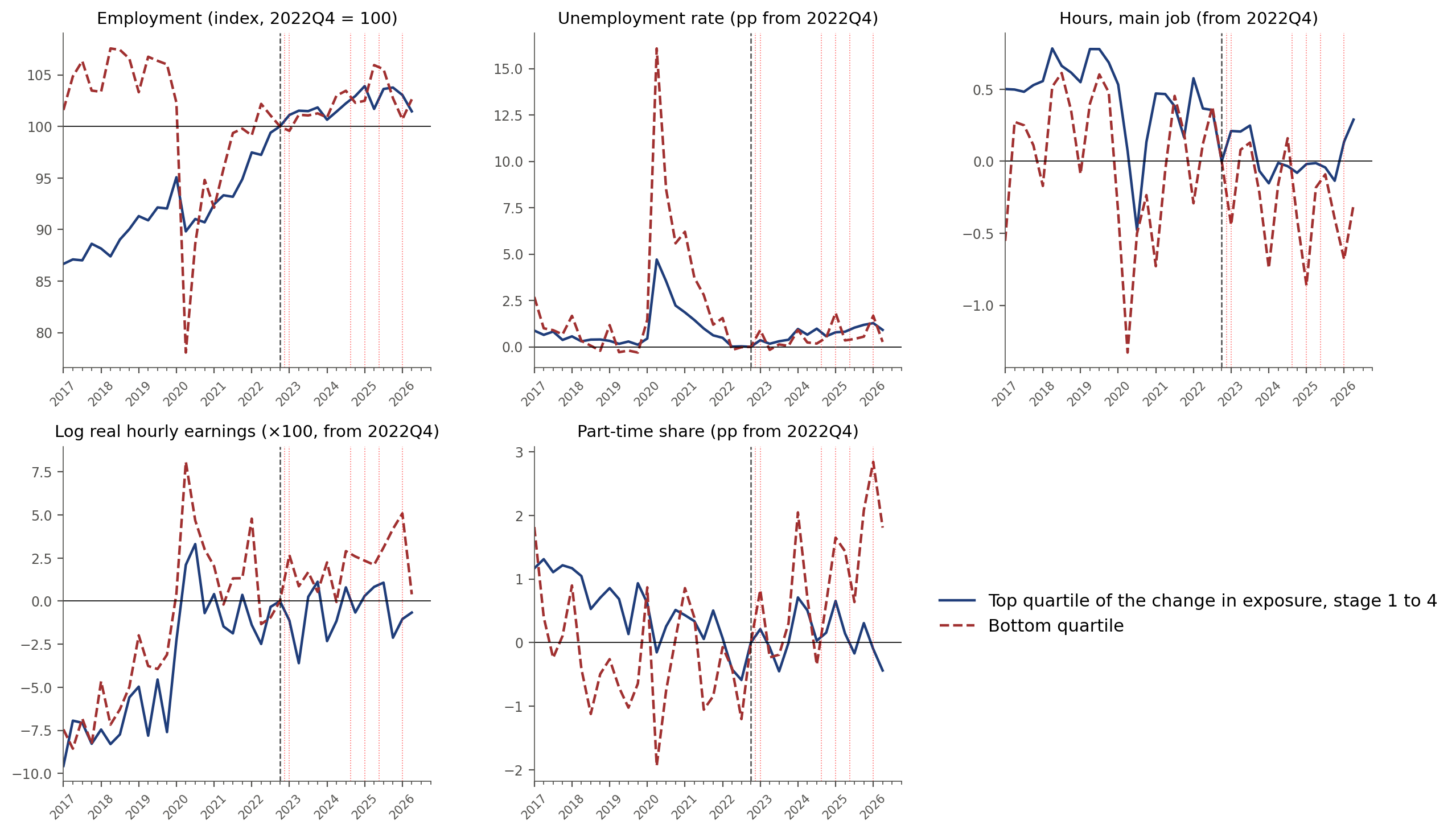}
\caption{Ten years of labor market outcomes, top versus bottom quartile of exposure change}
\label{fig:outcomes}
\begin{minipage}{0.92\textwidth}
\notesize
Notes: Quartiles of the stage-1 to stage-4 change across the harmonized occupations. Each series is normalized to its value in 2022Q4, the ChatGPT quarter: employment as an index equal to 100 in that quarter; the unemployment rate, hours, log real hourly earnings ($\times$100), and the part-time share as deviations from their 2022Q4 values. Dotted red lines mark the release and inflection dates of the three generations. Employment is normalized for the number of months surveyed per quarter.
\end{minipage}
\end{figure}

\begin{figure}[htbp]
\centering
\includegraphics[width=\linewidth]{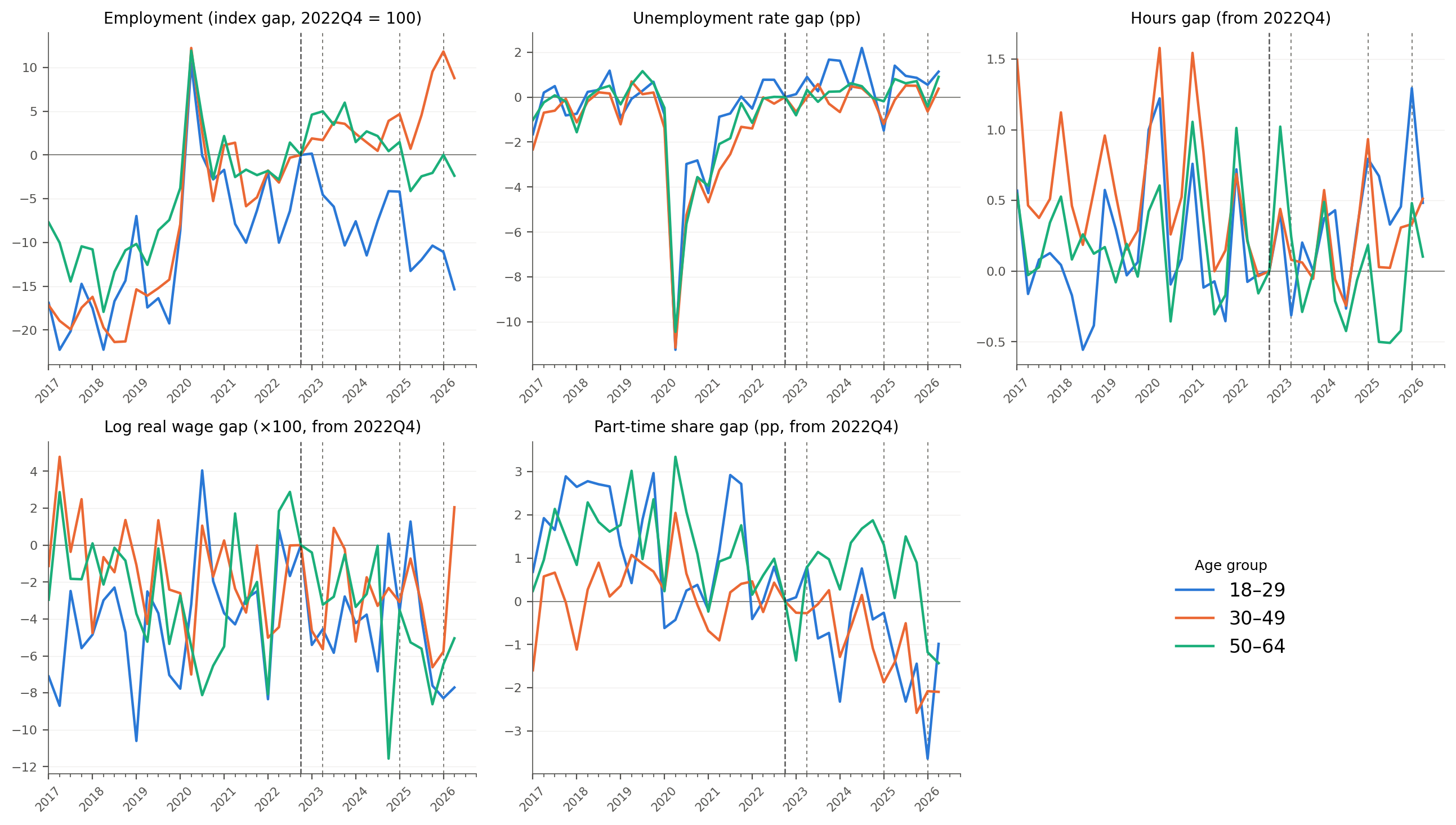}
\caption{The exposure gap in outcomes, by age group}
\label{fig:gapsage}
\begin{minipage}{0.92\textwidth}
\notesize
Notes: Top-quartile minus bottom-quartile value for the five outcomes of Figure~\ref{fig:outcomes}, quartiles of the stage-1 to stage-4 change, each series normalized to its 2022Q4 value within the age group (employment as an index, the rest as deviations). The dashed black line marks the ChatGPT quarter and the gray dashed lines the inflection quarters.
\end{minipage}
\end{figure}

\begin{figure}[htbp]
\centering
\includegraphics[width=\linewidth]{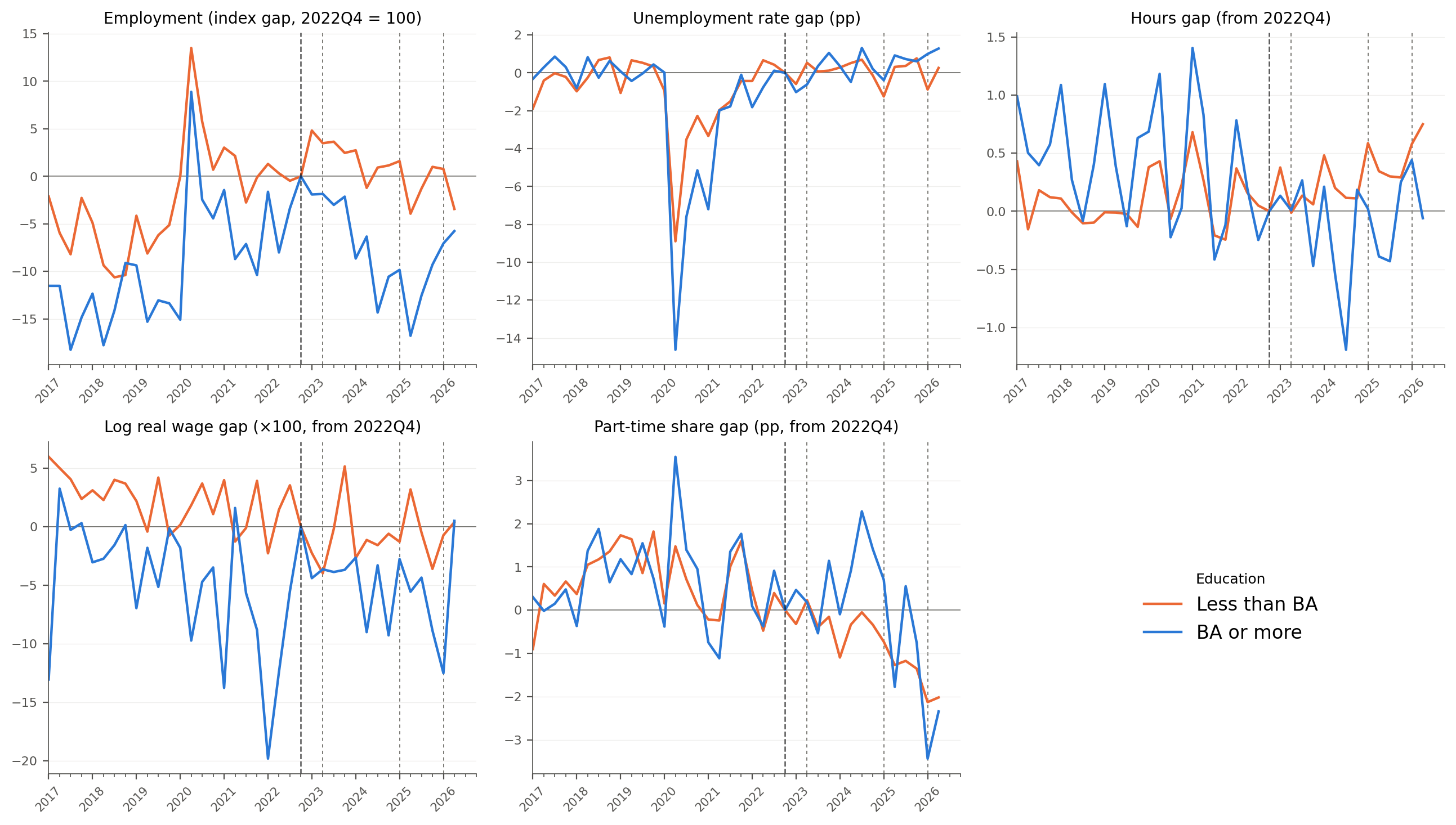}
\caption{The exposure gap in outcomes, by education}
\label{fig:gapsedu}
\begin{minipage}{0.92\textwidth}
\notesize
Notes: As Figure~\ref{fig:gapsage}, by education.
\end{minipage}
\end{figure}

Figure~\ref{fig:frontier} ranks workers from the least to the most exposed occupation and plots the exposure score of the worker's occupation at each stage.
Table~\ref{tab:characteristics} reports the characteristics of the workforce by quartile of the change in exposure from stage 1 to stage 4, which Section~\ref{sec:cps} of the paper summarizes.
Figure~\ref{fig:outcomes} plots the five outcomes from 2017 to 2026 for the top and bottom quartiles of the change in exposure.
Figures~\ref{fig:gapsage} and~\ref{fig:gapsedu} plot the gap between the two quartiles by age group and by education.

\begin{table}[htbp]
\centering
\begin{threeparttable}
\caption{Occupation characteristics by quartile of exposure change}
\label{tab:characteristics}
\footnotesize\setlength{\tabcolsep}{5pt}
\begin{spacing}{1}\scriptsize
\begin{tabular}{@{}lrrrr@{}}
\toprule
 & Bottom quartile & All occupations & Top quartile & Top minus bottom \\
\midrule
Exposure change, stage 1 to 4 & 16.4 & 32.5 & 47.5 & 31.1 \\
Exposure change, stage 1 to 3 & 11.0 & 23.6 & 35.6 & 24.6 \\
Exposure level, stage 1 & 6.1 & 12.4 & 14.3 & 8.2 \\
Exposure level, stage 4 & 22.5 & 44.9 & 61.8 & 39.3 \\
Occupations & 113 & 451 & 113 &  \\
\addlinespace
\multicolumn{5}{@{}l}{\emph{Outcomes, 2023}} \\
Employment (millions) & 32.4 & 147.5 & 41.1 & 8.7 \\
Unemployment rate (\%) & 5.4 & 3.3 & 2.3 & $-$3.1 \\
Hours worked last week & 36.8 & 39.1 & 40.3 & 3.5 \\
Real hourly wage (January 2020 dollars) & 20.8 & 34.5 & 46.0 & 25.3 \\
Part-time share (\%) & 22.9 & 14.2 & 7.9 & $-$15.1 \\
Part-time employment (millions) & 7.4 & 20.9 & 3.2 & $-$4.2 \\
Full-time employment (millions) & 25.0 & 126.6 & 37.9 & 12.9 \\
\addlinespace
\multicolumn{5}{@{}l}{\emph{Composition, 2023 (percent of employment)}} \\
Female & 40.8 & 47.0 & 47.5 & 6.6 \\
Under 30 & 32.2 & 24.3 & 18.9 & $-$13.3 \\
Aged 50 to 64 & 25.3 & 28.6 & 30.8 & 5.5 \\
White & 74.3 & 75.8 & 75.2 & 0.9 \\
Black & 15.8 & 13.3 & 10.7 & $-$5.1 \\
Asian & 4.9 & 7.1 & 10.8 & 5.9 \\
High school or less & 61.0 & 32.9 & 15.3 & $-$45.7 \\
Bachelor's degree & 9.2 & 26.2 & 41.9 & 32.7 \\
Post-bachelor's degree & 2.1 & 15.1 & 20.9 & 18.8 \\
\addlinespace
\multicolumn{5}{@{}l}{\emph{Change, first half of 2023 to first half of 2026}} \\
Log employment (log points) & 1.3 & 0.7 & 0.9 & $-$0.4 \\
Unemployment rate (percentage points) & 0.7 & 0.6 & 0.8 & 0.1 \\
Hours worked last week & $-$0.2 & $-$0.2 & 0.0 & 0.2 \\
Log real hourly wage (log points) & 1.3 & 1.3 & 0.9 & $-$0.5 \\
Part-time share (percentage points) & 1.5 & 0.7 & $-$0.2 & $-$1.7 \\
Log part-time employment (log points) & 9.8 & 6.9 & $-$3.6 & $-$13.4 \\
Log full-time employment (log points) & $-$1.4 & $-$0.3 & 1.3 & 2.6 \\
\bottomrule
\end{tabular}
\end{spacing}
\begin{tablenotes}\notesize
\item Notes: Quartiles of the stage-1 to stage-4 change in exposure across the harmonized census occupations, as in Figures~\ref{fig:outcomes} and~\ref{fig:gapsage}. Levels are 2023 averages, the first year of the generative AI era and the base year of the three-year window in Section~\ref{sec:findings}, weighted by 2023 employment within each group; employment counts are monthly averages. The outcomes are those of Section~\ref{sec:cps} and the composition shares are the controls of Section~\ref{sec:data}. The change rows compare the first half of 2026 with the first half of 2023, the headline window, i.e., 2026Q1 and 2026Q2 against 2023Q1 and 2023Q2 as in Table~\ref{tab:cell_fd_main_1}; rates and means use fixed 2023 employment weights and employment is the group total. Wages are in January 2020 dollars.
\end{tablenotes}
\end{threeparttable}
\end{table}


\section{Aggregate Results in Full}
\label{app:occupation}

This appendix reports the aggregate estimates of Section~\ref{sec:fd-results} of the paper in full.
Appendix~\ref{app:windows} describes the six-month cells and the stacked windows of the three-year first difference.
Appendix~\ref{app:aggregate} reports the three-year differences for all workers at both units of time, the fixed-effects estimates at the occupation and individual level, and the changes within the part-time and within the full-time workforce.

\subsection{The Six-Month Cells and the Windows}
\label{app:windows}

We estimate Eq.~(\ref{eq:group}) of the paper at two units of time.
The occupation-quarter cell is the finer unit and provides the most windows.
However, the median cell contains about a hundred respondents, and a single quarter's change within an occupation is mostly sampling noise.
Hence, we also estimate the regression using six-month averages, which halves the sampling noise in each cell.
Specifically, for end quarter $t$ we average the outcome over quarters $t-1$ and $t$, summing counts and weighting rates and means by the quarter counts, and compare it with the same average three years earlier.
Six-month windows shift one quarter at a time, so that consecutive windows overlap by one quarter, and clustering by occupation accounts for the overlap.

We name each window by its end and base periods.
A quarterly window is written as ``2026Q1 vs 2023Q1'', and a six-month window as ``avg(2025Q4--2026Q1) vs avg(2022Q4--2023Q1)'', i.e., the average of the two quarters compared with the average of the same two quarters three years earlier.
A stacked estimate lists its windows, e.g., the three-year headline estimate stacks ``2026Q1 vs 2023Q1 and 2026Q2 vs 2023Q2'', with a period fixed effect for each.
Table~\ref{tab:windows} lists the stacked three-year windows.
The main window compares the first half of 2026 with the first half of 2023, which is the last window whose base period is after the release of ChatGPT and outside the pandemic, and whose end period is within the data.
One thing to note is that the base period straddles the chatbot inflection, so the estimate assumes that outcomes in early 2023 had not yet responded to the first generation.
Figure~\ref{fig:timing} of the paper relaxes this assumption with a window whose base period is 2022.
The placebo window has both the base and end periods before the release of ChatGPT and a base period before the pandemic.
The underlying reasoning is that if the design finds an effect of the technology in the years before the technology existed, the design is not valid, regardless of what it finds afterwards.
We do not interpret any estimate from the pre-period as an effect of AI.

\begin{table}[htbp]
\centering
\begin{threeparttable}
\caption{The stacked three-year windows}
\label{tab:windows}
\footnotesize\setlength{\tabcolsep}{4pt}
\begin{spacing}{1}
\begin{tabular}{@{}>{\raggedright\arraybackslash}p{3.2cm}>{\raggedright\arraybackslash}p{6.4cm}>{\raggedright\arraybackslash}p{5.4cm}@{}}
\toprule
Horizon and unit & Windows stacked & Role \\
\midrule
Three years, quarters & 2026Q1 vs 2023Q1 and 2026Q2 vs 2023Q2 & Headline: the cumulative change through the reasoning and agent generations \\
Three years, six-month averages & avg(2025Q4--2026Q1) vs avg(2022Q4--2023Q1) and avg(2026Q1--Q2) vs avg(2023Q1--Q2) & Headline at the coarser unit; the first window's base includes the ChatGPT quarter \\
Three-year placebo, quarters & 2021Q4 vs 2018Q4, \ldots{}, 2022Q3 vs 2019Q3 (four windows) & Both ends before the chatbot inflection, base before the pandemic \\
Three-year placebo, six-month averages & avg(2021Q3--Q4) vs avg(2018Q3--Q4), \ldots{}, avg(2022Q2--Q3) vs avg(2019Q2--Q3) (four windows) & The same at the coarser unit \\
\bottomrule
\end{tabular}
\end{spacing}
\begin{tablenotes}\notesize
\item Notes: Each row is one estimate of equation~(\ref{eq:group}); the windows listed enter with a period effect each.
\end{tablenotes}
\end{threeparttable}
\end{table}

The stacked estimate summarizes the cumulative change between the base and end periods, but does not indicate when the change occurred.
Hence, we also estimate Eq.~(\ref{eq:group}) for each window separately, i.e., with a single end quarter $t$, shifting the window one quarter at a time:
\begin{equation}
\Delta_h Y_{ot} = \beta_{h,t}\,\Delta E_o + X_{o,t-h}'\gamma_t + \varepsilon_{ot},\qquad \text{one regression for each } t ,
\label{eq:fdwindow}
\end{equation}
where $h$ is the horizon in quarters.
For example, the coefficient estimate for 2025Q3 with $h = 12$ indicates how much larger the three-year change ending in 2025Q3 was in an occupation with a one standard deviation larger change in exposure.
If the technology affected labor market outcomes and the design is valid, we would expect the coefficient estimates for the placebo windows to be close to zero and the coefficient estimates to move away from zero only after the relevant capability stage arrives.
Figure~\ref{fig:byquarter} of the paper presents the single-window estimates for workers under 30, and Figure~\ref{fig:fd-groups-byquarter} presents them for the three age groups and the five main outcomes.

\subsection{The Aggregate Estimates}
\label{app:aggregate}

Table~\ref{tab:cell_fd_main_1} reports the three-year differences for all workers in full.
Panels A and C are the headline and placebo estimates that Table~\ref{tab:headline} of the paper shows, and Panels B and D repeat them at the six-month unit.
Each panel also reports the estimate with the change from stage 1 to stage 3 as the treatment.
Table~\ref{tab:fe-main} reports the fixed-effects estimates of Eq.~(\ref{eq:fe}) of the paper and the era split of Eq.~(\ref{eq:feshift}) at the occupation level, of which Table~\ref{tab:fe-body} of the paper shows Panel A, and at the individual level.
Appendix~\ref{sec:person} describes the individual-level design.
Table~\ref{tab:cell_fd_composition_1} reports the three-year differences within the part-time and within the full-time workforce, which Section~\ref{sec:who} of the paper discusses.

\begin{table}[htbp]
\centering
\caption{Three-Year First Differences: Headline and Placebo Comparisons}
\label{tab:cell_fd_main_1}
\fontsize{8.3}{9.8}\selectfont
\setlength{\tabcolsep}{2.15pt}
\renewcommand{\arraystretch}{1.0}
\begin{adjustbox}{max width=\linewidth}
\begin{tabularx}{\linewidth}{@{}>{\raggedright\arraybackslash}p{0.225\linewidth}*{5}{>{\centering\arraybackslash}X}@{}}
\toprule
 & \shortstack{Log\\employment} & \shortstack{Unemployment\\(pp)} & Hours & \shortstack{Log real hourly\\earnings} & \shortstack{Part-time share\\(pp)} \\
\midrule
\addlinespace[0.16em]
\multicolumn{6}{@{}l}{\textbf{\textit{A. Main comparison: occupation-quarter cells}}} \\
\multicolumn{6}{@{}l}{\parbox[t]{0.94\linewidth}{\fontsize{7.4}{8.8}\selectfont\itshape 2026Q1 vs. 2023Q1 and 2026Q2 vs. 2023Q2}} \\[0.14em]
S4--S1\hspace{0.55em} & -1.89 (1.45) & 0.16 (0.12) & 0.20\textsuperscript{**} (0.09) & 1.94\textsuperscript{**} (0.80) & -0.74\textsuperscript{***} (0.25) \\
\(R^{2}\) & 0.04 & 0.01 & 0.03 & 0.03 & 0.05 \\
\cmidrule(lr){1-6}
S3--S1\hspace{0.55em} & -1.82 (1.56) & 0.16 (0.13) & 0.22\textsuperscript{**} (0.09) & 2.19\textsuperscript{**} (0.86) & -0.72\textsuperscript{***} (0.26) \\
\(R^{2}\) & 0.04 & 0.01 & 0.03 & 0.03 & 0.04 \\
\cmidrule(lr){1-6}
Mean outcome change & -1.21 & 0.60 & -0.23 & 1.61 & 0.67 \\
\(N\) & 900 & 900 & 900 & 846 & 900 \\
\specialrule{0.075em}{0.20em}{0.14em}
\multicolumn{6}{@{}l}{\textbf{\textit{B. Main comparison: six-month averages}}} \\
\multicolumn{6}{@{}l}{\parbox[t]{0.94\linewidth}{\fontsize{7.4}{8.8}\selectfont\itshape 2025Q4--2026Q1 vs. 2022Q4--2023Q1; 2026Q1--Q2 vs. 2023Q1--Q2}} \\[0.14em]
S4--S1\hspace{0.55em} & -1.84 (1.59) & 0.29\textsuperscript{**} (0.13) & 0.16\textsuperscript{*} (0.08) & 0.49 (0.60) & -0.65\textsuperscript{***} (0.23) \\
\(R^{2}\) & 0.05 & 0.02 & 0.02 & 0.04 & 0.05 \\
\cmidrule(lr){1-6}
S3--S1\hspace{0.55em} & -1.63 (1.68) & 0.27\textsuperscript{**} (0.14) & 0.17\textsuperscript{**} (0.09) & 0.54 (0.66) & -0.60\textsuperscript{**} (0.24) \\
\(R^{2}\) & 0.05 & 0.02 & 0.02 & 0.04 & 0.05 \\
\cmidrule(lr){1-6}
Mean outcome change & -0.22 & 0.65 & -0.23 & 1.47 & 0.67 \\
\(N\) & 898 & 898 & 898 & 884 & 898 \\
\specialrule{0.075em}{0.20em}{0.14em}
\multicolumn{6}{@{}l}{\textbf{\textit{C. Pre-AI placebo: occupation-quarter cells}}} \\
\multicolumn{6}{@{}l}{\parbox[t]{0.94\linewidth}{\fontsize{7.4}{8.8}\selectfont\itshape 2021Q4 vs. 2018Q4 through 2022Q3 vs. 2019Q3}} \\[0.14em]
S4--S1\hspace{0.55em} & 0.49 (1.25) & -0.02 (0.11) & -0.08 (0.07) & 0.69 (0.61) & -0.00 (0.20) \\
\(R^{2}\) & 0.06 & 0.02 & 0.05 & 0.02 & 0.04 \\
\cmidrule(lr){1-6}
S3--S1\hspace{0.55em} & 0.55 (1.35) & -0.04 (0.12) & -0.07 (0.08) & 0.87 (0.66) & -0.05 (0.21) \\
\(R^{2}\) & 0.06 & 0.02 & 0.05 & 0.02 & 0.04 \\
\cmidrule(lr){1-6}
Mean outcome change & -2.27 & 0.25 & -0.29 & 3.08 & -0.45 \\
\(N\) & 1,794 & 1,794 & 1,793 & 1,733 & 1,794 \\
\specialrule{0.075em}{0.20em}{0.14em}
\multicolumn{6}{@{}l}{\textbf{\textit{D. Pre-pandemic-baseline placebo: six-month averages}}} \\
\multicolumn{6}{@{}l}{\parbox[t]{0.94\linewidth}{\fontsize{7.4}{8.8}\selectfont\itshape 2021Q3--Q4 vs. 2018Q3--Q4 through 2022Q2--Q3 vs. 2019Q2--Q3}} \\[0.14em]
S4--S1\hspace{0.55em} & 0.55 (1.24) & -0.08 (0.11) & -0.06 (0.07) & 0.34 (0.56) & -0.00 (0.20) \\
\(R^{2}\) & 0.07 & 0.08 & 0.07 & 0.04 & 0.06 \\
\cmidrule(lr){1-6}
S3--S1\hspace{0.55em} & 0.66 (1.35) & -0.11 (0.12) & -0.06 (0.08) & 0.47 (0.63) & -0.05 (0.21) \\
\(R^{2}\) & 0.07 & 0.08 & 0.07 & 0.04 & 0.06 \\
\cmidrule(lr){1-6}
Mean outcome change & -2.28 & 0.43 & -0.28 & 3.61 & -0.45 \\
\(N\) & 1,793 & 1,793 & 1,793 & 1,783 & 1,793 \\
\bottomrule
\end{tabularx}\end{adjustbox}
\par\vspace{0.50em}\noindent
\begin{minipage}{\linewidth}
\notesize \textit{Notes:} Coefficients per one standard deviation of the named occupation-level exposure change, with standard errors clustered by occupation. S4--S1 is the cumulative change from pre-generative AI to agents; S3--S1 runs through reasoning models. Outcomes are differences over the stated horizons. All specifications use base-period occupation employment weights, twelve base-period full-occupation workforce-composition controls, and later-quarter fixed effects. Log outcomes and indicator shares are multiplied by 100 (log points and percentage points); hours are unscaled. Outcome means are base-employment-weighted means of outcome differences in the relevant model sample (displayed units); the $R^{2}$ rows correspond to the coefficient above each row; $N$ is the number of occupation-window observations. Panels A--B are the first-half-2026 headline at quarterly and rolling six-month units; panels C--D repeat the design on historical placebo periods. The ``Mean outcome change'' row is the base-period-employment-weighted sample mean of the first-differenced dependent variable (the later-period outcome minus the earlier-period outcome), not the mean level of the outcome. A negative entry indicates an average decline over the comparison window; logged outcomes and shares use the display scaling described above. \mbox{\textsuperscript{*}\,$p<0.10$}, \mbox{\textsuperscript{**}\,$p<0.05$}, \mbox{\textsuperscript{***}\,$p<0.01$}.
\end{minipage}

\end{table}

\subsubsection{The individual-level design}
\label{sec:person}

The average wage of an occupation can increase either because the same workers are paid more or because the composition of workers in the occupation changed.
For example, if exposed occupations stopped hiring young, low-paid entrants, the average wage of the occupation would increase even if no individual worker's pay changed.
The same applies to hours worked and part-time status.
To examine whether the occupation-level changes reflect changes for a given worker or changes in the composition of the workforce, we estimate Eq.~(\ref{eq:fe}) of the paper at the individual level, holding worker characteristics constant:
\begin{equation}
Y_{im} = \alpha_{o(i)} + \delta_m + \beta_S\,E^{step}_{o(i)}(m) + X_i'\gamma + \varepsilon_{im},\qquad m \in S ,
\label{eq:person}
\end{equation}
where $Y_{im}$ is the outcome of individual $i$ in month $m$, $o(i)$ is the occupation of individual $i$, $\alpha_{o(i)}$ and $\delta_m$ denote the occupation and month fixed effects, and $X_i$ is the set of individual characteristics, i.e., age and its square, sex, race, education, and region.
The outcomes are an indicator for being employed, estimated on the labor force sample, weekly hours worked, the log real hourly wage, estimated for outgoing rotation group respondents using the earnings weights, and an indicator for usually working part-time.
The step exposure score switches at the monthly inflection dates, i.e., March 2023, January 2025, and January 2026, and the era split of Eq.~(\ref{eq:feshift}) is estimated at January 2025.
The sample window is the monthly version of the 2023Q1 to 2026Q2 window of Section~\ref{sec:design} of the paper.
Regressions use the CPS person weights and standard errors are clustered at the occupation level.
The individual-level regression uses individual observations with occupation fixed effects and controls for observed worker characteristics, but it does not track workers over time.
Hence, it indicates whether the occupation-level relationships persist for workers of given characteristics, although the difference between the two estimates does not isolate changes in the composition of the workforce.

\begin{table}[htbp]
\centering
\caption{Fixed-effects estimates at the occupation and individual level}
\label{tab:fe-main}
\fontsize{8.3}{9.8}\selectfont
\setlength{\tabcolsep}{2.3pt}
\renewcommand{\arraystretch}{1.08}
\begin{tabular*}{\linewidth}{@{\extracolsep{\fill}}l*{5}{c}@{}}
\toprule
 & \shortstack{Log\\employment} & \shortstack{Unemployment\\rate (pp)} & \shortstack{Main-job\\hours} & \shortstack{Log real hourly\\earnings} & \shortstack{Part-time\\share (pp)} \\
\midrule
\multicolumn{6}{@{}l}{\textbf{\textit{A. Occupation level, 2023Q1 to 2026Q2}}} \\[1pt]
Step & -0.60 (1.35) & 0.09 (0.13) & 0.01 (0.08) & -0.12 (0.73) & -0.78\textsuperscript{***} (0.28) \\
\cmidrule(lr){1-6}
Before 2025Q1 & -0.51 (1.74) & 0.27 (0.22) & -0.25\textsuperscript{**} (0.12) & -0.59 (1.09) & -0.14 (0.42) \\
Shift at 2025Q1 & -0.06 (0.96) & -0.10 (0.11) & 0.14\textsuperscript{***} (0.05) & 0.27 (0.46) & -0.37\textsuperscript{*} (0.19) \\
From 2025Q1 & -0.56 (1.28) & 0.16 (0.15) & -0.10 (0.09) & -0.32 (0.80) & -0.50 (0.31) \\
\cmidrule(lr){1-6}
$N$ & 6,295 & 6,295 & 6,295 & 6,108 & 6,295 \\
\addlinespace[3pt]
\multicolumn{6}{@{}l}{\textbf{\textit{B. Occupation level, 2021Q3 to 2026Q2}}} \\[1pt]
Step & 0.68 (1.24) & 0.30\textsuperscript{***} (0.11) & 0.05 (0.06) & -1.38\textsuperscript{***} (0.50) & -0.70\textsuperscript{***} (0.19) \\
\cmidrule(lr){1-6}
Before 2025Q1 & 1.96 (1.35) & 0.49\textsuperscript{***} (0.16) & -0.05 (0.08) & -2.50\textsuperscript{***} (0.67) & -0.39 (0.31) \\
Shift at 2025Q1 & -1.09 (0.89) & -0.16 (0.10) & 0.08\textsuperscript{*} (0.05) & 0.96\textsuperscript{**} (0.41) & -0.26 (0.20) \\
From 2025Q1 & 0.87 (1.20) & 0.32\textsuperscript{***} (0.11) & 0.03 (0.06) & -1.54\textsuperscript{***} (0.50) & -0.65\textsuperscript{***} (0.20) \\
\cmidrule(lr){1-6}
$N$ & 8,989 & 8,989 & 8,988 & 8,728 & 8,989 \\
\midrule
 & & \shortstack{Employed\\(pp)} & \shortstack{Main-job\\hours} & \shortstack{Log real hourly\\earnings} & \shortstack{Part-time\\(pp)} \\
\midrule
\multicolumn{6}{@{}l}{\textbf{\textit{C. Individual level, January 2023 to June 2026}}} \\[1pt]
Step &  & -0.13 (0.13) & -0.00 (0.07) & -0.09 (0.64) & -0.63\textsuperscript{**} (0.27) \\
\cmidrule(lr){1-6}
Before January 2025 &  & -0.36 (0.22) & -0.22\textsuperscript{*} (0.11) & -0.01 (0.99) & -0.15 (0.37) \\
Shift at January 2025 &  & 0.13 (0.09) & 0.11\textsuperscript{**} (0.05) & -0.04 (0.41) & -0.26 (0.17) \\
From January 2025 &  & -0.24 (0.16) & -0.10 (0.08) & -0.05 (0.73) & -0.41 (0.28) \\
\cmidrule(lr){1-6}
$N$ &  & 1,748,004 & 1,629,014 & 243,749 & 1,690,309 \\
\addlinespace[3pt]
\multicolumn{6}{@{}l}{\textbf{\textit{D. Individual level, October 2021 to June 2026}}} \\[1pt]
Step &  & -0.23\textsuperscript{**} (0.10) & 0.01 (0.05) & -1.23\textsuperscript{***} (0.44) & -0.54\textsuperscript{***} (0.18) \\
\cmidrule(lr){1-6}
Before January 2025 &  & -0.39\textsuperscript{***} (0.14) & -0.06 (0.07) & -1.89\textsuperscript{***} (0.61) & -0.33 (0.27) \\
Shift at January 2025 &  & 0.13 (0.08) & 0.06 (0.05) & 0.56 (0.36) & -0.17 (0.17) \\
From January 2025 &  & -0.26\textsuperscript{**} (0.10) & -0.00 (0.05) & -1.33\textsuperscript{***} (0.44) & -0.51\textsuperscript{***} (0.18) \\
\cmidrule(lr){1-6}
$N$ &  & 2,419,155 & 2,254,124 & 335,970 & 2,340,561 \\
\bottomrule
\end{tabular*}
\vspace{0.2em}
\begin{minipage}{\linewidth}
\notesize
\textit{Notes:} Each coefficient is for a one-standard-deviation increase in the exposure step, standardized within the estimation window, with standard errors clustered by occupation in parentheses. Log employment and log real hourly earnings coefficients are multiplied by 100; the unemployment rate, employment, and part-time coefficients are in percentage points; hours are in hours. Panels A and B are occupation-quarter regressions, Eq.~(\ref{eq:fe}), with occupation and quarter fixed effects and fixed 2019 employment weights, on harmonized OCC2010 codes; the step switches in 2023Q2, 2025Q1, and 2026Q1. Panels C and D are individual-month regressions, Eq.~(\ref{eq:person}), with occupation, month, and region fixed effects, controls for age, age squared, sex, race, and education, and CPS person weights (earnings weights for wages); employed is the probability that a worker in the labor force is employed, and the step switches in March 2023, January 2025, and January 2026. The era split, Eq.~(\ref{eq:feshift}), reports the coefficient before the reasoning model inflection, the shift at it, and their sum afterward. \textsuperscript{*} $p<0.10$, \textsuperscript{**} $p<0.05$, \textsuperscript{***} $p<0.01$.
\end{minipage}
\end{table}

\begin{table}[htbp]
\centering
\caption{What changed within the part-time and within the full-time workforce}
\label{tab:cell_fd_composition_1}
\fontsize{9.5}{11.4}\selectfont
\setlength{\tabcolsep}{3pt}
\renewcommand{\tabularxcolumn}[1]{m{#1}}
\renewcommand{\arraystretch}{1.18}
\begin{tabularx}{\linewidth}{@{}>{\raggedright\arraybackslash}m{0.26\linewidth}*{6}{>{\centering\arraybackslash}X}@{}}
\toprule
 & \shortstack{Log\\count} & \shortstack{Mean\\hours} & \shortstack{Log real\\hourly wage} & \shortstack{Under-30\\share (pp)} & \shortstack{BA+\\share (pp)} & \shortstack{Economic-\\reasons\\share (pp)} \\
\midrule
\addlinespace[0.3em]
\multicolumn{7}{@{}l}{\textbf{A. Part-time workers}} \\
\addlinespace[0.25em]
Three years & \shortstack[c]{\(-7.34^{***}\)\\[1pt]\((2.51)\)} & \shortstack[c]{\(0.10\)\\[1pt]\((0.20)\)} & \shortstack[c]{\(-4.20\)\\[1pt]\((3.46)\)} & \shortstack[c]{\(1.00\)\\[1pt]\((1.14)\)} & \shortstack[c]{\(-0.81\)\\[1pt]\((1.00)\)} & \shortstack[c]{\(1.52^{**}\)\\[1pt]\((0.71)\)} \\
\addlinespace[0.3em]
Three-year placebo across pandemic & \shortstack[c]{\(0.16\)\\[1pt]\((2.17)\)} & \shortstack[c]{\(-0.01\)\\[1pt]\((0.13)\)} & \shortstack[c]{\(2.67\)\\[1pt]\((2.78)\)} & \shortstack[c]{\(-0.12\)\\[1pt]\((0.81)\)} & \shortstack[c]{\(0.18\)\\[1pt]\((0.68)\)} & \shortstack[c]{\(0.17\)\\[1pt]\((0.70)\)} \\
\addlinespace[0.3em]
\addlinespace[0.2em]
N, three-year & 710 & 701 & 410 & 710 & 710 & 701 \\
\midrule
\addlinespace[0.3em]
\multicolumn{7}{@{}l}{\textbf{B. Full-time workers}} \\
\addlinespace[0.25em]
Three years & \shortstack[c]{\(-0.87\)\\[1pt]\((1.43)\)} & \shortstack[c]{\(0.07\)\\[1pt]\((0.07)\)} & \shortstack[c]{\(2.58^{***}\)\\[1pt]\((0.80)\)} & \shortstack[c]{\(-1.59^{***}\)\\[1pt]\((0.36)\)} & \shortstack[c]{\(1.76^{***}\)\\[1pt]\((0.35)\)} & \textemdash{} \\
\addlinespace[0.3em]
Three-year placebo across pandemic & \shortstack[c]{\(0.47\)\\[1pt]\((1.21)\)} & \shortstack[c]{\(-0.12^{**}\)\\[1pt]\((0.05)\)} & \shortstack[c]{\(1.00\)\\[1pt]\((0.74)\)} & \shortstack[c]{\(-0.21\)\\[1pt]\((0.31)\)} & \shortstack[c]{\(0.53^{*}\)\\[1pt]\((0.30)\)} & \textemdash{} \\
\addlinespace[0.3em]
\addlinespace[0.2em]
N, three-year & 899 & 899 & 835 & 899 & 899 & \textemdash{} \\
\bottomrule
\end{tabularx}
\vspace{0.15em}
\begin{minipage}{\linewidth}
\notesize \textit{Notes:} Each coefficient is for a one-standard-deviation increase in standardized S4--S1 occupation-level exposure; standard errors (in parentheses) are clustered by OCC2010 occupation. Eq.~(\ref{eq:group}) is estimated using occupation-quarter first differences separately for usual part-time and full-time workers, full-occupation base-quarter employment analytical weights, twelve base-quarter workforce-composition controls, and later-quarter effects. Log outcomes are multiplied by 100, shares are in percentage points, and mean main-job hours are unscaled. Within-status hours and shares use CPS final person weights; wages are weighted means of individual log real hourly earnings among outgoing-rotation respondents using CPS earnings weights. The part-time economic-reasons share is conditional on observed reason classification; respondents without an observed reason are excluded only from this outcome. The full-time economic-reasons column and its $N$ are structurally not applicable (em dashes). $N$ counts stacked occupation-quarter observations in the three-year comparison, not distinct occupations. The two comparisons (later minus base) are the three-year window, 2026Q1/Q2 versus 2023Q1/Q2, and the pre-AI placebo window, 2021Q4--2022Q3 versus 2018Q4--2019Q3. Stars use unrounded two-sided cluster-$t$ p-values. \mbox{\textsuperscript{*}\,\(p<0.10\), \textsuperscript{**}\,\(p<0.05\), \textsuperscript{***}\,\(p<0.01\).}
\end{minipage}
\end{table}

\section{Results by Age and Education in Full}
\label{app:heterogeneity}

This appendix reports the estimates by age and education that Section~\ref{sec:who} of the paper discusses.
Table~\ref{tab:cell_fd_main_5} reports the three-year differences for the six age-by-education cells on the five main outcomes, which Figure~\ref{fig:who} of the paper plots together with the age and education groups of Table~\ref{tab:cell_fd_main_4} of the paper.
Table~\ref{tab:cell_fd_secondary_5} reports the same six cells for the part-time and full-time counts, second-job holding, total hours, second-job hours, and weekly earnings.
Table~\ref{tab:cell_fd_main_6} repeats Table~\ref{tab:cell_fd_main_5} with the change from stage 1 to stage 3 as the treatment.
Table~\ref{tab:cellfe-polished-primary-occ2010-groups} reports the occupation fixed-effects estimates of Eq.~(\ref{eq:fe}) of the paper by age group and by education, with the era split of Eq.~(\ref{eq:feshift}) at 2025Q1, of which Table~\ref{tab:fe-body} of the paper shows the pooled coefficients.
The placebo estimates for these groups are in Appendix~\ref{app:placebo}.
Table~\ref{tab:flows} reports the three-year changes in the monthly flows of workers into and out of occupations, by age and by age and education.
The flows are constructed by linking consecutive monthly interviews of the same respondent in the CPS rotating panel.

\begin{table}[htbp]
\centering
\caption{Three-year differences in the main outcomes, by age and education jointly}
\label{tab:cell_fd_main_5}
\fontsize{7.7}{9.0}\selectfont
\setlength{\tabcolsep}{2pt}
\renewcommand{\arraystretch}{1.04}
\begin{adjustbox}{max width=\linewidth}
\begin{tabularx}{\linewidth}{@{}>{\raggedright\arraybackslash}p{0.236\linewidth}*{5}{>{\centering\arraybackslash}X}@{}}
\toprule
 & \shortstack{Log\\employment} & \shortstack{Unemployment\\(pp)} & Hours & \shortstack{Log real hourly\\earnings} & \shortstack{Part-time share\\(pp)} \\
\midrule
\multicolumn{6}{@{}l}{\parbox[t]{0.97\linewidth}{\itshape 2026Q1 vs. 2023Q1 and 2026Q2 vs. 2023Q2}} \\[0.2em]
Under 30, less than a bachelor's degree & -9.80\textsuperscript{***} (2.58) & -0.54 (0.40) & 0.22 (0.29) & -0.33 (1.97) & -0.48 (0.92) \\
\hspace{0.5em}$R^2$ & 0.09 & 0.01 & 0.02 & 0.04 & 0.03 \\
\hspace{0.5em}Mean outcome change & -3.98 & 0.75 & -0.55 & 3.62 & 2.25 \\
\hspace{0.5em}$N$ & 723 & 729 & 722 & 453 & 723 \\
\cmidrule(lr){1-6}
Under 30, bachelor's degree or more & -2.87 (3.46) & 0.54 (0.52) & 0.21 (0.26) & -2.32 (2.73) & 1.04 (0.89) \\
\hspace{0.5em}$R^2$ & 0.06 & 0.03 & 0.03 & 0.05 & 0.02 \\
\hspace{0.5em}Mean outcome change & -10.72 & 1.42 & -0.35 & 2.40 & 0.73 \\
\hspace{0.5em}$N$ & 588 & 599 & 581 & 350 & 588 \\
\cmidrule(lr){1-6}
\addlinespace[0.30em]
30 to 49, less than a bachelor's degree & -3.08 (2.32) & 0.23 (0.25) & 0.23 (0.15) & 1.23 (1.52) & -0.57 (0.40) \\
\hspace{0.5em}$R^2$ & 0.10 & 0.03 & 0.02 & 0.03 & 0.05 \\
\hspace{0.5em}Mean outcome change & -6.74 & 0.59 & -0.13 & 3.47 & 0.39 \\
\hspace{0.5em}$N$ & 811 & 811 & 811 & 596 & 811 \\
\cmidrule(lr){1-6}
30 to 49, bachelor's degree or more & 4.49\textsuperscript{**} (2.13) & 0.52\textsuperscript{**} (0.21) & 0.37\textsuperscript{**} (0.17) & 4.21\textsuperscript{**} (1.80) & -1.52\textsuperscript{**} (0.60) \\
\hspace{0.5em}$R^2$ & 0.09 & 0.03 & 0.04 & 0.04 & 0.04 \\
\hspace{0.5em}Mean outcome change & 1.42 & 0.79 & -0.21 & -1.70 & 0.40 \\
\hspace{0.5em}$N$ & 743 & 744 & 741 & 538 & 743 \\
\cmidrule(lr){1-6}
\addlinespace[0.30em]
50 to 64, less than a bachelor's degree & -1.32 (2.32) & -0.02 (0.25) & -0.07 (0.15) & -0.65 (2.30) & 0.37 (0.55) \\
\hspace{0.5em}$R^2$ & 0.11 & 0.02 & 0.03 & 0.03 & 0.02 \\
\hspace{0.5em}Mean outcome change & -10.93 & 0.47 & -0.00 & 0.09 & 0.61 \\
\hspace{0.5em}$N$ & 778 & 780 & 778 & 544 & 778 \\
\cmidrule(lr){1-6}
50 to 64, bachelor's degree or more & 3.05 (2.99) & 0.52 (0.33) & 0.05 (0.25) & -3.43 (2.74) & -2.03\textsuperscript{**} (0.84) \\
\hspace{0.5em}$R^2$ & 0.09 & 0.03 & 0.03 & 0.04 & 0.04 \\
\hspace{0.5em}Mean outcome change & -6.76 & 0.84 & -0.26 & -0.07 & 0.07 \\
\hspace{0.5em}$N$ & 677 & 682 & 676 & 444 & 677 \\
\bottomrule
\end{tabularx}\end{adjustbox}
\par\vspace{0.45em}\noindent
\begin{minipage}{\linewidth}
\notesize \textit{Notes:} Each row reports a separate occupation-by-age-by-education-group regression for the indicated three-year first differences. Exposure changes are standardized at the aggregate occupation level; S4--S1 is the cumulative pre-generative-to-agent change and S3--S1 the cumulative pre-generative-to-reasoning change. Models use group-specific base-period employment weights, twelve full-occupation baseline workforce-composition controls, later-quarter fixed effects, and occupation-clustered standard errors. Coefficient standard errors appear in parentheses. The mean outcome change is the group-weighted mean of the first-differenced dependent variable in that model sample, not an outcome level; $N$ is occupation-window observations. Log outcomes and shares are displayed in log points and percentage points; hours are in hours. Second-job hours are conditional on holding a second job. Significance: \textsuperscript{*}$p<0.10$, \textsuperscript{**}$p<0.05$, \textsuperscript{***}$p<0.01$.
\end{minipage}
\end{table}

\begin{table}[htbp]
\centering
\caption{Three-year differences in the main outcomes, by age and education jointly (Secondary Outcomes)}
\label{tab:cell_fd_secondary_5}
\fontsize{7.05}{8.35}\selectfont
\setlength{\tabcolsep}{2pt}
\renewcommand{\arraystretch}{1.04}
\begin{adjustbox}{max width=\linewidth}
\begin{tabularx}{\linewidth}{@{}>{\raggedright\arraybackslash}p{0.205\linewidth}*{6}{>{\centering\arraybackslash}X}@{}}
\toprule
 & \shortstack{Log PT\\employment} & \shortstack{Log FT\\employment} & \shortstack{Second job\\(pp)} & \shortstack{Total\\hours} & \shortstack{Second-job\\hours} & \shortstack{Log weekly\\earnings} \\
\midrule
\multicolumn{7}{@{}l}{\parbox[t]{0.97\linewidth}{\itshape 2026Q1 vs. 2023Q1 and 2026Q2 vs. 2023Q2}} \\[0.2em]
Under 30, less than a bachelor's degree & -15.67\textsuperscript{***} (4.76) & -9.35\textsuperscript{***} (3.10) & -0.08 (0.30) & 0.25 (0.28) & 0.23 (0.64) & -0.42 (2.75) \\
\hspace{0.5em}$R^2$ & 0.06 & 0.07 & 0.03 & 0.03 & 0.06 & 0.04 \\
\hspace{0.5em}Mean outcome change & 5.15 & -7.57 & 0.12 & -0.55 & 0.02 & 0.39 \\
\hspace{0.5em}$N$ & 422 & 691 & 722 & 722 & 227 & 451 \\
\cmidrule(lr){1-7}
Under 30, bachelor's degree or more & 19.08\textsuperscript{*} (9.70) & -3.66 (3.59) & 0.86 (0.68) & 0.28 (0.27) & -1.91\textsuperscript{**} (0.84) & -0.72 (3.36) \\
\hspace{0.5em}$R^2$ & 0.08 & 0.05 & 0.02 & 0.03 & 0.11 & 0.06 \\
\hspace{0.5em}Mean outcome change & 3.71 & -12.49 & 0.09 & -0.42 & -0.20 & 2.75 \\
\hspace{0.5em}$N$ & 245 & 564 & 581 & 581 & 183 & 350 \\
\cmidrule(lr){1-7}
\addlinespace[0.30em]
30 to 49, less than a bachelor's degree & -6.17 (5.47) & -2.37 (2.21) & -0.35 (0.31) & 0.20 (0.16) & 0.92 (0.65) & 2.50\textsuperscript{*} (1.35) \\
\hspace{0.5em}$R^2$ & 0.09 & 0.08 & 0.02 & 0.02 & 0.07 & 0.05 \\
\hspace{0.5em}Mean outcome change & -5.28 & -7.38 & 0.43 & -0.08 & -0.47 & 2.63 \\
\hspace{0.5em}$N$ & 433 & 802 & 811 & 811 & 357 & 596 \\
\cmidrule(lr){1-7}
30 to 49, bachelor's degree or more & -12.55 (7.63) & 5.74\textsuperscript{**} (2.26) & 0.44 (0.42) & 0.44\textsuperscript{**} (0.19) & 0.74 (0.57) & 3.80\textsuperscript{**} (1.90) \\
\hspace{0.5em}$R^2$ & 0.05 & 0.10 & 0.03 & 0.05 & 0.06 & 0.04 \\
\hspace{0.5em}Mean outcome change & 9.78 & 1.24 & 0.56 & -0.16 & -0.20 & -0.88 \\
\hspace{0.5em}$N$ & 365 & 730 & 741 & 741 & 341 & 538 \\
\cmidrule(lr){1-7}
\addlinespace[0.30em]
50 to 64, less than a bachelor's degree & 3.49 (6.71) & -2.14 (2.49) & 0.56 (0.37) & 0.03 (0.16) & -0.07 (0.84) & -1.75 (2.03) \\
\hspace{0.5em}$R^2$ & 0.05 & 0.10 & 0.04 & 0.03 & 0.11 & 0.05 \\
\hspace{0.5em}Mean outcome change & 0.12 & -11.47 & 0.19 & 0.06 & 0.74 & 0.65 \\
\hspace{0.5em}$N$ & 392 & 765 & 778 & 778 & 294 & 543 \\
\cmidrule(lr){1-7}
50 to 64, bachelor's degree or more & -14.51\textsuperscript{*} (8.02) & 5.67\textsuperscript{*} (3.34) & 0.69 (0.52) & 0.19 (0.25) & 0.78 (0.80) & -2.33 (2.48) \\
\hspace{0.5em}$R^2$ & 0.08 & 0.07 & 0.04 & 0.03 & 0.16 & 0.02 \\
\hspace{0.5em}Mean outcome change & -2.77 & -7.01 & 0.14 & -0.16 & 0.71 & -1.32 \\
\hspace{0.5em}$N$ & 335 & 656 & 675 & 675 & 245 & 443 \\
\bottomrule
\end{tabularx}\end{adjustbox}
\par\vspace{0.45em}\noindent
\begin{minipage}{\linewidth}
\notesize \textit{Notes:} Each row reports a separate occupation-by-age-by-education-group regression for the indicated three-year first differences. Exposure changes are standardized at the aggregate occupation level; S4--S1 is the cumulative pre-generative-to-agent change and S3--S1 the cumulative pre-generative-to-reasoning change. Models use group-specific base-period employment weights, twelve full-occupation baseline workforce-composition controls, later-quarter fixed effects, and occupation-clustered standard errors. Coefficient standard errors appear in parentheses. The mean outcome change is the group-weighted mean of the first-differenced dependent variable in that model sample, not an outcome level; $N$ is occupation-window observations. Log outcomes and shares are displayed in log points and percentage points; hours are in hours. Second-job hours are conditional on holding a second job. Significance: \textsuperscript{*}$p<0.10$, \textsuperscript{**}$p<0.05$, \textsuperscript{***}$p<0.01$.
\end{minipage}
\end{table}

\begin{table}[htbp]
\centering
\caption{Three-year differences in the main outcomes, by age and education jointly (Stage 1 to 3)}
\label{tab:cell_fd_main_6}
\fontsize{7.7}{9.0}\selectfont
\setlength{\tabcolsep}{2pt}
\renewcommand{\arraystretch}{1.04}
\begin{adjustbox}{max width=\linewidth}
\begin{tabularx}{\linewidth}{@{}>{\raggedright\arraybackslash}p{0.236\linewidth}*{5}{>{\centering\arraybackslash}X}@{}}
\toprule
 & \shortstack{Log\\employment} & \shortstack{Unemployment\\(pp)} & Hours & \shortstack{Log real hourly\\earnings} & \shortstack{Part-time share\\(pp)} \\
\midrule
\multicolumn{6}{@{}l}{\parbox[t]{0.97\linewidth}{\itshape 2026Q1 vs. 2023Q1 and 2026Q2 vs. 2023Q2}} \\[0.2em]
Under 30, less than a bachelor's degree & -10.65\textsuperscript{***} (2.78) & -0.60 (0.43) & 0.17 (0.31) & -0.22 (2.23) & -0.28 (0.98) \\
\hspace{0.5em}$R^2$ & 0.08 & 0.01 & 0.02 & 0.04 & 0.03 \\
\hspace{0.5em}Mean outcome change & -3.98 & 0.75 & -0.55 & 3.62 & 2.25 \\
\hspace{0.5em}$N$ & 723 & 729 & 722 & 453 & 723 \\
\cmidrule(lr){1-6}
Under 30, bachelor's degree or more & -4.03 (3.75) & 0.44 (0.59) & 0.30 (0.26) & -3.05 (2.76) & 0.99 (0.87) \\
\hspace{0.5em}$R^2$ & 0.06 & 0.03 & 0.03 & 0.05 & 0.02 \\
\hspace{0.5em}Mean outcome change & -10.72 & 1.42 & -0.35 & 2.40 & 0.73 \\
\hspace{0.5em}$N$ & 588 & 599 & 581 & 350 & 588 \\
\cmidrule(lr){1-6}
\addlinespace[0.30em]
30 to 49, less than a bachelor's degree & -2.81 (2.50) & 0.29 (0.27) & 0.26 (0.16) & 1.81 (1.72) & -0.53 (0.45) \\
\hspace{0.5em}$R^2$ & 0.10 & 0.03 & 0.02 & 0.03 & 0.05 \\
\hspace{0.5em}Mean outcome change & -6.74 & 0.59 & -0.13 & 3.47 & 0.39 \\
\hspace{0.5em}$N$ & 811 & 811 & 811 & 596 & 811 \\
\cmidrule(lr){1-6}
30 to 49, bachelor's degree or more & 4.89\textsuperscript{**} (2.34) & 0.47\textsuperscript{**} (0.22) & 0.41\textsuperscript{**} (0.17) & 4.05\textsuperscript{**} (1.88) & -1.49\textsuperscript{**} (0.59) \\
\hspace{0.5em}$R^2$ & 0.09 & 0.02 & 0.04 & 0.04 & 0.03 \\
\hspace{0.5em}Mean outcome change & 1.42 & 0.79 & -0.21 & -1.70 & 0.40 \\
\hspace{0.5em}$N$ & 743 & 744 & 741 & 538 & 743 \\
\cmidrule(lr){1-6}
\addlinespace[0.30em]
50 to 64, less than a bachelor's degree & -1.86 (2.52) & 0.02 (0.28) & -0.12 (0.16) & 0.06 (2.54) & 0.46 (0.59) \\
\hspace{0.5em}$R^2$ & 0.11 & 0.02 & 0.03 & 0.03 & 0.02 \\
\hspace{0.5em}Mean outcome change & -10.93 & 0.47 & -0.00 & 0.09 & 0.61 \\
\hspace{0.5em}$N$ & 778 & 780 & 778 & 544 & 778 \\
\cmidrule(lr){1-6}
50 to 64, bachelor's degree or more & 3.40 (3.12) & 0.35 (0.33) & -0.01 (0.25) & -3.53 (2.95) & -1.88\textsuperscript{**} (0.85) \\
\hspace{0.5em}$R^2$ & 0.09 & 0.03 & 0.03 & 0.04 & 0.04 \\
\hspace{0.5em}Mean outcome change & -6.76 & 0.84 & -0.26 & -0.07 & 0.07 \\
\hspace{0.5em}$N$ & 677 & 682 & 676 & 444 & 677 \\
\bottomrule
\end{tabularx}\end{adjustbox}
\par\vspace{0.45em}\noindent
\begin{minipage}{\linewidth}
\notesize \textit{Notes:} Each row reports a separate occupation-by-age-by-education-group regression for the indicated three-year first differences. Exposure changes are standardized at the aggregate occupation level; S4--S1 is the cumulative pre-generative-to-agent change and S3--S1 the cumulative pre-generative-to-reasoning change. Models use group-specific base-period employment weights, twelve full-occupation baseline workforce-composition controls, later-quarter fixed effects, and occupation-clustered standard errors. Coefficient standard errors appear in parentheses. The mean outcome change is the group-weighted mean of the first-differenced dependent variable in that model sample, not an outcome level; $N$ is occupation-window observations. Log outcomes and shares are displayed in log points and percentage points; hours are in hours. Second-job hours are conditional on holding a second job. Significance: \textsuperscript{*}$p<0.10$, \textsuperscript{**}$p<0.05$, \textsuperscript{***}$p<0.01$.
\end{minipage}
\end{table}

\begin{table}[htbp]
\centering
\caption{Occupation-level fixed-effects regressions by group: main results}
\label{tab:cellfe-polished-primary-occ2010-groups}
\fontsize{7.40}{8.45}\selectfont
\setlength{\tabcolsep}{1.65pt}
\renewcommand{\arraystretch}{1.02}
\begin{tabular*}{\linewidth}{@{\extracolsep{\fill}}l*{5}{c}@{}}
\toprule
 & \multicolumn{1}{c}{\shortstack{\textbf{Log}\\\textbf{employment}}} & \multicolumn{1}{c}{\shortstack{\textbf{Unemployment}\\\textbf{(pp)}}} & \multicolumn{1}{c}{\textbf{Hours}} & \multicolumn{1}{c}{\shortstack{\textbf{Log real hourly}\\\textbf{earnings}}} & \multicolumn{1}{c}{\shortstack{\textbf{Part-time}\\\textbf{(pp)}}} \\
\midrule
\multicolumn{6}{l}{\textit{Estimation window: 2023Q1--2026Q2}} \\
\specialrule{0.075em}{0.10em}{0.045em}
\multicolumn{6}{l}{\textbf{\textit{A. By age group}}} \\
\multicolumn{6}{l}{\textit{Full-period estimates}} \\
\textbf{18--29} & -5.05\textsuperscript{**} (2.11) & -0.26 (0.31) & 0.08 (0.19) & 0.32 (1.39) & -0.17 (0.64) \\
\addlinespace[0.08em]
\textbf{30--49} & 3.03\textsuperscript{*} (1.55) & 0.09 (0.16) & 0.02 (0.09) & -0.31 (1.08) & -0.85\textsuperscript{***} (0.28) \\
\addlinespace[0.08em]
\textbf{50--64} & -2.87 (1.80) & 0.49\textsuperscript{***} (0.17) & -0.36\textsuperscript{***} (0.12) & -1.54 (1.32) & -0.14 (0.42) \\
\cmidrule(lr){1-6}
\textbf{18--29}: Before 2025Q1 & -3.46 (2.31) & 0.28 (0.47) & -0.38 (0.24) & 1.84 (1.96) & 0.62 (0.95) \\
\quad Shift at 2025Q1 & -0.89 (1.38) & -0.30 (0.23) & 0.26\textsuperscript{**} (0.11) & -0.85 (0.88) & -0.44 (0.43) \\
\addlinespace[0.08em]
\textbf{30--49}: Before 2025Q1 & 0.62 (2.25) & 0.12 (0.23) & -0.28\textsuperscript{**} (0.13) & -1.14 (1.69) & 0.30 (0.38) \\
\quad Shift at 2025Q1 & 1.39 (1.04) & -0.02 (0.11) & 0.17\textsuperscript{***} (0.06) & 0.48 (0.73) & -0.67\textsuperscript{***} (0.19) \\
\addlinespace[0.08em]
\textbf{50--64}: Before 2025Q1 & 1.59 (2.52) & 0.61\textsuperscript{*} (0.31) & -0.42\textsuperscript{**} (0.17) & -1.78 (1.98) & -0.24 (0.55) \\
\quad Shift at 2025Q1 & -2.55\textsuperscript{*} (1.46) & -0.06 (0.15) & 0.03 (0.06) & 0.13 (0.87) & 0.06 (0.22) \\
\cmidrule(lr){1-6}
\(R^{2}\) & 0.98 & 0.42 & 0.86 & 0.85 & 0.89 \\
Outcome mean & 12.52 & 3.70 & 38.14 & 3.16 & 15.16 \\
\(N\) & 18,406 & 18,449 & 18,395 & 15,935 & 18,406 \\
\specialrule{0.075em}{0.10em}{0.045em}
\multicolumn{6}{l}{\textbf{\textit{B. By education}}} \\
\multicolumn{6}{l}{\textit{Full-period estimates}} \\
\textbf{Less than BA} & -3.53\textsuperscript{*} (1.81) & -0.25 (0.19) & -0.02 (0.12) & 0.87 (1.03) & -0.59 (0.39) \\
\addlinespace[0.08em]
\textbf{BA or more} & -0.79 (2.06) & 0.42\textsuperscript{*} (0.23) & 0.16 (0.10) & -0.05 (1.28) & -0.91\textsuperscript{**} (0.38) \\
\cmidrule(lr){1-6}
\textbf{Less than BA}: Before 2025Q1 & -2.25 (1.72) & 0.16 (0.31) & -0.21 (0.17) & 1.31 (1.58) & -0.17 (0.57) \\
\quad Shift at 2025Q1 & -0.65 (0.94) & -0.21\textsuperscript{*} (0.12) & 0.10 (0.07) & -0.22 (0.51) & -0.22 (0.26) \\
\addlinespace[0.08em]
\textbf{BA or more}: Before 2025Q1 & -4.33 (3.57) & -0.06 (0.32) & 0.01 (0.15) & -0.68 (1.77) & -0.13 (0.50) \\
\quad Shift at 2025Q1 & 2.28 (1.87) & 0.31\textsuperscript{**} (0.15) & 0.10 (0.08) & 0.41 (0.80) & -0.51\textsuperscript{**} (0.23) \\
\cmidrule(lr){1-6}
\(R^{2}\) & 0.99 & 0.49 & 0.89 & 0.89 & 0.92 \\
Outcome mean & 13.09 & 3.71 & 38.14 & 3.15 & 15.13 \\
\(N\) & 11,846 & 11,863 & 11,838 & 10,507 & 11,846 \\
\bottomrule
\end{tabular*}
\vspace{0.18em}
\begin{minipage}{\linewidth}
\notesize \textit{Notes:} Occupation-by-demographic-group-by-quarter regressions over 2023Q1--2026Q2. Exposure is standardized using the aggregate occupation-quarter window. Coefficients for log employment and log real hourly earnings are multiplied by 100, while their outcome means are reported in unscaled log units; unemployment and part-time shares are percentage points; main-job hours are in hours. Age regressions absorb occupation-by-age and age-by-quarter fixed effects and use fixed 2019 occupation-by-age employment weights. Education regressions absorb occupation-by-education and education-by-quarter fixed effects and use fixed 2019 occupation-by-education employment weights. Standard errors are clustered by occupation. Full-period coefficients come from the pooled step specification; the before-2025Q1 and shift rows come from the era-split specification. The displayed \(R^{2}\) and outcome mean come from the pooled step specification, and \(N\) is reported once per demographic panel. Outcome means are weighted means of the dependent-variable level (unscaled for logged outcomes) over the corresponding regression sample, pooled over the displayed demographic groups rather than calculated separately for each subgroup. \textsuperscript{*} \(p<0.10\), \textsuperscript{**} \(p<0.05\), \textsuperscript{***} \(p<0.01\).
\end{minipage}
\end{table}

\begin{table}[htbp]
\centering
\begin{threeparttable}
\caption{Three-year changes in the monthly flows of workers into and out of occupations, by age and by age and education}
\label{tab:flows}
\setstretch{1}\fontsize{8.3}{9.8}\selectfont\setlength{\tabcolsep}{2.5pt}
\begin{tabular}{@{}l*{8}{c}@{}}
\toprule
 & \multicolumn{3}{c}{Entrants} & \multicolumn{4}{c}{Exits} & \\
\cmidrule(lr){2-4}\cmidrule(lr){5-8}
 & \shortstack{From\\non-\\employment} & \shortstack{From\\another\\occupation} & \shortstack{All\\sources} & \shortstack{To\\non-\\employment} & \shortstack{To\\unem-\\ployment} & \shortstack{To\\another\\occupation} & \shortstack{All\\destinations} & \shortstack{Net:\\entrants\\minus exits} \\
\midrule
\multicolumn{9}{@{}l}{\emph{A. Headline window: 2026Q1--Q2 vs. 2023Q1--Q2}} \\
Under 30 & $-$0.50\textsuperscript{*} & 0.02 & $-$0.48 & $-$0.63\textsuperscript{**} & 0.11 & 0.42 & $-$0.21 & $-$0.37 \\
 & (0.27) & (0.36) & (0.45) & (0.26) & (0.17) & (0.43) & (0.52) & (0.53) \\
30 to 49 & 0.03 & 0.21 & 0.23 & 0.05 & $-$0.08 & 0.09 & 0.15 & 0.08 \\
 & (0.14) & (0.24) & (0.32) & (0.14) & (0.07) & (0.23) & (0.26) & (0.31) \\
50 to 64 & 0.30\textsuperscript{**} & $-$0.55\textsuperscript{**} & $-$0.25 & $-$0.24 & 0.00 & $-$0.19 & $-$0.43 & 0.15 \\
 & (0.14) & (0.25) & (0.28) & (0.18) & (0.08) & (0.31) & (0.39) & (0.41) \\
All workers & $-$0.03 & 0.00 & $-$0.03 & $-$0.19\textsuperscript{*} & $-$0.02 & 0.04 & $-$0.14 & 0.13 \\
 & (0.10) & (0.19) & (0.22) & (0.10) & (0.06) & (0.19) & (0.21) & (0.26) \\
\emph{Under 30, by education} &  &  &  &  &  &  &  &  \\
\quad Less than a bachelor's degree & $-$1.04\textsuperscript{**} & $-$0.42 & $-$1.45\textsuperscript{**} & $-$1.15\textsuperscript{***} & 0.23 & 0.38 & $-$0.77 & $-$0.83 \\
 & (0.42) & (0.45) & (0.63) & (0.39) & (0.20) & (0.48) & (0.61) & (0.69) \\
\quad Bachelor's degree or more & 0.72\textsuperscript{**} & 0.40 & 1.11 & 0.33 & $-$0.01 & 1.02 & 1.34 & $-$0.23 \\
 & (0.36) & (0.66) & (0.75) & (0.41) & (0.30) & (0.70) & (0.83) & (0.71) \\
\addlinespace
\multicolumn{9}{@{}l}{\emph{B. Placebo window: 2021Q4--2022Q3 vs. 2018Q4--2019Q3}} \\
Under 30 & $-$0.02 & 0.20 & 0.18 & 0.03 & 0.02 & 0.25 & 0.28 & $-$0.09 \\
 & (0.23) & (0.28) & (0.40) & (0.26) & (0.11) & (0.28) & (0.44) & (0.29) \\
30 to 49 & $-$0.14 & 0.45\textsuperscript{***} & 0.31 & $-$0.18\textsuperscript{**} & $-$0.06 & 0.17 & $-$0.01 & 0.32 \\
 & (0.10) & (0.17) & (0.20) & (0.09) & (0.05) & (0.16) & (0.20) & (0.19) \\
50 to 64 & 0.05 & 0.24 & 0.30 & $-$0.13 & 0.05 & 0.17 & 0.04 & 0.24 \\
 & (0.11) & (0.20) & (0.23) & (0.13) & (0.06) & (0.18) & (0.22) & (0.24) \\
All workers & $-$0.05 & 0.33\textsuperscript{**} & 0.29 & $-$0.10 & $-$0.02 & 0.21 & 0.11 & 0.18 \\
 & (0.09) & (0.13) & (0.18) & (0.10) & (0.04) & (0.13) & (0.19) & (0.13) \\
\emph{Under 30, by education} &  &  &  &  &  &  &  &  \\
\quad Less than a bachelor's degree & $-$0.09 & 0.18 & 0.09 & $-$0.09 & $-$0.02 & 0.01 & $-$0.08 & 0.16 \\
 & (0.30) & (0.36) & (0.51) & (0.31) & (0.15) & (0.36) & (0.58) & (0.40) \\
\quad Bachelor's degree or more & $-$0.14 & $-$0.01 & $-$0.15 & $-$0.12 & 0.15 & 0.84\textsuperscript{*} & 0.72 & $-$0.91\textsuperscript{**} \\
 & (0.33) & (0.42) & (0.57) & (0.34) & (0.14) & (0.46) & (0.53) & (0.46) \\
\bottomrule
\end{tabular}
\begin{tablenotes}\notesize
\item Notes: Each cell is the coefficient of Eq.~(\ref{eq:group}) with a monthly flow of workers in place of the log count of workers: the three-year difference, per standard deviation of the stage-1 to stage-4 change in exposure, in the number of workers entering or leaving the occupation each month per 100 base-period workers of the group in the occupation, with the twelve base-period composition shares, base-period group employment weights, a period effect, and standard errors clustered by occupation in parentheses. Movers are identified by linking consecutive monthly interviews of the same respondent in the CPS rotating panel. Non-employment is unemployment plus being out of the labor force. An entrant from non-employment was unemployed or out of the labor force in the previous month, and an entrant from another occupation was employed under a different occupation code; exits are the reverse, and exits to unemployment are the part of exits to non-employment whose destination is unemployment. The last column is the change in entrants from all sources minus the change in exits to all destinations, estimated on the cells where both are observed. The monthly rate from the linked sample is multiplied by the group's employment in the occupation and quarter to obtain the count, so that quarters with no movers remain in the sample. Moves between occupation codes include coding noise: about 7 percent of linked workers employed in both months carry a different occupation code, against 2 percent among those who report the same employer, so the non-employment columns are the cleaner measure. A net coefficient of $-$0.37 in the first row cumulates over 36 months to a decline of about 13 percent, against the 8.20 log point decline in the number of workers under 30 over the same window (Table~\ref{tab:cell_fd_main_4}), but its standard error is wide. $N$ is between 502 and 899 occupation-quarter cells in Panel A and between 1,045 and 1,792 in Panel B. \textsuperscript{*} $p<0.10$, \textsuperscript{**} $p<0.05$, \textsuperscript{***} $p<0.01$.
\end{tablenotes}
\end{threeparttable}
\end{table}

\section{The Decomposition by Occupation}
\label{app:decomposition}

\begin{figure}[htbp]
\centering
\includegraphics[width=\linewidth]{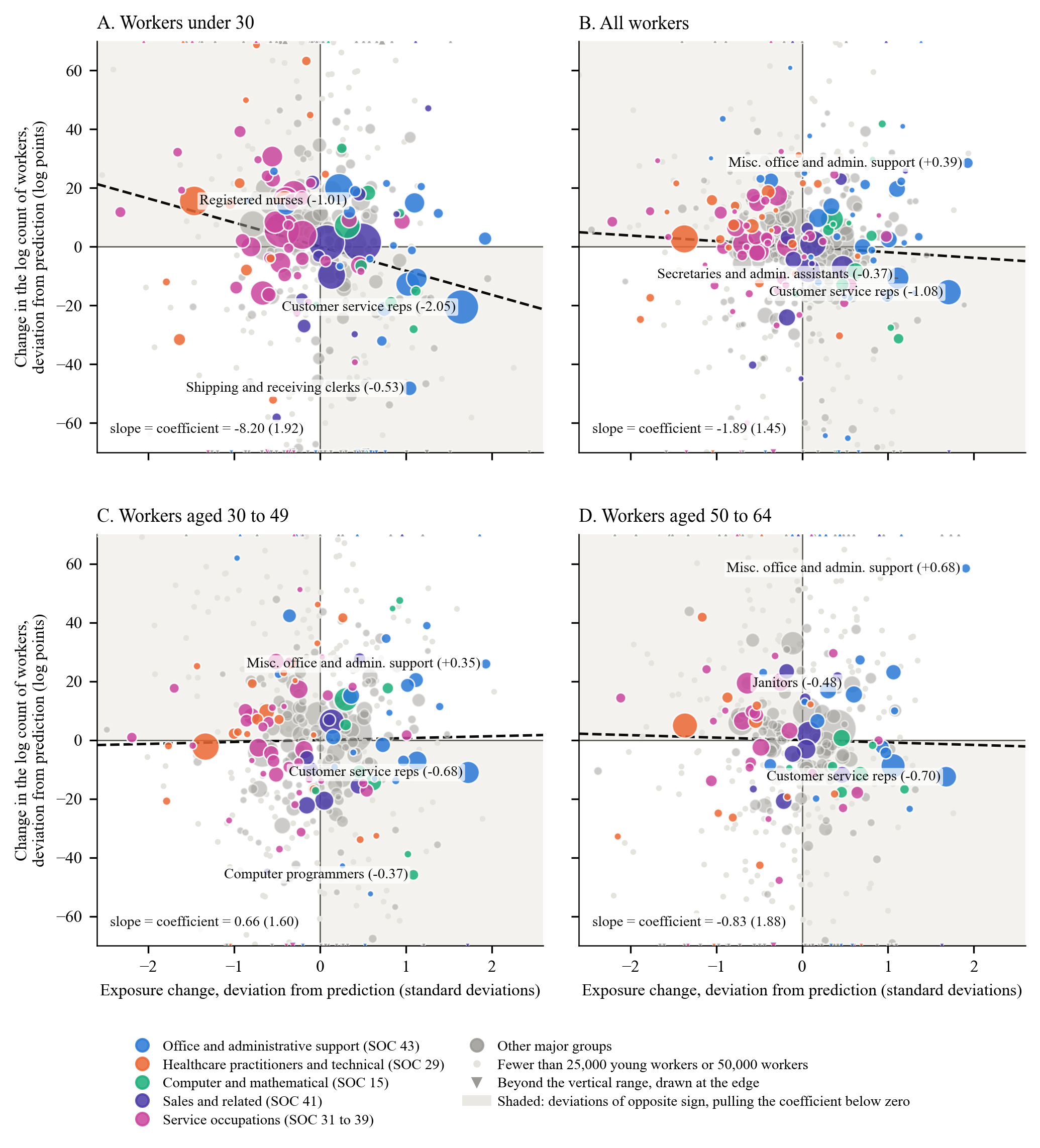}
\caption{The decomposition of Figure~\ref{fig:decomposition} of the paper, by age group}
\label{fig:decomposition-ages}
\begin{minipage}{\linewidth}
\notesize
Notes: As Figure~\ref{fig:decomposition} of the paper, with the same colors, for workers under 30 (Panel A), all workers (Panel B), and workers aged 30 to 49 (Panel C) and 50 to 64 (Panel D). Each bubble is an occupation, with area proportional to the group's base-period employment, the regression weight. The axes are the deviations of the occupation's stage-1 to stage-4 exposure change and of its three-year change in the group's log count from the values predicted by the twelve composition controls and the period effects. The dashed line's slope is the group's coefficient of Eq.~(\ref{eq:group}), from Table~\ref{tab:cell_fd_main_4} of the paper for the age groups and Table~\ref{tab:cell_fd_main_1} for all workers.
\end{minipage}
\end{figure}

This appendix reports the exhibits behind Section~\ref{sec:decomposition} of the paper.
Table~\ref{tab:decomposition} reports the decomposition of Eq.~(\ref{eq:decomp}) of the paper by major occupation group,\footnote{The 22 major occupation groups are the major groups of the 2018 Standard Occupational Classification (SOC 11 to 53) as coded in the census occupation classification that the IPUMS OCC2010 codes follow: management (OCC2010 codes 10 to 430), business and financial operations (500 to 950), computer and mathematical (1000 to 1240), architecture and engineering (1300 to 1560), life, physical, and social science (1600 to 1980), community and social service (2000 to 2060), legal (2100 to 2160), educational instruction and library (2200 to 2550), arts, design, entertainment, sports, and media (2600 to 2960), healthcare practitioners and technical (3000 to 3540), healthcare support (3600 to 3655), protective service (3700 to 3955), food preparation and serving related (4000 to 4160), building and grounds cleaning and maintenance (4200 to 4250), personal care and service (4300 to 4650), sales and related (4700 to 4965), office and administrative support (5000 to 5940), farming, fishing, and forestry (6005 to 6130), construction and extraction (6200 to 6940), installation, maintenance, and repair (7000 to 7630), production (7700 to 8965), and transportation and material moving (9000 to 9750). The military group (SOC 55) does not appear in the employed civilian sample. The figures of the paper color four of these groups, office and administrative support, healthcare practitioners and technical, computer and mathematical, and sales and related, and the service occupations, which the Bureau of Labor Statistics defines as SOC 31 to 39, i.e., healthcare support, protective service, food preparation and serving related, building and grounds cleaning and maintenance, and personal care and service.} which Figure~\ref{fig:decompgroups} of the paper plots.
Figure~\ref{fig:decomposition-ages} repeats Panel A of Figure~\ref{fig:decomposition} of the paper for all workers and for workers aged 30 to 49 and 50 to 64.
Table~\ref{tab:topoccs} lists the occupations in the most exposed quartile with the largest declines in young workers between the first half of 2023 and the first half of 2026, for workers under 30 and for young workers without a bachelor's degree.

\begin{table}[htbp]
\centering
\begin{threeparttable}
\caption{Where the decline in young workers comes from: contributions to the coefficient by major occupation group}
\label{tab:decomposition}
\footnotesize\setlength{\tabcolsep}{2pt}
\begin{spacing}{1}\scriptsize
\begin{tabular}{@{}l*{7}{r}@{}}
\toprule
 & & \multicolumn{3}{c}{Workers under 30} & \multicolumn{3}{c}{Under 30, less than a bachelor's degree} \\
\cmidrule(lr){3-5}\cmidrule(lr){6-8}
Major occupation group & \shortstack[r]{Mean\\exposure\\change} & \shortstack[r]{Contribution\\(log\\points)} & \shortstack[r]{Share of\\coefficient\\(\%)} & \shortstack[r]{Share of\\young\\employment (\%)} & \shortstack[r]{Contribution\\(log\\points)} & \shortstack[r]{Share of\\coefficient\\(\%)} & \shortstack[r]{Share of\\young\\employment (\%)} \\
\midrule
Office and administrative support & 38.0 & $-$3.46 & 42.2 & 12.8 & $-$2.65 & 27.1 & 14.5 \\
Healthcare practitioners and technical & 27.2 & $-$1.01 & 12.3 & 5.3 & $-$1.54 & 15.8 & 3.0 \\
Arts, design, entertainment, sports, and media & 40.7 & $-$0.90 & 11.0 & 2.2 & $-$0.40 & 4.1 & 1.4 \\
Transportation and material moving & 22.0 & $-$0.71 & 8.7 & 6.4 & $-$0.69 & 7.1 & 8.5 \\
Protective service & 26.0 & $-$0.54 & 6.6 & 1.8 & $-$0.69 & 7.1 & 2.0 \\
Food preparation and serving related & 19.3 & $-$0.49 & 5.9 & 10.0 & $-$0.43 & 4.4 & 13.2 \\
Healthcare support & 25.8 & $-$0.39 & 4.8 & 3.2 & $-$0.38 & 3.9 & 3.8 \\
Installation, maintenance, and repair & 22.7 & $-$0.38 & 4.6 & 3.2 & $-$0.65 & 6.7 & 4.3 \\
Educational instruction and library & 33.5 & $-$0.36 & 4.4 & 5.3 & $-$0.37 & 3.7 & 2.8 \\
Business and financial operations & 46.0 & $-$0.30 & 3.7 & 4.6 & $-$0.16 & 1.6 & 1.7 \\
Architecture and engineering & 47.2 & $-$0.28 & 3.4 & 2.3 & $-$0.50 & 5.1 & 0.7 \\
Computer and mathematical & 52.6 & $-$0.20 & 2.4 & 3.8 & $-$0.44 & 4.5 & 1.2 \\
Life, physical, and social science & 41.0 & $-$0.14 & 1.7 & 1.1 & $-$0.07 & 0.7 & 0.3 \\
Production & 24.0 & $-$0.12 & 1.5 & 5.5 & $-$0.03 & 0.3 & 7.2 \\
Construction and extraction & 17.8 & $-$0.03 & 0.4 & 5.5 & $-$0.11 & 1.1 & 7.6 \\
Farming, fishing, and forestry & 21.1 & $-$0.01 & 0.1 & 0.6 & $-$0.07 & 0.7 & 0.8 \\
Personal care and service & 23.4 & 0.00 & 0.0 & 4.5 & $-$0.49 & 5.0 & 5.3 \\
Community and social service & 32.0 & 0.07 & $-$0.9 & 1.3 & $-$0.11 & 1.2 & 0.5 \\
Legal & 44.0 & 0.20 & $-$2.4 & 0.6 & $-$0.17 & 1.7 & 0.2 \\
Building and grounds cleaning and maintenance & 21.9 & 0.22 & $-$2.7 & 3.1 & 0.03 & $-$0.3 & 4.2 \\
Management & 41.6 & 0.30 & $-$3.7 & 6.5 & 0.15 & $-$1.5 & 4.4 \\
Sales and related & 37.6 & 0.33 & $-$4.0 & 10.6 & $-$0.02 & 0.2 & 12.3 \\
\midrule
Coefficient (sum of contributions) & & $-$8.20 & 100.0 & 100.0 & $-$9.80 & 100.0 & 100.0 \\
Standard error & & (1.92) & & & (2.58) & & \\
Occupation-windows [occupations] & & 831 [432] & & & 723 [387] & & \\
\bottomrule
\end{tabular}
\end{spacing}
\begin{tablenotes}\notesize
\item Notes: The coefficient is the three-year first difference of Eq.~(\ref{eq:group}) of the paper for the log count of young workers, 2026Q1 vs.\ 2023Q1 and 2026Q2 vs.\ 2023Q2 stacked, with the twelve base-period composition controls, base-period young employment weights, and standard errors clustered by occupation, i.e., the under-30 row of Table~\ref{tab:cell_fd_main_4} of the paper and the young non-graduate cell of Table~\ref{tab:cell_fd_main_5}, decomposed in Figures~\ref{fig:decomposition} and~\ref{fig:decompgroups} of the paper. After the controls and the period effect are partialled out, the coefficient equals the weighted sum over occupations of the product of the exposure-change deviation and the outcome deviation, divided by the weighted sum of squared exposure-change deviations, and each occupation's term is its contribution. Rows sum the contributions within the 22 major census occupation groups. A negative contribution comes from occupations with above-average exposure change whose young employment fell relative to prediction, or from occupations with below-average exposure change whose young employment rose relative to prediction. Mean exposure change is the unweighted mean of the stage-1 to stage-4 change across the group's occupations; share of young employment is the group's share of the base-period weights.
\end{tablenotes}
\end{threeparttable}
\end{table}

\begin{table}[htbp]
\centering
\begin{threeparttable}
\caption{Occupations in the most exposed quartile with the largest declines in young workers, 2023H1 to 2026H1}
\label{tab:topoccs}
\footnotesize\setlength{\tabcolsep}{4pt}
\begin{spacing}{1}\scriptsize
\begin{tabular}{@{}lrrrr@{}}
\toprule
Occupation & \shortstack[r]{Exposure\\change} & \shortstack[r]{Young workers\\in 2023H1 (thousands)} & \shortstack[r]{Change in log\\count (log points)} & \shortstack[r]{Share of the quartile's\\net loss (\%)} \\
\midrule
\multicolumn{5}{@{}l}{\textit{Under 30}} \\
Customer service representatives & 50 & 946 & $-$26 & 34.6 \\
Computer and information systems managers & 43 & 76 & $-$112 & 12.1 \\
Physical scientists, all other & 44 & 96 & $-$53 & 7.2 \\
Human resources workers & 42 & 184 & $-$23 & 6.0 \\
Management analysts & 48 & 171 & $-$23 & 5.5 \\
Secretaries and administrative assistants & 48 & 327 & $-$11 & 4.9 \\
Mechanical engineers & 50 & 119 & $-$26 & 4.4 \\
Miscellaneous managers & 44 & 697 & $-$4 & 3.6 \\
Miscellaneous mathematical science occupations & 53 & 120 & $-$18 & 3.1 \\
Web developers & 56 & 55 & $-$39 & 3.0 \\
\addlinespace
\multicolumn{5}{@{}l}{\textit{Under 30, less than a bachelor's degree}} \\
Customer service representatives & 50 & 770 & $-$25 & 34.8 \\
Secretaries and administrative assistants & 48 & 219 & $-$20 & 8.0 \\
Human resources workers & 42 & 62 & $-$55 & 6.2 \\
Bookkeeping, accounting, and auditing clerks & 42 & 132 & $-$20 & 4.9 \\
Miscellaneous managers & 44 & 310 & $-$8 & 4.7 \\
Software developers, applications and systems software & 57 & 60 & $-$19 & 2.1 \\
Construction managers & 44 & 76 & $-$13 & 1.8 \\
Computer support specialists & 49 & 61 & $-$14 & 1.6 \\
Computer occupations, all other & 54 & 76 & $-$6 & 0.8 \\
\addlinespace
\bottomrule
\end{tabular}
\end{spacing}
\begin{tablenotes}\notesize
\item Notes: Occupations are ranked by their employment-weighted decline in young workers, i.e., the base-period count times the change in the log count between the first half of 2023 and the first half of 2026 (the average of the 2026Q1 vs.\ 2023Q1 and 2026Q2 vs.\ 2023Q2 differences), as a share of the net loss across all occupations in the most exposed quartile of the stage-1 to stage-4 change. Shares are raw changes, not deviations from a mean, and do not use the composition controls. Occupations with fewer than 50 weighted young workers in the base period (monthly average) are not ranked. The employment-weighted mean change in the log count of young workers is $-$9.3 log points in the most exposed quartile and 3.0 in the least exposed for workers under 30, and $-$16.2 and 1.9 for young workers without a bachelor's degree. Counts are CPS weighted counts of employed workers, monthly averages.
\end{tablenotes}
\end{threeparttable}
\end{table}

\section{Robustness in Full}
\label{app:robustness}

This appendix reports the robustness checks of Section~\ref{sec:robustness} of the paper for the other outcomes and designs.
Table~\ref{tab:cell_fd_spec_main_1} reports the specification checks of the three-year difference for the five main outcomes for all workers.
Table~\ref{tab:cell_fd_spec_secondary_1} reports them for the part-time count, the full-time count, and the number of workers under 30, and Figure~\ref{fig:alternatives} of the paper plots its last column.
Table~\ref{tab:cell_fe_spec_main_1} applies the same checks to the occupation fixed-effects design.
Tables~\ref{tab:cell_fd_spec_logo_3} and~\ref{tab:cell_fd_spec_logo_4} report the three-year difference with one major occupation group omitted at a time, for the same three outcomes.
Tables~\ref{tab:cell_fd_telework_1} and~\ref{tab:cell_fd_telework_age_18_29_1} add the remote-work measures to the three-year difference for all workers and for workers under 30, and Table~\ref{tab:cellfe-tw-aggregate-main} adds them to the occupation fixed-effects design.
Tables~\ref{tab:cell_fd_main_14} and~\ref{tab:cell_fd_main_17} report the three-year differences on the data built with the 2018 census codes, for all workers and by age and by education.
The 2018-code file begins in 2020 and hence has no pre-AI placebo window.

\subsection{Specification Checks}
\label{app:logo}

\begin{table}[htbp]
\centering
\caption{Specification Robustness: Primary Outcomes}
\label{tab:cell_fd_spec_main_1}
\noindent{\footnotesize  2026Q1 vs. 2023Q1 and 2026Q2 vs. 2023Q2}\par\vspace{6pt}
\fontsize{9.4}{11.1}\selectfont
\setlength{\tabcolsep}{4pt}
\renewcommand{\arraystretch}{0.92}
\begin{tabularx}{\linewidth}{@{}>{\raggedright\arraybackslash}p{0.31\linewidth}*{5}{>{\centering\arraybackslash}X}@{}}
\toprule
Specification & {\fontsize{9}{10.7}\selectfont\shortstack{Log\\employment}} & {\fontsize{9}{10.7}\selectfont\shortstack{Unemployment\\rate (pp)}} & {\fontsize{9}{10.7}\selectfont\shortstack{Main-job\\hours}} & {\fontsize{9}{10.7}\selectfont\shortstack{Log real hourly\\earnings}} & {\fontsize{9}{10.7}\selectfont\shortstack{Part-time\\share (pp)}} \\
\midrule
\addlinespace[3pt]\multicolumn{6}{@{}l}{\textbf{\textit{A. Reference, weighting, and additional controls}}} \\[2pt]
Baseline & $-1.89$ & $0.16$ & $0.20^{**}$ & $1.94^{**}$ & $-0.74^{***}$ \\
 & $(1.45)$ & $(0.12)$ & $(0.09)$ & $(0.80)$ & $(0.25)$ \\[2pt]
Unweighted & $-1.46$ & $0.56$ & $0.16$ & $0.09$ & $-0.78$ \\
 & $(2.50)$ & $(0.37)$ & $(0.18)$ & $(1.53)$ & $(0.55)$ \\[2pt]
Major group $\times$ quarter effects & $-4.58^{*}$ & $0.23$ & $0.10$ & $1.97$ & $-0.49$ \\
 & $(2.67)$ & $(0.26)$ & $(0.13)$ & $(1.43)$ & $(0.37)$ \\[2pt]
Add base-quarter industry shares & $-1.26$ & $0.16$ & $0.20^{**}$ & $1.33$ & $-0.87^{***}$ \\
 & $(1.65)$ & $(0.15)$ & $(0.09)$ & $(0.92)$ & $(0.33)$ \\[2pt]
Industry shares $\times$ quarter effects & $-1.25$ & $0.17$ & $0.20^{**}$ & $1.33$ & $-0.88^{***}$ \\
 & $(1.66)$ & $(0.15)$ & $(0.10)$ & $(0.92)$ & $(0.33)$ \\[2pt]
Region shares $\times$ quarter effects & $-1.88$ & $0.16$ & $0.20^{**}$ & $1.94^{**}$ & $-0.75^{***}$ \\
 & $(1.45)$ & $(0.12)$ & $(0.09)$ & $(0.80)$ & $(0.25)$ \\[2pt]
\addlinespace[3pt]\multicolumn{6}{@{}l}{\textbf{\textit{B. Sample restrictions}}} \\[2pt]
Drop computer/math occupations & $-1.83$ & $0.13$ & $0.20^{**}$ & $2.19^{***}$ & $-0.79^{***}$ \\
 & $(1.48)$ & $(0.13)$ & $(0.09)$ & $(0.80)$ & $(0.26)$ \\[2pt]
Drop computer/math and science & $-1.77$ & $0.10$ & $0.19^{**}$ & $2.22^{***}$ & $-0.79^{***}$ \\
 & $(1.49)$ & $(0.13)$ & $(0.09)$ & $(0.80)$ & $(0.27)$ \\[2pt]
Trim 1\% exposure tails & $-2.22$ & $0.20$ & $0.19^{**}$ & $2.37^{***}$ & $-0.78^{***}$ \\
 & $(1.39)$ & $(0.13)$ & $(0.09)$ & $(0.77)$ & $(0.26)$ \\[2pt]
At least 30 base-quarter respondents & $-1.69$ & $0.16$ & $0.20^{**}$ & $1.91^{**}$ & $-0.76^{***}$ \\
 & $(1.50)$ & $(0.13)$ & $(0.09)$ & $(0.82)$ & $(0.26)$ \\[2pt]
One-to-one harmonized codes only & $-1.09$ & $0.09$ & $0.08$ & $1.89^{**}$ & $-0.53^{*}$ \\
 & $(1.74)$ & $(0.13)$ & $(0.09)$ & $(0.85)$ & $(0.29)$ \\[2pt]
\midrule
\addlinespace[3pt]\multicolumn{6}{@{}l}{\textbf{\textit{C. Leave-one-major-group-out coefficient ranges}}} \\[2pt]
Minimum & $-2.42$ & $0.12$ & $0.08$ & $1.36$ & $-0.91$ \\
Maximum & $-1.62$ & $0.20$ & $0.28$ & $2.74$ & $-0.40$ \\
\midrule
\addlinespace[3pt]\multicolumn{6}{@{}l}{\textbf{\textit{D. Sample sizes: observations [occupations]}}} \\[2pt]
Baseline sample & 900 [451] & 900 [451] & 900 [451] & 846 [438] & 900 [451] \\
Drop computer/math occupations & 872 [437] & 872 [437] & 872 [437] & 818 [424] & 872 [437] \\
Drop computer/math and science & 838 [420] & 838 [420] & 838 [420] & 785 [407] & 838 [420] \\
Trim 1\% exposure tails & 880 [441] & 880 [441] & 880 [441] & 826 [428] & 880 [441] \\
At least 30 base-quarter respondents & 728 [377] & 728 [377] & 728 [377] & 716 [372] & 728 [377] \\
One-to-one harmonized codes only & 816 [409] & 816 [409] & 816 [409] & 762 [396] & 816 [409] \\
\bottomrule\end{tabularx}
\par\vspace{7pt}\begin{minipage}{\linewidth}\notesize \textit{Notes:} Coefficients per standard deviation of S4--S1, with occupation-clustered standard errors in parentheses. Differences stack the two comparisons shown. All specifications retain the original exposure standardization and base-quarter full-occupation composition controls. The baseline uses base-quarter employment weights and later-quarter effects; unweighted omits weights, and major-group $\times$ quarter effects replace later-quarter effects. Industry/region interactions use base-quarter shares. Log outcomes are in log points; rates/shares in percentage points; hours are unscaled. Panel D reports occupation-quarter observations [distinct occupations]. The 30-person rule uses full-occupation base-quarter CPS respondent counts. Panel C gives coefficient ranges across 22 separate exclusions, not confidence intervals; individual estimates and samples appear in the supplement. \mbox{$^{*}p<0.10$}, \mbox{$^{**}p<0.05$}, \mbox{$^{***}p<0.01$}.\end{minipage}
\end{table}

\begin{table}[htbp]
\centering
\caption{Specification Robustness: Additional Employment Outcomes}
\label{tab:cell_fd_spec_secondary_1}
\noindent{\footnotesize  2026Q1 vs. 2023Q1 and 2026Q2 vs. 2023Q2}\par\vspace{6pt}
\fontsize{9.4}{11.1}\selectfont
\setlength{\tabcolsep}{4pt}
\renewcommand{\arraystretch}{0.92}
\begin{tabularx}{\linewidth}{@{}>{\raggedright\arraybackslash}p{0.31\linewidth}*{3}{>{\centering\arraybackslash}X}@{}}
\toprule
Specification & {\fontsize{9}{10.7}\selectfont\shortstack{Log part-time\\employment}} & {\fontsize{9}{10.7}\selectfont\shortstack{Log full-time\\employment}} & {\fontsize{9}{10.7}\selectfont\shortstack{Under-30 log\\employment}} \\
\midrule
\addlinespace[3pt]\multicolumn{4}{@{}l}{\textbf{\textit{A. Reference, weighting, and additional controls}}} \\[2pt]
Baseline & $-7.34^{***}$ & $-0.87$ & $-8.20^{***}$ \\
 & $(2.51)$ & $(1.43)$ & $(1.92)$ \\[2pt]
Unweighted & $-7.68$ & $0.97$ & $-10.41^{**}$ \\
 & $(4.76)$ & $(2.48)$ & $(4.11)$ \\[2pt]
Major group $\times$ quarter effects & $-9.17^{*}$ & $-3.83$ & $-9.72^{***}$ \\
 & $(4.88)$ & $(2.62)$ & $(3.27)$ \\[2pt]
Add base-quarter industry shares & $-7.77^{**}$ & $0.07$ & $-6.57^{***}$ \\
 & $(3.35)$ & $(1.70)$ & $(2.18)$ \\[2pt]
Industry shares $\times$ quarter effects & $-7.71^{**}$ & $0.09$ & $-6.57^{***}$ \\
 & $(3.36)$ & $(1.71)$ & $(2.19)$ \\[2pt]
Region shares $\times$ quarter effects & $-7.47^{***}$ & $-0.85$ & $-8.07^{***}$ \\
 & $(2.52)$ & $(1.43)$ & $(1.92)$ \\[2pt]
\addlinespace[3pt]\multicolumn{4}{@{}l}{\textbf{\textit{B. Sample restrictions}}} \\[2pt]
Drop computer/math occupations & $-8.37^{***}$ & $-0.76$ & $-8.45^{***}$ \\
 & $(2.60)$ & $(1.46)$ & $(2.00)$ \\[2pt]
Drop computer/math and science & $-8.29^{***}$ & $-0.71$ & $-8.15^{***}$ \\
 & $(2.62)$ & $(1.47)$ & $(2.01)$ \\[2pt]
Trim 1\% exposure tails & $-8.49^{***}$ & $-1.15$ & $-8.55^{***}$ \\
 & $(2.52)$ & $(1.36)$ & $(1.97)$ \\[2pt]
At least 30 base-quarter respondents & $-7.05^{***}$ & $-0.64$ & $-7.91^{***}$ \\
 & $(2.54)$ & $(1.47)$ & $(1.96)$ \\[2pt]
One-to-one harmonized codes only & $-4.75$ & $-0.30$ & $-8.40^{***}$ \\
 & $(2.93)$ & $(1.73)$ & $(2.16)$ \\[2pt]
\midrule
\addlinespace[3pt]\multicolumn{4}{@{}l}{\textbf{\textit{C. Leave-one-major-group-out coefficient ranges}}} \\[2pt]
Minimum & $-8.37$ & $-1.38$ & $-8.96$ \\
Maximum & $-6.41$ & $-0.37$ & $-6.23$ \\
\midrule
\addlinespace[3pt]\multicolumn{4}{@{}l}{\textbf{\textit{D. Sample sizes: observations [occupations]}}} \\[2pt]
Baseline sample & 710 [385] & 899 [451] & 831 [432] \\
Drop computer/math occupations & 687 [372] & 871 [437] & 804 [418] \\
Drop computer/math and science & 666 [358] & 837 [420] & 773 [402] \\
Trim 1\% exposure tails & 696 [377] & 880 [441] & 811 [422] \\
At least 30 base-quarter respondents & 646 [350] & 728 [377] & 708 [372] \\
One-to-one harmonized codes only & 628 [343] & 815 [409] & 750 [391] \\
\bottomrule\end{tabularx}
\par\vspace{7pt}\begin{minipage}{\linewidth}\notesize \textit{Notes:} Coefficients per standard deviation of S4--S1, with occupation-clustered standard errors in parentheses. Differences stack the two comparisons shown. All specifications retain the original exposure standardization and base-quarter full-occupation composition controls. The baseline uses base-quarter employment weights and later-quarter effects; unweighted omits weights, and major-group $\times$ quarter effects replace later-quarter effects. Industry/region interactions use base-quarter shares. Log outcomes are in log points; rates/shares in percentage points; hours are unscaled. Panel D reports occupation-quarter observations [distinct occupations]. The 30-person rule uses full-occupation base-quarter CPS respondent counts. Panel C gives coefficient ranges across 22 separate exclusions, not confidence intervals; individual estimates and samples appear in the supplement. Under-30 employment uses ages 18--29 and full-occupation composition controls. \mbox{$^{*}p<0.10$}, \mbox{$^{**}p<0.05$}, \mbox{$^{***}p<0.01$}.\end{minipage}
\end{table}

\begin{table}[htbp]
\centering
\caption{Fixed-Effects Specification Robustness: Primary Outcomes}
\label{tab:cell_fe_spec_main_1}
\noindent{\footnotesize  2025Q1 SHIFT in the standardized step-exposure slope | 2023Q1--2026Q2}\par\vspace{6pt}
\fontsize{9.4}{11.1}\selectfont\setlength{\tabcolsep}{4pt}\renewcommand{\arraystretch}{0.92}
\begin{tabularx}{\linewidth}{@{}>{\raggedright\arraybackslash}p{.31\linewidth}*{5}{>{\centering\arraybackslash}X}@{}}
\toprule
Specification & {\fontsize{9}{10.7}\selectfont\shortstack{Log\\employment}} & {\fontsize{9}{10.7}\selectfont\shortstack{Unemployment\\rate (pp)}} & {\fontsize{9}{10.7}\selectfont\shortstack{Main-job\\hours}} & {\fontsize{9}{10.7}\selectfont\shortstack{Log real hourly\\earnings}} & {\fontsize{9}{10.7}\selectfont\shortstack{Part-time\\share (pp)}} \\
\midrule
\addlinespace[3pt]\multicolumn{6}{@{}l}{\textbf{\textit{A. Reference, weighting, and additional controls}}} \\[2pt]
Approved FE baseline & $-0.06$ & $-0.10$ & $0.14^{***}$ & $0.27$ & $-0.37^{*}$ \\
 & $(0.96)$ & $(0.11)$ & $(0.05)$ & $(0.46)$ & $(0.19)$ \\[2pt]
Unweighted & $2.41$ & $-0.09$ & $0.16$ & $0.72$ & $-0.55^{*}$ \\
 & $(1.92)$ & $(0.21)$ & $(0.10)$ & $(1.05)$ & $(0.33)$ \\[2pt]
Major group $\times$ quarter effects & $-4.28^{*}$ & $-0.12$ & $0.05$ & $0.81$ & $-0.32$ \\
 & $(2.18)$ & $(0.16)$ & $(0.09)$ & $(0.89)$ & $(0.27)$ \\[2pt]
2019 industry shares $\times$ quarter & $-0.62$ & $0.01$ & $0.11^{*}$ & $0.96^{*}$ & $-0.44^{**}$ \\
 & $(1.17)$ & $(0.11)$ & $(0.06)$ & $(0.54)$ & $(0.19)$ \\[2pt]
2019 region shares $\times$ quarter & $-0.01$ & $-0.08$ & $0.14^{***}$ & $0.25$ & $-0.36^{*}$ \\
 & $(0.97)$ & $(0.10)$ & $(0.05)$ & $(0.46)$ & $(0.20)$ \\[2pt]
\addlinespace[3pt]\multicolumn{6}{@{}l}{\textbf{\textit{B. Sample restrictions}}} \\[2pt]
Drop computer/math occupations & $-0.09$ & $-0.10$ & $0.14^{***}$ & $0.25$ & $-0.36^{*}$ \\
 & $(0.98)$ & $(0.11)$ & $(0.05)$ & $(0.49)$ & $(0.20)$ \\[2pt]
Drop computer/math and science & $-0.05$ & $-0.11$ & $0.14^{***}$ & $0.25$ & $-0.35^{*}$ \\
 & $(0.98)$ & $(0.11)$ & $(0.05)$ & $(0.49)$ & $(0.20)$ \\[2pt]
Trim 1\% exposure tails & $-0.30$ & $-0.07$ & $0.15^{***}$ & $0.21$ & $-0.36^{*}$ \\
 & $(0.97)$ & $(0.11)$ & $(0.05)$ & $(0.48)$ & $(0.20)$ \\[2pt]
Occupation-quarter count $\geq30$ & $-0.36$ & $-0.10$ & $0.15^{***}$ & $0.30$ & $-0.35^{*}$ \\
 & $(0.93)$ & $(0.11)$ & $(0.05)$ & $(0.47)$ & $(0.20)$ \\[2pt]
One-to-one harmonized codes only & $0.51$ & $-0.06$ & $0.14^{**}$ & $0.65$ & $-0.35$ \\
 & $(0.84)$ & $(0.10)$ & $(0.06)$ & $(0.50)$ & $(0.22)$ \\[2pt]
\midrule
\addlinespace[3pt]\multicolumn{6}{@{}l}{\textbf{\textit{C. Leave-one-major-group-out coefficient ranges}}} \\[2pt]
Minimum & $-0.75$ & $-0.21$ & $0.08$ & $-0.57$ & $-0.46$ \\
Maximum & $0.43$ & $-0.04$ & $0.17$ & $0.54$ & $-0.10$ \\
\midrule
\addlinespace[3pt]\multicolumn{6}{@{}l}{\textbf{\textit{D. Sample sizes: observations [occupation clusters]}}} \\[2pt]
Baseline sample & 6,295 [450] & 6,295 [450] & 6,295 [450] & 6,108 [450] & 6,295 [450] \\
Drop computer/math occupations & 6,099 [436] & 6,099 [436] & 6,099 [436] & 5,912 [436] & 6,099 [436] \\
Drop computer/math and science & 5,861 [419] & 5,861 [419] & 5,861 [419] & 5,681 [419] & 5,861 [419] \\
Trim 1\% exposure tails & 6,155 [440] & 6,155 [440] & 6,155 [440] & 5,974 [440] & 6,155 [440] \\
Occupation-quarter count $\geq30$ & 4,932 [383] & 4,932 [383] & 4,932 [383] & 4,915 [383] & 4,932 [383] \\
One-to-one harmonized codes only & 5,707 [408] & 5,707 [408] & 5,707 [408] & 5,525 [408] & 5,707 [408] \\
\bottomrule\end{tabularx}
\par\vspace{7pt}\begin{minipage}{\linewidth}\notesize \textit{Notes:} Coefficients are changes at 2025Q1 in the slope on standardized step exposure, with occupation-clustered SEs in parentheses. They are not post-2025Q1 slope levels. Log coefficients/SEs are multiplied by 100 (log points); rates/shares use percentage points; hours are unscaled. Exposure normalization is fixed at the approved W2023 baseline. Baseline models absorb occupation and quarter effects and use fixed 2019 employment analytical weights; unweighted omits weights. Major-group effects are added to the baseline effects. Industry/region shares are fixed in 2019 and interacted with quarter. The first quarter (2023Q1) is the reference for these added slopes. Panel D reports model observations [occupation clusters]; Residual inference degrees of freedom equal clusters minus one. Trimming uses strict-interior 1st/99th percentiles of raw S4--S1 over 450 baseline occupations (440 retained). The 30-record rule uses contemporaneous aggregate occupation-quarter person-month records, including in age models; it is not a base-quarter rule or a count of distinct persons. Panel C gives coefficient extrema across 22 exclusions, not confidence intervals. \mbox{$^{*}p<0.10$}, \mbox{$^{**}p<0.05$}, \mbox{$^{***}p<0.01$}.\end{minipage}
\end{table}

\providecommand{\result}[3]{\begin{tabular}[t]{@{}c@{}}{\fontsize{9.4}{11.8}\selectfont #1}\\{\fontsize{9.1}{10.8}\selectfont #2}\\{\fontsize{8.5}{10.2}\selectfont #3}\end{tabular}}
\begin{table}[htbp]
\centering
\caption{Major-Group Exclusions: Additional Employment Outcomes (Groups 1--11)}
\label{tab:cell_fd_spec_logo_3}
\noindent{\footnotesize  2026Q1 vs. 2023Q1 and 2026Q2 vs. 2023Q2}\par\vspace{6pt}
\fontsize{9.4}{11.8}\selectfont
\setlength{\tabcolsep}{4pt}
\renewcommand{\arraystretch}{0.92}
\begin{tabularx}{\linewidth}{@{}>{\raggedright\arraybackslash}p{0.34\linewidth}*{3}{>{\centering\arraybackslash}X}@{}}
\toprule
Omitted major group & {\fontsize{8.6}{10.7}\selectfont\shortstack{Log part-time\\employment}} & {\fontsize{8.6}{10.7}\selectfont\shortstack{Log full-time\\employment}} & {\fontsize{8.6}{10.7}\selectfont\shortstack{Under-30 log\\employment}} \\
\midrule
\textbf{1.} Management\newline {\fontsize{8.2}{9.7}\selectfont OCC2010 0010--0430} & \result{$-7.46^{***}$}{$(2.54)$}{669 [363]} & \result{$-1.26$}{$(1.46)$}{847 [425]} & \result{$-8.68^{***}$}{$(1.93)$}{785 [408]} \\
\addlinespace[1pt]\midrule[0.25pt]\addlinespace[2pt]
\textbf{2.} Business and financial operations\newline {\fontsize{8.2}{9.7}\selectfont OCC2010 0500--0950} & \result{$-7.41^{***}$}{$(2.51)$}{663 [360]} & \result{$-1.22$}{$(1.47)$}{843 [423]} & \result{$-8.24^{***}$}{$(2.01)$}{777 [404]} \\
\addlinespace[1pt]\midrule[0.25pt]\addlinespace[2pt]
\textbf{3.} Computer and mathematical\newline {\fontsize{8.2}{9.7}\selectfont OCC2010 1000--1240} & \result{$-8.37^{***}$}{$(2.60)$}{687 [372]} & \result{$-0.76$}{$(1.46)$}{871 [437]} & \result{$-8.45^{***}$}{$(2.00)$}{804 [418]} \\
\addlinespace[1pt]\midrule[0.25pt]\addlinespace[2pt]
\textbf{4.} Architecture and engineering\newline {\fontsize{8.2}{9.7}\selectfont OCC2010 1300--1560} & \result{$-6.41^{***}$}{$(2.46)$}{687 [372]} & \result{$-0.89$}{$(1.45)$}{863 [433]} & \result{$-8.05^{***}$}{$(1.97)$}{796 [414]} \\
\addlinespace[1pt]\midrule[0.25pt]\addlinespace[2pt]
\textbf{5.} Life, physical, and social science\newline {\fontsize{8.2}{9.7}\selectfont OCC2010 1600--1980} & \result{$-7.30^{***}$}{$(2.51)$}{689 [371]} & \result{$-0.83$}{$(1.44)$}{865 [434]} & \result{$-8.09^{***}$}{$(1.93)$}{800 [416]} \\
\addlinespace[1pt]\midrule[0.25pt]\addlinespace[2pt]
\textbf{6.} Community and social service\newline {\fontsize{8.2}{9.7}\selectfont OCC2010 2000--2060} & \result{$-7.14^{***}$}{$(2.53)$}{697 [378]} & \result{$-0.88$}{$(1.45)$}{885 [444]} & \result{$-8.33^{***}$}{$(1.93)$}{817 [425]} \\
\addlinespace[1pt]\midrule[0.25pt]\addlinespace[2pt]
\textbf{7.} Legal\newline {\fontsize{8.2}{9.7}\selectfont OCC2010 2100--2160} & \result{$-7.17^{***}$}{$(2.54)$}{703 [381]} & \result{$-0.82$}{$(1.46)$}{891 [447]} & \result{$-8.51^{***}$}{$(1.90)$}{823 [428]} \\
\addlinespace[1pt]\midrule[0.25pt]\addlinespace[2pt]
\textbf{8.} Education, training, and library\newline {\fontsize{8.2}{9.7}\selectfont OCC2010 2200--2550} & \result{$-7.77^{***}$}{$(2.53)$}{688 [374]} & \result{$-1.21$}{$(1.46)$}{877 [440]} & \result{$-8.30^{***}$}{$(1.97)$}{810 [421]} \\
\addlinespace[1pt]\midrule[0.25pt]\addlinespace[2pt]
\textbf{9.} Arts, design, entertainment,\newline sports, and media\newline {\fontsize{8.2}{9.7}\selectfont OCC2010 2600--2960} & \result{$-6.94^{***}$}{$(2.57)$}{676 [367]} & \result{$-0.91$}{$(1.45)$}{863 [433]} & \result{$-7.44^{***}$}{$(1.93)$}{795 [414]} \\
\addlinespace[1pt]\midrule[0.25pt]\addlinespace[2pt]
\textbf{10.} Healthcare practitioners\newline and technical\newline {\fontsize{8.2}{9.7}\selectfont OCC2010 3000--3540} & \result{$-8.09^{***}$}{$(3.02)$}{649 [354]} & \result{$-0.37$}{$(1.69)$}{837 [420]} & \result{$-8.42^{***}$}{$(2.12)$}{778 [405]} \\
\addlinespace[1pt]\midrule[0.25pt]\addlinespace[2pt]
\textbf{11.} Healthcare support\newline {\fontsize{8.2}{9.7}\selectfont OCC2010 3600--3655} & \result{$-6.89^{***}$}{$(2.53)$}{688 [374]} & \result{$-0.69$}{$(1.44)$}{877 [440]} & \result{$-7.94^{***}$}{$(1.96)$}{809 [421]} \\
\bottomrule\end{tabularx}
\par\vspace{7pt}\begin{minipage}{\linewidth}\notesize \textit{Notes:} Each block reports the S4--S1 coefficient, its occupation-clustered standard error in parentheses, and observations [occupations] on the third line. Occupational labels align with coefficients; thin rules separate omitted groups. $N$ counts occupation-quarter observations; brackets count distinct occupations. Each regression omits the indicated major group, defined by the listed OCC2010 code range, and retains the baseline controls, base-quarter occupation employment weights, later-quarter effects, and original exposure standardization. Log outcomes use log points, rates/shares percentage points, and hours are unscaled. The under-30 sample uses ages 18--29 and full-occupation controls. All 22 exclusions are reported for each outcome across this supplement. \mbox{$^{*}p<0.10$}, \mbox{$^{**}p<0.05$}, \mbox{$^{***}p<0.01$}.\end{minipage}
\end{table}

\providecommand{\result}[3]{\begin{tabular}[t]{@{}c@{}}{\fontsize{9.4}{11.8}\selectfont #1}\\{\fontsize{9.1}{10.8}\selectfont #2}\\{\fontsize{8.5}{10.2}\selectfont #3}\end{tabular}}
\begin{table}[htbp]
\centering
\caption{Major-Group Exclusions: Additional Employment Outcomes (Groups 12--22)}
\label{tab:cell_fd_spec_logo_4}
\noindent{\footnotesize  2026Q1 vs. 2023Q1 and 2026Q2 vs. 2023Q2}\par\vspace{6pt}
\fontsize{9.4}{11.8}\selectfont
\setlength{\tabcolsep}{4pt}
\renewcommand{\arraystretch}{0.92}
\begin{tabularx}{\linewidth}{@{}>{\raggedright\arraybackslash}p{0.34\linewidth}*{3}{>{\centering\arraybackslash}X}@{}}
\toprule
Omitted major group & {\fontsize{8.6}{10.7}\selectfont\shortstack{Log part-time\\employment}} & {\fontsize{8.6}{10.7}\selectfont\shortstack{Log full-time\\employment}} & {\fontsize{8.6}{10.7}\selectfont\shortstack{Under-30 log\\employment}} \\
\midrule
\textbf{12.} Protective service\newline {\fontsize{8.2}{9.7}\selectfont OCC2010 3700--3955} & \result{$-6.42^{**}$}{$(2.51)$}{690 [373]} & \result{$-0.58$}{$(1.50)$}{867 [435]} & \result{$-7.80^{***}$}{$(1.96)$}{807 [418]} \\
\addlinespace[1pt]\midrule[0.25pt]\addlinespace[2pt]
\textbf{13.} Food preparation and serving\newline {\fontsize{8.2}{9.7}\selectfont OCC2010 4000--4160} & \result{$-7.50^{***}$}{$(2.54)$}{688 [374]} & \result{$-1.12$}{$(1.45)$}{877 [440]} & \result{$-7.79^{***}$}{$(1.95)$}{809 [421]} \\
\addlinespace[1pt]\midrule[0.25pt]\addlinespace[2pt]
\textbf{14.} Building/grounds cleaning\newline and maintenance\newline {\fontsize{8.2}{9.7}\selectfont OCC2010 4200--4250} & \result{$-7.64^{***}$}{$(2.53)$}{698 [379]} & \result{$-0.53$}{$(1.48)$}{887 [445]} & \result{$-8.66^{***}$}{$(1.87)$}{819 [426]} \\
\addlinespace[1pt]\midrule[0.25pt]\addlinespace[2pt]
\textbf{15.} Personal care and service\newline {\fontsize{8.2}{9.7}\selectfont OCC2010 4300--4650} & \result{$-7.80^{***}$}{$(2.71)$}{677 [368]} & \result{$-1.38$}{$(1.48)$}{865 [434]} & \result{$-8.96^{***}$}{$(2.04)$}{798 [415]} \\
\addlinespace[1pt]\midrule[0.25pt]\addlinespace[2pt]
\textbf{16.} Sales and related\newline {\fontsize{8.2}{9.7}\selectfont OCC2010 4700--4965} & \result{$-7.66^{***}$}{$(2.56)$}{678 [369]} & \result{$-0.96$}{$(1.46)$}{865 [434]} & \result{$-8.87^{***}$}{$(1.92)$}{797 [415]} \\
\addlinespace[1pt]\midrule[0.25pt]\addlinespace[2pt]
\textbf{17.} Office and administrative support\newline {\fontsize{8.2}{9.7}\selectfont OCC2010 5000--5940} & \result{$-6.79^{*}$}{$(3.70)$}{634 [345]} & \result{$-1.16$}{$(1.39)$}{808 [405]} & \result{$-6.23^{**}$}{$(2.47)$}{746 [387]} \\
\addlinespace[1pt]\midrule[0.25pt]\addlinespace[2pt]
\textbf{18.} Farming, fishing, and forestry\newline {\fontsize{8.2}{9.7}\selectfont OCC2010 6005--6130} & \result{$-7.31^{***}$}{$(2.52)$}{698 [379]} & \result{$-0.84$}{$(1.44)$}{885 [444]} & \result{$-8.22^{***}$}{$(1.93)$}{819 [426]} \\
\addlinespace[1pt]\midrule[0.25pt]\addlinespace[2pt]
\textbf{19.} Construction and extraction\newline {\fontsize{8.2}{9.7}\selectfont OCC2010 6200--6940} & \result{$-7.86^{***}$}{$(2.54)$}{668 [361]} & \result{$-1.13$}{$(1.42)$}{837 [420]} & \result{$-8.35^{***}$}{$(1.93)$}{772 [402]} \\
\addlinespace[1pt]\midrule[0.25pt]\addlinespace[2pt]
\textbf{20.} Installation, maintenance,\newline and repair\newline {\fontsize{8.2}{9.7}\selectfont OCC2010 7000--7630} & \result{$-7.60^{***}$}{$(2.50)$}{671 [361]} & \result{$-0.63$}{$(1.47)$}{837 [420]} & \result{$-7.93^{***}$}{$(1.98)$}{772 [402]} \\
\addlinespace[1pt]\midrule[0.25pt]\addlinespace[2pt]
\textbf{21.} Production\newline {\fontsize{8.2}{9.7}\selectfont OCC2010 7700--8965} & \result{$-7.12^{***}$}{$(2.57)$}{640 [345]} & \result{$-0.59$}{$(1.48)$}{787 [394]} & \result{$-8.17^{***}$}{$(1.94)$}{738 [381]} \\
\addlinespace[1pt]\midrule[0.25pt]\addlinespace[2pt]
\textbf{22.} Transportation and material moving\newline {\fontsize{8.2}{9.7}\selectfont OCC2010 9000--9750} & \result{$-7.50^{***}$}{$(2.62)$}{672 [365]} & \result{$-0.83$}{$(1.50)$}{845 [424]} & \result{$-7.67^{***}$}{$(2.01)$}{780 [405]} \\
\bottomrule\end{tabularx}
\par\vspace{7pt}\begin{minipage}{\linewidth}\notesize \textit{Notes:} Each block reports the S4--S1 coefficient, its occupation-clustered standard error in parentheses, and observations [occupations] on the third line. Occupational labels align with coefficients; thin rules separate omitted groups. $N$ counts occupation-quarter observations; brackets count distinct occupations. Each regression omits the indicated major group, defined by the listed OCC2010 code range, and retains the baseline controls, base-quarter occupation employment weights, later-quarter effects, and original exposure standardization. Log outcomes use log points, rates/shares percentage points, and hours are unscaled. The under-30 sample uses ages 18--29 and full-occupation controls. All 22 exclusions are reported for each outcome across this supplement. \mbox{$^{*}p<0.10$}, \mbox{$^{**}p<0.05$}, \mbox{$^{***}p<0.01$}.\end{minipage}
\end{table}

\clearpage
\subsection{Remote Work}
\label{app:remote}

\begin{table}[htbp]
\centering
\caption{Three-Year First Differences: Remote-Work Robustness}
\label{tab:cell_fd_telework_1}
\fontsize{7.9}{9.45}\selectfont
\setlength{\tabcolsep}{2.0pt}
\renewcommand{\arraystretch}{1.14}
\begin{adjustbox}{max width=\linewidth}
\begin{tabularx}{\linewidth}{@{}>{\raggedright\arraybackslash}p{0.315\linewidth}*{5}{>{\centering\arraybackslash}X}@{}}
\toprule
 & \shortstack{Log\\employment} & \shortstack{Unemployment\\(pp)} & Hours & \shortstack{Log real hourly\\earnings} & \shortstack{Part-time share\\(pp)} \\
\midrule
\addlinespace[0.16em]
\multicolumn{6}{@{}l}{\textbf{\textit{2026Q1 vs. 2023Q1 and 2026Q2 vs. 2023Q2}}} \\[0.14em]
\addlinespace[0.05em]
\multicolumn{6}{@{}l}{\textbf{\textit{A. CPS remote-work controls}}} \\
Baseline & -1.89 (1.45) & 0.16 (0.12) & 0.20\textsuperscript{**} (0.09) & 1.94\textsuperscript{**} (0.80) & -0.74\textsuperscript{***} (0.25) \\
\addlinespace[0.10em]
Add base-quarter telework share & -2.61 (1.95) & 0.10 (0.16) & 0.31\textsuperscript{***} (0.10) & 2.02\textsuperscript{*} (1.04) & -1.02\textsuperscript{***} (0.35) \\
\addlinespace[0.10em]
Add change in telework share & -1.92 (1.44) & 0.16 (0.12) & 0.21\textsuperscript{**} (0.09) & 1.75\textsuperscript{**} (0.81) & -0.77\textsuperscript{***} (0.25) \\
\addlinespace[0.10em]
Add base share and change & -2.66 (1.96) & 0.09 (0.16) & 0.33\textsuperscript{***} (0.10) & 1.68 (1.05) & -1.08\textsuperscript{***} (0.35) \\
\addlinespace[0.10em]
Add hours share: base and change & -2.46 (1.92) & 0.13 (0.16) & 0.30\textsuperscript{***} (0.10) & 1.93\textsuperscript{*} (1.05) & -0.99\textsuperscript{***} (0.35) \\
\addlinespace[0.10em]
\specialrule{0.075em}{0.20em}{0.16em}
\multicolumn{6}{@{}l}{\textbf{\textit{B. Dingel--Neiman teleworkability}}} \\
Baseline, Dingel--Neiman sample & -2.00 (1.44) & 0.16 (0.12) & 0.19\textsuperscript{**} (0.09) & 1.94\textsuperscript{**} (0.80) & -0.72\textsuperscript{***} (0.26) \\
\addlinespace[0.10em]
Add Dingel--Neiman teleworkability & -1.28 (1.66) & 0.10 (0.17) & 0.23\textsuperscript{**} (0.10) & 1.79\textsuperscript{*} (0.96) & -0.64\textsuperscript{*} (0.36) \\
\addlinespace[0.10em]
\cmidrule(lr){1-6}
Add all three remote-work controls & -2.14 (1.89) & 0.06 (0.18) & 0.34\textsuperscript{***} (0.11) & 1.56 (1.10) & -0.99\textsuperscript{**} (0.39) \\
\addlinespace[0.10em]
\specialrule{0.075em}{0.20em}{0.16em}
\multicolumn{6}{@{}l}{\textbf{\textit{C. Remote-work coefficients from augmented regressions}}} \\
Base-quarter telework share & 1.32 (2.01) & 0.11 (0.18) & -0.22\textsuperscript{**} (0.10) & 0.11 (1.17) & 0.55\textsuperscript{*} (0.31) \\
Change in telework share & 0.57 (2.24) & 0.10 (0.19) & -0.29\textsuperscript{***} (0.10) & 3.78\textsuperscript{***} (1.28) & 0.65\textsuperscript{**} (0.30) \\
Base-quarter remote-hours share & 0.95 (2.01) & 0.06 (0.17) & -0.18\textsuperscript{*} (0.09) & 0.09 (1.20) & 0.45 (0.32) \\
Change in remote-hours share & -0.40 (2.19) & 0.09 (0.21) & -0.29\textsuperscript{***} (0.11) & 2.38\textsuperscript{*} (1.34) & 0.64\textsuperscript{**} (0.32) \\
Dingel--Neiman teleworkability & -1.27 (1.43) & 0.10 (0.16) & -0.06 (0.09) & 0.26 (0.82) & -0.14 (0.31) \\
\specialrule{0.075em}{0.20em}{0.14em}
\multicolumn{6}{@{}l}{\textbf{\textit{D. Sample statistics}}} \\
\multicolumn{6}{@{}l}{\textit{Baseline / CPS specifications}} \\
Mean outcome change & -1.21 & 0.60 & -0.23 & 1.61 & 0.67 \\
$N$ (occupation-quarter cells) & 900 & 900 & 900 & 846 & 900 \\
Occupations & 451 & 451 & 451 & 438 & 451 \\
\cmidrule(lr){1-6}
\multicolumn{6}{@{}l}{\textit{Dingel--Neiman-available specifications}} \\
Mean outcome change & -1.30 & 0.60 & -0.24 & 1.60 & 0.69 \\
$N$ (occupation-quarter cells) & 894 & 894 & 894 & 840 & 894 \\
Occupations & 448 & 448 & 448 & 435 & 448 \\
\bottomrule
\end{tabularx}\end{adjustbox}
\par\vspace{0.45em}\noindent
\begin{minipage}{\linewidth}
\notesize \textit{Notes:} Log outcomes are in log points, unemployment and part-time shares are in percentage points, and hours are in hours. Each regression relates the three-year difference in the outcome to the one-standard-deviation occupation-level S4--S1 AI-exposure change. Specifications use base-quarter occupation employment weights, twelve base-quarter occupation-composition controls, later-quarter fixed effects, and standard errors clustered by OCC2010. Remote-work controls are standardized within their regression samples. CPS measures are the base-quarter telework share and its change, or corresponding shares of hours worked remotely. Dingel--Neiman is the occupation-level work-from-home feasibility index. Section C reports remote-work controls' own coefficients from the CPS base-share-and-change regression, the remote-hours base-and-change regression, and the Dingel--Neiman-only regression, respectively; these are not AI-exposure coefficients. The baseline restricted to available CPS telework data is identical to the unrestricted baseline and is omitted. Section D reports base-employment-weighted mean \emph{outcome changes}, not outcome levels, and occupation-quarter observation counts; counts for specifications including Dingel--Neiman coincide with the Dingel--Neiman-available baseline. The Dingel--Neiman specification can lose occupations for which no index exists. Significance: \mbox{\textsuperscript{*}\,$p<0.10$}, \mbox{\textsuperscript{**}\,$p<0.05$}, \mbox{\textsuperscript{***}\,$p<0.01$}.
\end{minipage}
\end{table}

\begin{table}[htbp]\centering
\caption{Three-Year First Differences: Remote-Work Robustness, Workers under 30}
\label{tab:cell_fd_telework_age_18_29_1}
\fontsize{9}{11}\selectfont
\setlength{\tabcolsep}{2pt}\renewcommand{\arraystretch}{1.15}
\begin{tabularx}{\linewidth}{@{}>{\raggedright\arraybackslash}p{0.32\linewidth}*{5}{>{\centering\arraybackslash}X}@{}}
\toprule
 & \shortstack{Log\\employment} & \shortstack{Unemployment\\(pp)} & Hours & \shortstack{Log real hourly\\earnings} & \shortstack{Part-time share\\(pp)} \\
\midrule
\multicolumn{6}{@{}l}{\textbf{\textit{2026Q1 vs. 2023Q1 and 2026Q2 vs. 2023Q2}}} \\
\multicolumn{6}{@{}l}{\textit{CPS remote-work controls}} \\
Baseline & -8.20\textsuperscript{***} (1.92) & -0.04 (0.30) & 0.18 (0.22) & 0.61 (1.76) & 0.13 (0.69) \\
Add base-quarter telework share & -5.67\textsuperscript{**} (2.69) & -0.13 (0.43) & 0.43\textsuperscript{*} (0.25) & -0.17 (2.27) & -0.65 (0.96) \\
Add change in telework share & -7.60\textsuperscript{***} (1.99) & -0.09 (0.30) & 0.20 (0.22) & 0.25 (1.70) & 0.12 (0.70) \\
Add base share and change & -4.70\textsuperscript{*} (2.79) & -0.20 (0.42) & 0.45\textsuperscript{*} (0.26) & -0.61 (2.23) & -0.67 (0.97) \\
Add hours share: base and change & -4.42 (2.71) & -0.18 (0.42) & 0.44\textsuperscript{*} (0.24) & -0.21 (2.24) & -0.53 (0.97) \\
\cmidrule(lr){1-6}
\multicolumn{6}{@{}l}{\textit{Dingel--Neiman teleworkability}} \\
Baseline, Dingel--Neiman sample & -8.22\textsuperscript{***} (1.93) & -0.05 (0.30) & 0.19 (0.22) & 0.60 (1.76) & 0.13 (0.69) \\
Add Dingel--Neiman teleworkability & -8.42\textsuperscript{***} (2.23) & -0.11 (0.35) & 0.15 (0.24) & 0.64 (1.92) & 0.29 (0.85) \\
\cmidrule(lr){1-6}
Add all three remote-work controls & -5.09\textsuperscript{*} (2.75) & -0.23 (0.43) & 0.41 (0.26) & -0.50 (2.29) & -0.50 (1.01) \\
\addlinespace[0.10em]
\cmidrule(lr){1-6}
\multicolumn{6}{@{}l}{\textit{Own coefficients on remote-work controls}} \\
Base-quarter telework share & -5.81 (3.62) & 0.24 (0.49) & -0.52\textsuperscript{**} (0.25) & 1.72 (2.46) & 1.59\textsuperscript{*} (0.95) \\
Change in telework share & -10.87\textsuperscript{***} (3.57) & 0.73 (0.56) & -0.27 (0.24) & 6.94\textsuperscript{**} (2.97) & 0.17 (0.90) \\
Base-quarter remote-hours share & -7.35\textsuperscript{**} (3.45) & 0.26 (0.44) & -0.49\textsuperscript{**} (0.23) & 1.70 (2.41) & 1.24 (0.93) \\
Change in remote-hours share & -11.03\textsuperscript{***} (3.86) & 0.58 (0.57) & -0.13 (0.28) & 5.24 (3.23) & -0.15 (1.05) \\
Dingel--Neiman teleworkability & 0.41 (2.03) & 0.12 (0.30) & 0.07 (0.21) & -0.08 (1.70) & -0.33 (0.79) \\
\bottomrule\end{tabularx}
\par\vspace{0.6em}\noindent\begin{minipage}{\linewidth}\notesize \textit{Notes:} This table uses separate regressions on occupation-by-age-by-quarter outcome cells for ages 18--29. CPS remote-work shares and changes are full-occupation measures, identical across age groups within an occupation and comparison period; the dependent variables and base-quarter employment weights remain age-specific. The twelve base-occupation composition covariates, later-quarter fixed effects, and OCC2010-clustered standard errors are retained. Coefficients in the first eight rows are for a one-standard-deviation change in occupation-level S4--S1 AI exposure. Own coefficients on remote-work controls are from the jointly entered CPS base-share-and-change specification, the jointly entered remote-hours base-and-change specification, or the Dingel--Neiman-only specification, respectively; they are not coefficients on AI exposure. Telework controls use the full-occupation reference scaling for CPS measures; Dingel--Neiman is standardized in the regression sample. A Dingel--Neiman-matched baseline is shown because some occupations lack the index. Sample counts and the CPS-available baseline are omitted from this age table. Log outcomes and shares are multiplied by 100; hours remain unscaled. Significance: \mbox{\textsuperscript{*}\,$p<0.10$}, \mbox{\textsuperscript{**}\,$p<0.05$}, \mbox{\textsuperscript{***}\,$p<0.01$}.
\end{minipage}\end{table}

\begin{table}[htbp]
\centering
\caption{Occupation-Level Fixed Effects: Remote-Work Robustness}\label{tab:cellfe-tw-aggregate-main}
\begingroup\fontsize{8.30}{9.4}\selectfont\setlength{\tabcolsep}{1.8pt}\renewcommand{\arraystretch}{1.17}
\begin{tabularx}{\linewidth}{@{}>{\raggedright\arraybackslash}p{0.340\linewidth}*{5}{>{\centering\arraybackslash}X}@{}}
\toprule
 & \shortstack{Log\\employment} & \shortstack{Unemployment\\(pp)} & Hours & \shortstack{Log real hourly\\earnings} & \shortstack{Part-time share\\(pp)} \\
\midrule
\multicolumn{6}{@{}l}{\textit{2023Q1--2026Q2}} \\
\addlinespace[0.10em]
\multicolumn{6}{@{}l}{\textbf{A. CPS remote-work controls}} \\
Baseline\,\textit{(pooled)} & -0.60 (1.35) & 0.09 (0.13) & 0.01 (0.08) & -0.12 (0.73) & -0.78\textsuperscript{***} (0.28) \\
\hspace*{1em}\textit{Shift at 2025Q1} & -0.06 (0.96) & -0.10 (0.11) & 0.14\textsuperscript{***} (0.05) & 0.27 (0.46) & -0.37\textsuperscript{*} (0.19) \\
\addlinespace[0.06em]
Add quarterly telework share\,\textit{(pooled)} & -0.63 (1.37) & 0.09 (0.13) & 0.01 (0.08) & -0.18 (0.73) & -0.78\textsuperscript{***} (0.28) \\
\hspace*{1em}\textit{Shift at 2025Q1} & -0.05 (0.96) & -0.10 (0.11) & 0.14\textsuperscript{***} (0.05) & 0.29 (0.46) & -0.37\textsuperscript{*} (0.19) \\
\addlinespace[0.06em]
Add remote-hours share\,\textit{(pooled)} & -0.45 (1.36) & 0.08 (0.13) & 0.00 (0.08) & 0.22 (0.73) & -0.77\textsuperscript{***} (0.28) \\
\hspace*{1em}\textit{Shift at 2025Q1} & -0.07 (0.96) & -0.10 (0.11) & 0.14\textsuperscript{***} (0.05) & 0.23 (0.45) & -0.37\textsuperscript{*} (0.19) \\
\addlinespace[0.25em]\midrule
\multicolumn{6}{@{}l}{\textbf{B. Dingel--Neiman teleworkability}} \\
Baseline, Dingel--Neiman sample\,\textit{(pooled)} & -0.67 (1.36) & 0.09 (0.13) & 0.00 (0.08) & -0.10 (0.73) & -0.77\textsuperscript{***} (0.28) \\
\hspace*{1em}\textit{Shift at 2025Q1} & -0.09 (0.97) & -0.10 (0.11) & 0.14\textsuperscript{***} (0.05) & 0.30 (0.46) & -0.37\textsuperscript{*} (0.19) \\
\addlinespace[0.06em]
Dingel--Neiman $\times$ 2025Q1 onward\,\textit{(pooled)} & -0.36 (1.43) & 0.20 (0.17) & -0.04 (0.10) & 1.20 (0.90) & -0.59\textsuperscript{*} (0.35) \\
\hspace*{1em}\textit{Shift at 2025Q1} & 0.01 (0.91) & -0.08 (0.11) & 0.14\textsuperscript{**} (0.06) & 0.77 (0.51) & -0.35 (0.23) \\
\addlinespace[0.06em]
Add both remote-work controls\,\textit{(pooled)} & -0.40 (1.44) & 0.21 (0.17) & -0.04 (0.10) & 1.12 (0.89) & -0.59\textsuperscript{*} (0.35) \\
\hspace*{1em}\textit{Shift at 2025Q1} & 0.01 (0.91) & -0.08 (0.11) & 0.14\textsuperscript{**} (0.06) & 0.78 (0.50) & -0.35 (0.23) \\
\addlinespace[0.25em]\midrule
\multicolumn{6}{@{}l}{\textbf{C. Estimation sample}} \\
$N$: Baseline / CPS sample & 6,295 & 6,295 & 6,295 & 6,108 & 6,295 \\
$N$: Dingel--Neiman sample & 6,253 & 6,253 & 6,253 & 6,067 & 6,253 \\
Occupations: Baseline / CPS & 450 & 450 & 450 & 450 & 450 \\
Occupations: Dingel--Neiman sample & 447 & 447 & 447 & 447 & 447 \\
\bottomrule
\end{tabularx}
\endgroup
\par\vspace{0.40em}\noindent\begin{minipage}{\linewidth}\notesize \textit{Notes:} The first row of each specification is the full-period pooled-step estimate; the indented second row is the incremental change at 2025Q1 from a separate era-split model (not the total coefficient from 2025Q1 onward). Each estimate corresponds to a one-standard-deviation AI-exposure change with OCC2010-clustered standard errors in parentheses. Logged outcomes are displayed in log points (coefficients multiplied by 100); unemployment and shares in percentage points; hours in hours. CPS controls use contemporaneous occupation-quarter shares of paid teleworkers or hours worked remotely, standardized within the estimation sample. The standardized Dingel--Neiman work-from-home feasibility index enters interacted with an indicator for 2025Q1 onward; its time-invariant level is absorbed by occupation fixed effects. The joint model includes the quarterly telework share and the Dingel--Neiman interaction, not the remote-hours alternative. Occupation and calendar-quarter fixed effects; fixed 2019 occupation-employment weights. The CPS-available baseline equals the unrestricted aggregate baseline and is omitted. Observation and occupation counts are outcome-specific; Dingel--Neiman-matched specifications can exclude occupations without that index. \textsuperscript{*} $p<0.10$, \textsuperscript{**} $p<0.05$, \textsuperscript{***} $p<0.01$.\end{minipage}
\end{table}

\clearpage
\subsection{The 2018 Occupation Classification}
\label{app:codes2018}

\begin{table}[htbp]
\centering
\caption{Three-Year First Differences: Headline and Placebo Comparisons (2018 Census Codes)}
\label{tab:cell_fd_main_14}
\fontsize{8.3}{9.8}\selectfont
\setlength{\tabcolsep}{2.15pt}
\renewcommand{\arraystretch}{1.11}
\begin{adjustbox}{max width=\linewidth}
\begin{tabularx}{\linewidth}{@{}>{\raggedright\arraybackslash}p{0.225\linewidth}*{5}{>{\centering\arraybackslash}X}@{}}
\toprule
 & \shortstack{Log\\employment} & \shortstack{Unemployment\\(pp)} & Hours & \shortstack{Log real hourly\\earnings} & \shortstack{Part-time share\\(pp)} \\
\midrule
\addlinespace[0.16em]
\multicolumn{6}{@{}l}{\textbf{\textit{A. Main comparison: occupation-quarter cells}}} \\
\multicolumn{6}{@{}l}{\parbox[t]{0.94\linewidth}{\fontsize{7.4}{8.8}\selectfont\itshape 2026Q1 vs. 2023Q1 and 2026Q2 vs. 2023Q2}} \\[0.14em]
S4--S1\hspace{0.55em} & -1.49 (1.47) & 0.08 (0.12) & 0.17\textsuperscript{**} (0.08) & 1.98\textsuperscript{**} (0.77) & -0.63\textsuperscript{**} (0.26) \\
\(R^{2}\) & 0.04 & 0.01 & 0.02 & 0.03 & 0.03 \\
\cmidrule(lr){1-6}
S3--S1\hspace{0.55em} & -1.31 (1.58) & 0.07 (0.13) & 0.18\textsuperscript{**} (0.08) & 2.17\textsuperscript{***} (0.83) & -0.58\textsuperscript{**} (0.27) \\
\(R^{2}\) & 0.04 & 0.01 & 0.02 & 0.03 & 0.03 \\
\cmidrule(lr){1-6}
Mean outcome change & -1.30 & 0.60 & -0.22 & 1.67 & 0.65 \\
\(N\) & 1,024 & 1,024 & 1,024 & 961 & 1,024 \\
\specialrule{0.075em}{0.20em}{0.14em}
\multicolumn{6}{@{}l}{\textbf{\textit{B. Main comparison: six-month averages}}} \\
\multicolumn{6}{@{}l}{\parbox[t]{0.94\linewidth}{\fontsize{7.4}{8.8}\selectfont\itshape 2025Q4--2026Q1 vs. 2022Q4--2023Q1; 2026Q1--Q2 vs. 2023Q1--Q2}} \\[0.14em]
S4--S1\hspace{0.55em} & -1.54 (1.57) & 0.21\textsuperscript{*} (0.12) & 0.13\textsuperscript{*} (0.07) & 0.53 (0.60) & -0.56\textsuperscript{**} (0.23) \\
\(R^{2}\) & 0.05 & 0.02 & 0.02 & 0.03 & 0.03 \\
\cmidrule(lr){1-6}
S3--S1\hspace{0.55em} & -1.24 (1.67) & 0.18 (0.13) & 0.13\textsuperscript{*} (0.08) & 0.58 (0.65) & -0.47\textsuperscript{**} (0.23) \\
\(R^{2}\) & 0.05 & 0.01 & 0.02 & 0.03 & 0.03 \\
\cmidrule(lr){1-6}
Mean outcome change & -0.23 & 0.64 & -0.23 & 1.59 & 0.66 \\
\(N\) & 1,022 & 1,022 & 1,022 & 1,006 & 1,022 \\
\bottomrule
\end{tabularx}\end{adjustbox}
\par\vspace{0.50em}\noindent
\begin{minipage}{\linewidth}
\notesize \textit{Notes:} Coefficients per one standard deviation of the named occupation-level exposure change, with standard errors clustered by occupation. S4--S1 is the cumulative change from pre-generative AI to agents; S3--S1 runs through reasoning models. Outcomes are differences over the stated horizons. All specifications use base-period occupation employment weights, twelve base-period full-occupation workforce-composition controls, and later-quarter fixed effects. Log outcomes and indicator shares are multiplied by 100 (log points and percentage points); hours are unscaled. Outcome means are base-employment-weighted means of outcome differences in the relevant model sample (displayed units); the $R^{2}$ rows correspond to the coefficient above each row; $N$ is the number of occupation-window observations. Clean pre-2020 placebos are unavailable for the native 2018-code panel; they appear in the OCC2010 results. Panels A--B report the first-half-2026 headline at quarterly and rolling six-month units. The ``Mean outcome change'' row is the base-period-employment-weighted sample mean of the first-differenced dependent variable (the later-period outcome minus the earlier-period outcome), not the mean level of the outcome. A negative entry indicates an average decline over the comparison window; logged outcomes and shares use the display scaling described above. \mbox{\textsuperscript{*}\,$p<0.10$}, \mbox{\textsuperscript{**}\,$p<0.05$}, \mbox{\textsuperscript{***}\,$p<0.01$}.
\end{minipage}

\end{table}

\begin{table}[htbp]
\centering
\caption{Three-Year First Differences: Age and Education Separately (2018 Census Codes)}
\label{tab:cell_fd_main_17}
\fontsize{8.3}{9.8}\selectfont
\setlength{\tabcolsep}{2.15pt}
\renewcommand{\arraystretch}{1.11}
\begin{adjustbox}{max width=\linewidth}
\begin{tabularx}{\linewidth}{@{}>{\raggedright\arraybackslash}p{0.225\linewidth}*{5}{>{\centering\arraybackslash}X}@{}}
\toprule
 & \shortstack{Log\\employment} & \shortstack{Unemployment\\(pp)} & Hours & \shortstack{Log real hourly\\earnings} & \shortstack{Part-time share\\(pp)} \\
\midrule
\addlinespace[0.16em]
\multicolumn{6}{@{}l}{\textbf{\textit{A. By age}}} \\[-0.06em]
Under 30 & -7.32\textsuperscript{***} (2.00) & -0.13 (0.28) & 0.11 (0.21) & 2.74 (2.18) & 0.40 (0.66) \\
\(R^{2}\) & 0.07 & 0.01 & 0.02 & 0.05 & 0.02 \\
\cmidrule(lr){1-6}
Mean outcome change & -3.96 & 0.82 & -0.46 & 2.48 & 1.69 \\
\(N\) & 934 & 943 & 932 & 649 & 934 \\
\cmidrule(lr){1-6}
30 to 49 & 0.96 (1.62) & 0.23 (0.16) & 0.26\textsuperscript{***} (0.10) & 2.58\textsuperscript{**} (1.09) & -0.96\textsuperscript{***} (0.29) \\
\(R^{2}\) & 0.14 & 0.02 & 0.02 & 0.03 & 0.02 \\
\cmidrule(lr){1-6}
Mean outcome change & -1.39 & 0.66 & -0.16 & 1.03 & 0.35 \\
\(N\) & 1,014 & 1,014 & 1,014 & 853 & 1,014 \\
\cmidrule(lr){1-6}
50 to 64 & -0.25 (1.91) & 0.14 (0.19) & -0.07 (0.12) & -1.23 (1.73) & -0.75\textsuperscript{*} (0.45) \\
\(R^{2}\) & 0.14 & 0.01 & 0.02 & 0.03 & 0.02 \\
\cmidrule(lr){1-6}
Mean outcome change & -7.44 & 0.53 & -0.10 & 0.17 & 0.27 \\
\(N\) & 986 & 988 & 986 & 739 & 986 \\
\specialrule{0.075em}{0.20em}{0.14em}
\multicolumn{6}{@{}l}{\textbf{\textit{B. By education}}} \\[-0.06em]
Less than a bachelor's degree & -4.89\textsuperscript{***} (1.67) & -0.07 (0.18) & 0.13 (0.10) & 0.22 (1.06) & -0.12 (0.35) \\
\(R^{2}\) & 0.05 & 0.01 & 0.02 & 0.02 & 0.03 \\
\cmidrule(lr){1-6}
Mean outcome change & -4.40 & 0.52 & -0.22 & 2.45 & 1.00 \\
\(N\) & 957 & 957 & 957 & 814 & 957 \\
\cmidrule(lr){1-6}
Bachelor's degree or more & 3.63\textsuperscript{**} (1.78) & 0.40\textsuperscript{*} (0.20) & 0.15 (0.13) & 2.35\textsuperscript{**} (1.18) & -0.91\textsuperscript{**} (0.39) \\
\(R^{2}\) & 0.05 & 0.02 & 0.01 & 0.03 & 0.03 \\
\cmidrule(lr){1-6}
Mean outcome change & 0.50 & 0.76 & -0.23 & -0.22 & 0.27 \\
\(N\) & 926 & 928 & 926 & 723 & 926 \\
\bottomrule
\end{tabularx}\end{adjustbox}
\par\vspace{0.50em}\noindent
\begin{minipage}{\linewidth}
\notesize \textit{Notes:} Three-year differences stack 2026Q1 versus 2023Q1 and 2026Q2 versus 2023Q2. Each group coefficient is obtained from its own occupation-by-group-by-quarter regression using that group's base-period employment weights and the twelve full-occupation baseline workforce-composition controls. Regressions use S4--S1 exposure standardized at the aggregate occupation level. Log outcomes and shares are multiplied by 100; hours are in hours. $N$ is the group-outcome estimation sample. The ``Mean outcome change'' row is the within-group, base-period-employment-weighted mean of the first-differenced dependent variable (the later-period outcome minus the earlier-period outcome), not the mean outcome level. Logged outcomes and shares use the display scaling described above. \mbox{\textsuperscript{*}\,$p<0.10$}, \mbox{\textsuperscript{**}\,$p<0.05$}, \mbox{\textsuperscript{***}\,$p<0.01$}.
\end{minipage}

\end{table}

\section{Placebo Windows}
\label{app:placebo}

\begin{figure}[htbp]
\centering
\includegraphics[width=0.9\linewidth]{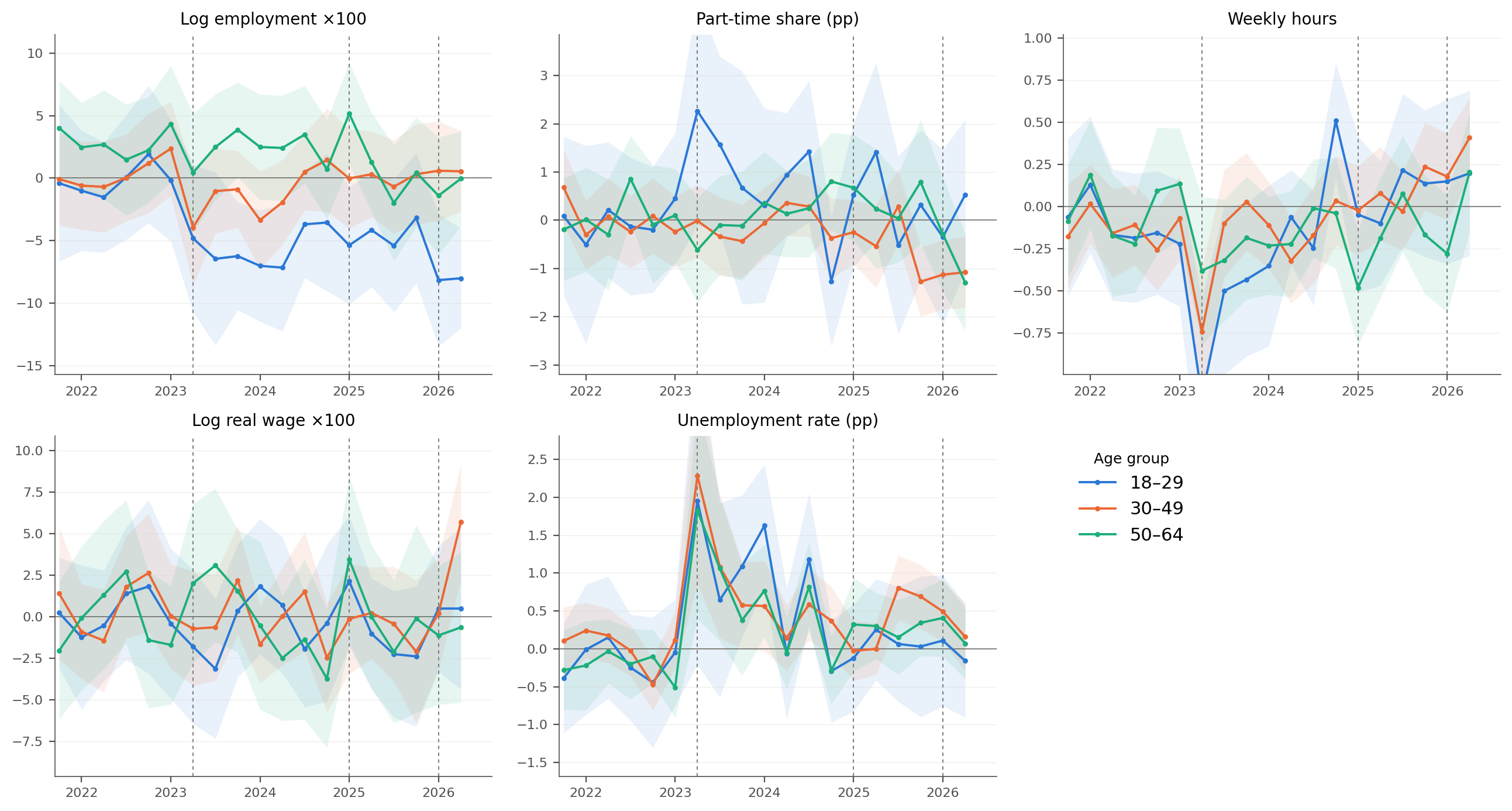}
\caption{Three-year differences by end quarter and age group}
\label{fig:fd-groups-byquarter}
\begin{minipage}{0.92\textwidth}
\notesize
Notes: Each point is a separate three-year difference of Eq.~(\ref{eq:group}) ending in the quarter shown, by age group, with 95 percent bands. The placebo windows end between 2021Q4 and 2022Q3 and compare each quarter with the same quarter three years earlier; the windows of the generative AI era end in 2025Q1 or later and have base periods in 2022 or later.
\end{minipage}
\end{figure}

This appendix reports the placebo estimates behind the figures of the paper.
The placebo window compares 2021Q4 through 2022Q3 with the same quarters three years earlier, so that both the base and the end periods precede the release of ChatGPT and the base period precedes the pandemic.
The placebo estimates for all workers are in Table~\ref{tab:headline} of the paper and Table~\ref{tab:cell_fd_main_1}.
Table~\ref{tab:cell_fd_main_11} reports the placebo estimates by age and by education, which Figures~\ref{fig:who} and~\ref{fig:timing} of the paper plot as the hollow markers and the first row.
Table~\ref{tab:cell_fd_main_7} reports the placebo estimates for the six age-by-education cells of Table~\ref{tab:cell_fd_main_5}.
Figure~\ref{fig:fd-groups-byquarter} plots the single-window three-year estimates of Eq.~(\ref{eq:fdwindow}) by end quarter for the three age groups and the five main outcomes, with the placebo windows at the left, and extends Figure~\ref{fig:byquarter} of the paper.

\begin{table}[htbp]
\centering
\caption{Three-year differences in the main outcomes, by age and by education: pre-AI placebo window}
\label{tab:cell_fd_main_11}
\fontsize{8.3}{9.8}\selectfont
\setlength{\tabcolsep}{2.15pt}
\renewcommand{\arraystretch}{1.11}
\begin{adjustbox}{max width=\linewidth}
\begin{tabularx}{\linewidth}{@{}>{\raggedright\arraybackslash}p{0.225\linewidth}*{5}{>{\centering\arraybackslash}X}@{}}
\toprule
 & \shortstack{Log\\employment} & \shortstack{Unemployment\\(pp)} & Hours & \shortstack{Log real hourly\\earnings} & \shortstack{Part-time share\\(pp)} \\
\midrule
\addlinespace[0.16em]
\multicolumn{6}{@{}l}{\textbf{\textit{A. By age}}} \\[-0.06em]
Under 30 & -0.70 (2.06) & -0.10 (0.26) & -0.07 (0.12) & -0.06 (1.06) & -0.11 (0.55) \\
\(R^{2}\) & 0.06 & 0.01 & 0.02 & 0.02 & 0.02 \\
\cmidrule(lr){1-6}
Mean outcome change & -10.32 & 0.50 & -0.15 & 5.82 & -0.38 \\
\(N\) & 1,699 & 1,706 & 1,698 & 1,324 & 1,699 \\
\cmidrule(lr){1-6}
30 to 49 & -0.26 (1.41) & 0.12 (0.10) & -0.11 (0.08) & 0.18 (0.86) & 0.04 (0.21) \\
\(R^{2}\) & 0.10 & 0.03 & 0.04 & 0.02 & 0.03 \\
\cmidrule(lr){1-6}
Mean outcome change & -1.05 & 0.38 & -0.44 & 2.19 & -0.23 \\
\(N\) & 1,775 & 1,776 & 1,773 & 1,593 & 1,775 \\
\cmidrule(lr){1-6}
50 to 64 & 2.64\textsuperscript{*} (1.46) & -0.18 (0.15) & -0.07 (0.09) & 0.64 (1.09) & 0.08 (0.28) \\
\(R^{2}\) & 0.14 & 0.02 & 0.03 & 0.02 & 0.02 \\
\cmidrule(lr){1-6}
Mean outcome change & -3.99 & 0.05 & -0.27 & 1.46 & -0.56 \\
\(N\) & 1,740 & 1,742 & 1,735 & 1,418 & 1,740 \\
\specialrule{0.075em}{0.20em}{0.14em}
\multicolumn{6}{@{}l}{\textbf{\textit{B. By education}}} \\[-0.06em]
Less than a bachelor's degree & -0.61 (1.36) & 0.11 (0.15) & -0.10 (0.08) & 0.48 (0.79) & 0.10 (0.26) \\
\(R^{2}\) & 0.03 & 0.02 & 0.03 & 0.02 & 0.04 \\
\cmidrule(lr){1-6}
Mean outcome change & -6.76 & 0.45 & -0.19 & 2.73 & -0.39 \\
\(N\) & 1,685 & 1,686 & 1,684 & 1,536 & 1,685 \\
\cmidrule(lr){1-6}
Bachelor's degree or more & 2.28 (1.71) & -0.28\textsuperscript{***} (0.11) & 0.01 (0.09) & 0.56 (0.98) & -0.26 (0.30) \\
\(R^{2}\) & 0.04 & 0.02 & 0.05 & 0.02 & 0.02 \\
\cmidrule(lr){1-6}
Mean outcome change & 2.50 & -0.00 & -0.48 & 2.36 & -0.42 \\
\(N\) & 1,644 & 1,645 & 1,640 & 1,322 & 1,644 \\
\bottomrule
\end{tabularx}\end{adjustbox}
\par\vspace{0.50em}\noindent
\begin{minipage}{\linewidth}
\notesize \textit{Notes:} The three-year historical placebo compares 2021Q4--2022Q3 with 2018Q4--2019Q3. Each group coefficient is obtained from its own occupation-by-group-by-quarter regression using that group's base-period employment weights and the twelve full-occupation baseline workforce-composition controls. Regressions use S4--S1 exposure standardized at the aggregate occupation level. Log outcomes and shares are multiplied by 100; hours are in hours. $N$ is the group-outcome estimation sample. The ``Mean outcome change'' row is the within-group, base-period-employment-weighted mean of the first-differenced dependent variable (the later-period outcome minus the earlier-period outcome), not the mean outcome level. Logged outcomes and shares use the display scaling described above. \mbox{\textsuperscript{*}\,$p<0.10$}, \mbox{\textsuperscript{**}\,$p<0.05$}, \mbox{\textsuperscript{***}\,$p<0.01$}.
\end{minipage}

\end{table}

\begin{table}[htbp]
\centering
\caption{Three-year differences in the main outcomes, by age and education jointly (Placebo Window)}
\label{tab:cell_fd_main_7}
\fontsize{7.7}{9.0}\selectfont
\setlength{\tabcolsep}{2pt}
\renewcommand{\arraystretch}{1.04}
\begin{adjustbox}{max width=\linewidth}
\begin{tabularx}{\linewidth}{@{}>{\raggedright\arraybackslash}p{0.236\linewidth}*{5}{>{\centering\arraybackslash}X}@{}}
\toprule
 & \shortstack{Log\\employment} & \shortstack{Unemployment\\(pp)} & Hours & \shortstack{Log real hourly\\earnings} & \shortstack{Part-time share\\(pp)} \\
\midrule
\multicolumn{6}{@{}l}{\parbox[t]{0.97\linewidth}{\itshape 2021Q4 vs. 2018Q4 through 2022Q3 vs. 2019Q3}} \\[0.2em]
Under 30, less than a bachelor's degree & -1.34 (2.27) & -0.01 (0.31) & -0.05 (0.16) & 0.56 (1.06) & 0.16 (0.72) \\
\hspace{0.5em}$R^2$ & 0.06 & 0.02 & 0.02 & 0.03 & 0.01 \\
\hspace{0.5em}Mean outcome change & -12.07 & 0.53 & -0.04 & 5.66 & -0.60 \\
\hspace{0.5em}$N$ & 1,488 & 1,495 & 1,483 & 1,021 & 1,488 \\
\cmidrule(lr){1-6}
Under 30, bachelor's degree or more & -1.39 (3.13) & -0.24 (0.31) & -0.00 (0.17) & -3.10 (2.41) & -0.33 (0.61) \\
\hspace{0.5em}$R^2$ & 0.06 & 0.02 & 0.01 & 0.03 & 0.03 \\
\hspace{0.5em}Mean outcome change & -13.62 & 0.54 & -0.41 & 4.75 & 0.13 \\
\hspace{0.5em}$N$ & 1,232 & 1,236 & 1,229 & 793 & 1,232 \\
\cmidrule(lr){1-6}
\addlinespace[0.30em]
30 to 49, less than a bachelor's degree & -1.64 (1.67) & 0.13 (0.16) & -0.18\textsuperscript{*} (0.10) & -0.70 (1.27) & 0.09 (0.31) \\
\hspace{0.5em}$R^2$ & 0.08 & 0.03 & 0.02 & 0.02 & 0.03 \\
\hspace{0.5em}Mean outcome change & -6.82 & 0.63 & -0.35 & 1.80 & 0.04 \\
\hspace{0.5em}$N$ & 1,620 & 1,624 & 1,618 & 1,300 & 1,620 \\
\cmidrule(lr){1-6}
30 to 49, bachelor's degree or more & 1.83 (1.98) & -0.12 (0.14) & 0.06 (0.11) & 0.56 (1.26) & -0.26 (0.34) \\
\hspace{0.5em}$R^2$ & 0.04 & 0.02 & 0.05 & 0.03 & 0.02 \\
\hspace{0.5em}Mean outcome change & 2.91 & 0.13 & -0.55 & 1.59 & -0.56 \\
\hspace{0.5em}$N$ & 1,516 & 1,520 & 1,509 & 1,138 & 1,516 \\
\cmidrule(lr){1-6}
\addlinespace[0.30em]
50 to 64, less than a bachelor's degree & 1.81 (1.78) & 0.11 (0.20) & -0.02 (0.12) & 0.90 (1.40) & 0.10 (0.36) \\
\hspace{0.5em}$R^2$ & 0.07 & 0.01 & 0.02 & 0.02 & 0.02 \\
\hspace{0.5em}Mean outcome change & -10.00 & 0.33 & -0.18 & 1.41 & -0.58 \\
\hspace{0.5em}$N$ & 1,574 & 1,577 & 1,568 & 1,138 & 1,574 \\
\cmidrule(lr){1-6}
50 to 64, bachelor's degree or more & 6.27\textsuperscript{***} (2.13) & -0.55\textsuperscript{**} (0.22) & -0.06 (0.17) & 0.99 (2.42) & -0.24 (0.51) \\
\hspace{0.5em}$R^2$ & 0.09 & 0.03 & 0.03 & 0.04 & 0.01 \\
\hspace{0.5em}Mean outcome change & 0.89 & -0.26 & -0.47 & -0.29 & -0.44 \\
\hspace{0.5em}$N$ & 1,394 & 1,398 & 1,389 & 901 & 1,394 \\
\bottomrule
\end{tabularx}\end{adjustbox}
\par\vspace{0.45em}\noindent
\begin{minipage}{\linewidth}
\notesize \textit{Notes:} Each row reports a separate occupation-by-age-by-education-group regression for the indicated three-year first differences. Exposure changes are standardized at the aggregate occupation level; S4--S1 is the cumulative pre-generative-to-agent change and S3--S1 the cumulative pre-generative-to-reasoning change. Models use group-specific base-period employment weights, twelve full-occupation baseline workforce-composition controls, later-quarter fixed effects, and occupation-clustered standard errors. Coefficient standard errors appear in parentheses. The mean outcome change is the group-weighted mean of the first-differenced dependent variable in that model sample, not an outcome level; $N$ is occupation-window observations. Log outcomes and shares are displayed in log points and percentage points; hours are in hours. Second-job hours are conditional on holding a second job. Significance: \textsuperscript{*}$p<0.10$, \textsuperscript{**}$p<0.05$, \textsuperscript{***}$p<0.01$.
The placebo uses pre-pandemic base quarters and 2021--2022 later quarters; it is pre-AI but not wholly pre-pandemic.
\end{minipage}
\end{table}

\end{document}